\newif\iffigs\figstrue
\documentclass[10pt,a4paper]{article}
\usepackage{latexsym,amssymb,lscape,graphics,setspace}
\usepackage{amsmath}
\usepackage{graphicx}        % pacchetto grafico standard LaTeX  per l'inclusione di file immagine
\usepackage{longtable}
\usepackage[vcentermath]{youngtab}
\usepackage{multirow}
\usepackage{booktabs}
\usepackage{multirow}
\usepackage{color}
\usepackage{slashed,epsfig}
\usepackage{amsfonts}
\usepackage{cite}
\usepackage{verbatim}
\usepackage{colortbl}
\usepackage[table]{xcolor}
\usepackage{hyperref}       % questo pacchetto e'  quello che produce un ipertesto
\usepackage{array}
\usepackage{fancyhdr}
\usepackage{tikz}
\usepackage{multicol}
\usepackage{lscape}
\graphicspath{{images/}}
\newcommand{\mathsym}[1]{{}}

\newtheorem{definizione}{Definition}[section]
\newtheorem{teorema}{Theorem}[section]

\newcommand{\bd}{\begin{definizione}}
\newcommand{\ed}{\end{definizione}}

\def\IC{\relax\,\hbox{$\inbar\kern-.3em{\rm C}$}}
\def\IG{\relax\,\hbox{$\inbar\kern-.3em{\rm G}$}}
\def\IB{\relax{\rm I\kern-.18em B}}
\def\ID{\relax{\rm I\kern-.18em D}}
\def\IL{\relax{\rm I\kern-.18em L}}
\def\IF{\relax{\rm I\kern-.18em F}}
\def\IH{\relax{\rm I\kern-.18em H}}
\def\II{\relax{\rm I\kern-.17em I}}
\def\IN{\relax{\rm I\kern-.18em N}}
\def\IP{\relax{\rm I\kern-.18em P}}
\def\IQ{\relax\,\hbox{$\inbar\kern-.3em{\rm Q}$}}
\def\bfzero{\relax\,\hbox{$\inbar\kern-.3em{\rm 0}$}}
\def\IK{\relax{\rm I\kern-.18em K}}
\def\IG{\relax\,\hbox{$\inbar\kern-.3em{\rm G}$}}
 \font\cmss=cmss10 \font\cmsss=cmss10 at 7pt
\def\IR{\relax{\rm I\kern-.18em R}}
\def\ZZ{\relax\ifmmode\mathchoice
{\hbox{\cmss Z\kern-.4em Z}}{\hbox{\cmss Z\kern-.4em Z}}
{\lower.9pt\hbox{\cmsss Z\kern-.4em Z}} {\lower1.2pt\hbox{\cmsss
Z\kern-.4em Z}}\else{\cmss Z\kern-.4em Z}\fi}
\def\bfone{\relax{\rm 1\kern-.35em 1}}

\def\inbar{\vrule height1.5ex width.4pt depth0pt}
\def\bfzero{\relax{\rm I\kern-.18em 0}}
\def\bfone{\relax{\rm 1\kern-.35em 1}}

\DeclareFontFamily{U}{rsf}{} \DeclareFontShape{U}{rsf}{m}{n}{
  <5> <6> rsfs5 <7> <8> <9> rsfs7 <10-> rsfs10}{}
\DeclareMathAlphabet\Scr{U}{rsf}{m}{n}

\newcommand{\ft}[2]{{\textstyle\frac{#1}{#2}}}
\def\tilde{\widetilde}

\def\1bar{1\hskip -.275cm -}
\def\2bar{2\hskip -.275cm -}
\def\3bar{3\hskip -.275cm -}

\newsavebox{\uuunit}
\sbox{\uuunit}
{\setlength{\unitlength}{0.825em}
\begin{picture}(0.6,0.7)
\thinlines
\put(0,0){\line(1,0){0.5}}
\put(0.15,0){\line(0,1){0.7}}
\put(0.35,0){\line(0,1){0.8}}
\multiput(0.3,0.8)(-0.04,-0.02){10}{\rule{0.5pt}{0.5pt}}
\end {picture}}

\makeatletter \@addtoreset{equation}{section} \makeatother

\def\bfone{\relax{\rm 1\kern-.35em 1}}

\def\bfone{\relax{\rm 1\kern-.35em 1}}
\font\cmss=cmss10 \font\cmsss=cmss10 at 7pt

\begin{document}
\begin{titlepage}
\begin{center}
\vskip 0.2cm
%%%%%%%%%%%%%%%%%%%%%%%%%%%%%%%%%%%%%%%%%%%%%%%%%%%%%%%%%%%%%%%%%%%%
%%%%%%%%%%%%%%%%%%%%%%%%%%%%%%%%%%%%%%%%%%%%%%%%%%%%%%%%%%%%%%%%%%%%
%%%%%%%%%%%%%%%%%%%%%%%%%%%%%%%%%%%%%%%%%%%%%%%%%%%%%%%%%%%%%%%%%%%%
{{\large {\sc   Axial Symmetric Navier Stokes equations \\
\vskip 0.2 cm and
 the Beltrami/anti Beltrami spectrum \\
 \vskip 0.2 cm in view of Physics Informed Neural Networks
  ${}^\dagger$}} }\\
 \vskip 1cm {\sc Pietro Fr\'e\,$^{a,b}$} \vskip 0.5cm
\smallskip
{\sl \small \frenchspacing ${}^a\,$ {\tt Emeritus Professor of}  Dipartimento di Fisica, 
Universit\`a di Torino, Via P. Giuria 1, I-10125 Torino, Italy \\[2pt]
${}^{b}\,${\tt Senior Consultant of } Additati\&Partners Consulting s.r.l, 
Via Filippo Pacini 36, I-51100 Pistoia, Italy \\[2pt]

E-mail:  {\tt pietro.fre@unito.it } }

\begin{abstract}
In this paper, I further continue an investigation on Beltrami Flows began in 2015 with A. Sorin and amply revised and developed in 2022 with M. Trigiante. Instead of a compact $3$-torus $T^3=\mathbb{R}^3/\Lambda$ where $\Lambda$ is a crystallographic lattice, as done in previous work, here I considered flows confined in a cylinder with identified opposite bases. In this topology I considered axial symmetric flows and found a complete basis of axial symmetric harmonic $1$-forms that, for each energy level, decomposes  into six components: two Beltrami, two anti-Beltrami and two closed forms. These objects, that are written in terms of trigonometric and Bessel functions, constitute a function basis for an  $L^2$ space of axial symmetric flows. I have presented a general scheme for the search of axial symmetric solutions of Navier Stokes equation by reducing the latter to an hierachy of quadratic relations on the development coefficients of the flow in the above described functional basis. It is proposed that the coefficients can be determined by means of a Physics Informed like Neural Network optimization recursive algorithm. Indeed the present paper provides the theoretical foundations for such a algorithmic construction that is planned for a future publication.
\end{abstract}
\vfill
\end{center}
\noindent \parbox{175mm}{\hrulefill}
\par
%\begin{align}
${}^\dagger$ P.G. Fr\'e acknowledges support by the Company \textit{Additati\&Partners 
Consulting s.r.l} during the development of the present research. 
%\end{align}
\\[5pt]
%\end{center}
\end{titlepage}
{\small \tableofcontents} \noindent {}
\newpage
\section{Introduction} 
\label{fluidostuff}
In a series of recent papers \cite{pgtstheory,TSnaviga,naviga,tassellandum}, together with my coauthors I have contributed to establish the new paradigm of Cartan Neural Networks (CaNN) based on the selection  of  the non-compact symmetric spaces $\mathrm{U/H}$ as the correct mathematical model of the hidden layers. In a forthcoming paper \cite{geotermico} that will appear within  the next couple of weeks and it is written  with my historical  coauthors and friends Alexander Sorin and Mario Trigiante, we clarify the issue of \textit{statistical distributions} and \textit{generalized temperatures} on the very same non-compact symmetric spaces that provide the mathematical modeling  of Cartan Neural Networks performing \textit{tasks} like that of \textit{classification}. There is however another direction in Machine Learning that goes under the name of \textbf{Physics Informed Neural Networks}. With such a name one refers to \textbf{algorithms based on an iterative updating of parameters that utilize the gradient of convenient \textit{loss functions}} and are deputed \textit{to find solutions of partial differential equations \textit{PDE}.s}, typically those non-linear of Mathematical Physics. In such a context one of the most famous and most important set of partial differential equations are the \textit{Navier Stokes equations} of \textit{Fluid Dynamics}. Due to their ubiquitousness in many technological sectors  and applied sciences (climatology, oceanology just to mention a couple of notable examples)  and to their hard type of non linearity, from the mathematical view-point, a huge amount of software has been constructed to deal with them so that the theoretical discipline of \textit{Fluid Dynamics} has been substantially replaced by so named \textit{CFD}, namely \textit{Computer Fluid Dynamics}. In CFD one typically gives up understanding the \textit{rational} behind phenomena and is happy with numerical solutions of NS equations, consistent with a given set of boundary conditions, that are obtained with a discretization of the PDE.s  These latter, once discretized,  cannot detect criticalities or general underlying  structures. 
\par
Yet the mathematical study of NS equations has continued through the years and some milestones were settled in particular by the great mathematician \textit{Vladimir Arnol'd}. One of the most relevant phenomena in the dynamics of fluids is the on-set of chaotic turbulence which is mathematically described as \textit{Lagrangian Chaos}: the stream-lines, \textit{i.e.} the actual trajectories of fluid elements that correspond to integral curves of the velocity field $\mathbf{U}(x,y,z)$, become capriciously disordered, filling the full available space in an almost stochastic way, while infinitesimal fluid elements that start at two very close points follow quite different paths and end up in quite distant from each other destinations after a finite amount of time.  Arnol'd, focusing on the case of flows confined in a compact region, modeled as a Riemannian three-manifold $(\mathcal{M},g)$ established his famous theorem, recalled in this paper in section (\ref{arnaldoseczia}). Expressed in simple words that theorem states that a necessary condition for the possible on-set of chaos in a certain region $\mathcal{D}$ of space is that in such a region the velocity field $\mathbf{U}$ should be a \textit{Beltrami} or \textit{anti-Beltrami} vector field, namely proportional with positive or negative sign  to its own rotor $\text{rot}\mathbf{U}$. Beltrami vector fields in three dimensions are naturally related with the concept of a contact-stucture defined by the existence of a contact $1$-form $\Omega$ of which the Beltrami field is the Reeb field (see below  section \ref{belatramocontatto}). Relying on this geometrical premises, in the two papers \cite{Fre:2015mla,Fr__2023} authored, in two different combinations, by the same triplet of authors of the above quoted paper \cite{geotermico}, we addressed the general construction of Beltrami fields on a compact three-manifold with the topology of a three torus $\mathrm{T}^{3} \sim\mathbb{ S}^1\times \mathbb{S}^1\times \mathbb{S}^1$. Furthermore we considered three torii realized as:
\begin{equation}\label{cortilecongallina}
  \mathrm{T}^{3}\sim \mathbb{R}^3/\boldsymbol{\Lambda}
\end{equation}
where $\boldsymbol{\Lambda}$ is a regular crystallographic three-dimensional lattice. Essentially this corresponds to impose periodicity conditions  on the solutions both of Navier Stokes and Beltrami equations:
\begin{equation}\label{periodandoquiela}
  \forall \mathbf{x}\in \mathbb{R}^3 \quad \forall \mathbf{n} \in \boldsymbol{\Lambda} \quad : \quad \mathbf{U}(\mathbf{x}+\mathbf{n}) \, = \, \mathbf{U}(\mathbf{x}) 
\end{equation}
\par
Using the point group $\mathfrak{P}_{\boldsymbol{\Lambda}}\subset \mathrm{SO(3)}$ of the lattice and the various translation symmetries inherent to $\boldsymbol{\Lambda}$ we where able to show that all Beltrami fields can be classified by irreducible representations of a big discrete finite group, the \textit{Universal Classifying Group} $\mathfrak{UG}_{\boldsymbol{\Lambda}}$ that is an extension of $\mathfrak{P}_{\boldsymbol{\Lambda}}$ and is an intrinsic property of the chosen lattice.
\par
The Beltrami fields constructed in this way are special linear combinations of the terms appearing in the expansion of a generic vector field in Fourier series of the three coordinates $(x,y,z)$. Hence taking the full Fourier expansion one retrieves all possible Beltrami and anti-Beltrami vector fields that, therefore constitute a basis for the functional space of square summable vector fields with the given boundary conditions. 
\par
Relying on this observation the
authors of \cite{beltraspectra} introduced the Beltrami index of a
stationary Navier Stokes  solution $\mathbf{U}$  by means of the
following formula:
\begin{equation}\label{balengus}
    \beta_{r_k}\left[\mathbf{U}\right] \, = \, \frac{\mid U_{r_k}^+\mid^2 -\mid
    U_{r_k}^-\mid^2}{\mid U_{r_k}^+\mid^2 +\mid
    U_{r_k}^-\mid^2}
\end{equation}
where $r_k$ is a compact notation for a momentum shell, namely for all the vectors in momentum lattice,
with the same norm and $\mid U_{r_k}^+\mid^2$ denotes the norm of the contribution to the vector field $\mathbf{U}$
from that shell, which is composed of Beltrami vectors, while $\mid
U_{r_k}^-\mid^2$ is the norm of the contribution to the same from of anti-Beltrami vectors. These two contribution exhaust the whole thing. The authors of \cite{beltraspectra}
partially proved, partially conjectured from the results of
computer simulations a set of properties of this chiral spectral
index. The word chiral is utilized because a space reflection
$\mathbf{x} \to - \mathbf{x}$ transforms Beltrami fields into anti
Beltrami ones and viceversa. What was not even envisaged in the very
interesting papers  \cite{beltraspectra} and  \cite{ondoso} is the
group theoretical structure underlying Beltrami (and anti Beltrami)
fields appearing in the Fourier expansions of Navier Stokes
solutions. Indeed that group theoretical structure, based on the new
conception of the \textit{Universal Classifying Group} was unveiled
only  in 2015 in \cite{Fre:2015mla}, starting from the observation
by Arnol'd of a hidden roto-translation symmetry in the so named AAA model,
which is isomorphic, as a group, to the relevant point group
$\mathrm{O_{24}}$.
\par
As I show in this paper in section \ref{baseintubata}, in the case of axial symmetric flows in a
tube, the Beltrami index becomes tripartite, since, physically, Beltrami flows are \textit{laevo-rotatory}, the
anti-Beltrami ones are \textit{dextro-rotatory} and there is a third case, mathematically provided by closed $1$-forms, that are irrotational flows.
\par
Notwithstanding the fascination of such a finite group theoretical structure, the Beltrami fields 
emerging from lattice periodic boundary conditions are not very handy as a functional basis for the construction
of generic flows. Furthermore the three torus topology is not the best in order to model flows in a tube, in a pipeline, in an autoclav or in a blood vessel. Much more to the point is the topology of a cylinder with opposite bases identified, which makes the topology of a portion of $\mathbb{R}^3$ bounded by an ordinary torus surface. 
In this case the only relevant point group is not a discrete rather a continuous Lie group $\mathrm{SO(2)}$: the rotation around the central axis of the tube. 
\paragraph{Axial Symmetric Flows} Hence a setup which reduces the  complexity inherent to three dimensional space, yet  provides a basis for understanding and can be the starting point for more complicated solutions when one  introduces, by steps, the dependence on the rotation angle $\phi$, consists of constructing full solutions of steady Navier Stokes equations that are axial symmetric, namely depend effectively only on two coordinates, the radial one $r$ and the longitudinal one $z$.  
\paragraph{A complete functional basis of Beltrami/anti Beltrami/closed hydro $1$-forms} The goal of the research plan which motivates the present paper is that outlined in the previous paragraph, namely constructing by means of a \textit{Physics Informed Neural Network} algorithm,  \textbf{axial symmetric solutions} of \textit{steady Navier Stokes Equations at constant Bernoulli function} (see below section \ref{riscrivitutto} and eq.(\ref{finocchionabiscotta}) for the relevant definitions). Here I do not accomplish such a task, that is postponed to a future publication 
\cite{matthewandme}, rather I prepare the toolbox to implement it in my own favourite setup. Differently from the approach which is standard  in numerical PDE solution algorithms that address the very differential operator and make it discrete  by reduction to finite differences, and differently, also, from existing \textit{Physics Informed Neural Networks} that, within the iterative updating algorithm,  utilize as well the very differential operator corresponding to the PDE.s, I prefer to reduce the problem to an algebraic one by using a suitable infinite orthogonal basis of functions for a properly defined $L^2$ space. The reason is three-fold: 
\begin{description}
  \item[a)] The symmetries one would like to impose on the solution are already built in through the choice of the basis functions.
  \item[b)] Inspecting the outcome of the Neural Network procedure for the coefficients of the solution expressed as a series (obviously truncated to some maximal order) in the chosen function basis, one can hope to detect regularities, intelligible structures, infer some scheme and, possibly, guess the rules underlying the analytic derivation of exact solutions.
   \item[c)] Successively one can break the imposed symmetry (in this case the axial one) by introducing step by stpep perturbations that are naturally organized into irreducible representations of the broken symmetry group. In this case the symmetry is $\mathrm{SO(2)}$ whose irreps are all $2$-dimensional  in a real basiss.     
\end{description}
For Physics Informed Neural Networks this strategy is inspired by the same philosophical posture that inspires the proposal of CaNN.s (\cite{pgtstheory,TSnaviga,naviga,tassellandum,geotermico}): the goal is to make Neural Network outcomes more intelligible  and  therefore useful to conquer some true scientific knowledge, not simply obtain a whatever result of pragmatic interest, possibly also scientific in another field. 
\par
Specifically in the present paper, after turning from standard Cartesian coordinates $(x,y,z)$ to cylindrical ones
$(r,\phi,z)$, I introduce the functional space $L^2_{tube}$ of square summable vector fields properly defined over 
the previously mentioned compact tube $\boldsymbol{\mathfrak{T}}$ (see fig. \ref{tubastro} and eq.(\ref{torone}) ) 
and within it I consider the proper functional subspace $L^2_{axial}\subset L^2_{tube} $ of those square summable vector fields that are \textit{axial symmetric}. In such a functional space I construct a complete basis 
of \textbf{Beltrami/anti-Beltrami/closed hydro $1$-forms}, that are all \textbf{harmonic} and are in the number of six for each energy shell (see section \ref{cimentoarmonia} and eq.s(\ref{frippone})).
\par
The energy shells, that substitute the spherical layers defined and utilized in \cite{Fr__2023}, are determined by two integer numbers $(n,k)$ that single out the Bessel function of $r$, the first, and the trigonometric functions of $z$, the second, entering as building blocks in the expressions of the six \textbf{Beltrami /anti-Beltrami/closed $1$-forms} composing the basis  (see eq.s (\ref{cunegonda})).
\par 
The essential result of this paper is that, schematically, naming $s=\{n,k,i\}$ ($i=1,\dots, 6$) the composite label that uniquely identifies any element of the basis, we can write the expansion of the vector field $\mathbf{U}$ as
\begin{equation}\label{dilatonato}
  \mathbf{U} \, = \, \sum_{s} \, c[s] \, \mathbf{U}_{s} \quad \Leftrightarrow \quad \Omega^{[\mathbf{U}]} 
  \, = \, \sum_{s} \, c[s] \, \Omega^{[\mathbf{U}_s]}
\end{equation}
where $\mathbf{U}_{s}$ are the basis vector fields (Beltrami, or anti Beltrami, or closed) and $c[s]$ their expansion coefficients: in such a basis the steady NS equation becomes:
\begin{equation}\label{biscottino}
 \nu \, \sum_s  \, \varpi^2(s) \, c[s] \, \mathbf{U}_{s} - \sum_{s1} \, \sum_{s2}  \, c[s_1]\, c[s_2] \,
  \left(\sigma(i_2)\,\varpi(s_2)  \,\mathbf{U}_{s_1}\diamond\mathbf{U}_{s_2} \right) \, = \, 0
\end{equation}
where $\nu$ is the viscosity,
\begin{equation}\label{variopi}
  \varpi(s)\, = \, \sqrt{\boldsymbol{\varepsilon}(s)}
\end{equation}
is the square-root of the eigenvalue of the appropriate Laplace Beltrami operator when acting  on the vector field $\mathbf{U}_{s}$ or on its dual $1$-form $\Omega^{[\mathbf{U}_s]}$:
\begin{equation}\label{rovagnati}
  \boldsymbol{\triangle}_{LB} \Omega^{[\mathbf{U}_s]}\, = \, \boldsymbol{\varepsilon}(s) \, \Omega^{[\mathbf{U}_s]} \quad ;
  \quad \boldsymbol{\triangle}_{LB} \,\mathbf{U}_s\, = \, \boldsymbol{\varepsilon}(s) \, \mathbf{U}_s 
\end{equation}
while $\sigma(s)$ is a signature:
\begin{equation}\label{segnando}
  \sigma(i) \, = \,\left\{\begin{array}{rcl}
                            +1 & \text{if} & \text{$s\, \Rightarrow \,$ Beltrami $1$-form} \\
                            -1& \text{if} & \text{$s\, \Rightarrow \,$ anti Beltrami $1$-form} \\
                            0 & \text{if} & \text{$s\, \Rightarrow \,$ closed $1$-form} 
                          \end{array}
   \right.
\end{equation}
whose rational is the following. As I explain in section \ref{cimentoarmonia}, on the flat compact manifold 
$\boldsymbol{\mathfrak{T}}$ (see fig. \ref{tubastro} and eq.(\ref{torone})), the eigen-spectrum of the Laplace-Beltrami operator on $1$-forms depends only on the integer numbers $(n,k)$, so that the
eigenvalue $\boldsymbol{\varepsilon}(s)$ (and hence its square root $\varpi(s)$) depend only on this part of the label $s$. For all levels the degeneracy of $\boldsymbol{\varepsilon}(s)$ is six and the eigenspace decomposes into an orthogonal pair of Beltrami $1$-forms plus an orthogonal pair of anti Beltrami ones plus an orthogonal pair of closed forms. Schematically one has:
\begin{equation}\label{scumatone}
  \text{$(n,k)$ - harmonic eigenspace} \, = \, \text{span}\left\{\text{$\Omega$A}_{\pm,0}(n,k), \, \text{$\Omega$B}_{\pm,0}(n,k)\, \right\}
\end{equation}
where the Beltrami operator $\Game$ introduced and discussed in the sequel (see sect.\ref{cimentoarmonia} and eq.(\ref{pulcinella})) has eigenvalue $\sigma(i)\varpi{(n,k)}$ on the two triplets of $1$-forms:
\begin{eqnarray}\label{purgaionica}
  \Game \, \text{$\Omega$A}_{\pm} \, = \, \pm \, \varpi(n,k) \text{$\Omega$A}_{\pm} & ; & \Game \, \text{$\Omega$A}_{0} \, = \, 0\nonumber\\
   \Game \, \text{$\Omega$B}_{\pm} \, = \, \pm \, \varpi(n,k) \text{$\Omega$B}_{\pm} & ; & \Game \, \text{$\Omega$B}_{0} \, = \, 0
\end{eqnarray}
Indeed since $\Game \sim \star_g \boldsymbol{\mathrm{d}}$ it annihilates all closed $1$-forms. 
\par 
Finally $\mathbf{U}_{s_1}\diamond\mathbf{U}_{s_2}$ denotes the  antisymmetric diamond product of  two vector fields (see eq.(\ref{regredo}) in sect.\ref{gibuti}) that is the appropriate generalization to an arbitrary $3$-manifold of the stantard vector product $\times $ of elementary three-dimensional euclidian geometry.
\par
The diamond product is brought into the game by the non-linear term of the Navier Stokes equation, as I explain in section  \ref{gibuti}. The diamond product of two basis vectors can be expanded in the same basis:
\begin{equation}\label{paleontologo}
  \mathbf{U}_{s_1}\diamond\mathbf{U}_{s_2} \, = \, \sum_{s} \, \mathbb{C}(s_1,s_2|s) \, \mathbf{U}_s
\end{equation}
In (\ref{paleontologo}) the triple index coefficients $\mathbb{C}(s_1,s_2|s)$ are the main structural building blocks of the non linear equation. Indeed selecting each value of $s$ we see that the coefficients $c[s]$ must satisfy the following hierarchy of quadratic equations:
\begin{equation}\label{tuttologo}
   c[s] \,  = \, \frac{1}{\nu \,  \varpi(s)^2 }\,  \sum_{s_1} \, \sum_{s_2} \,\varpi(s_2)  \, c[s_1]\, c[s_2] \,
  \mathbb{C}(s_1,s_2|s) 
  \end{equation}
Eq.s(\ref{tuttologo}) are those that we plan to solve by means of a recursive updating algorithm in the next paper \cite{matthewandme}. An overview of the possible strategy is provided in sect.\ref{ricorrenzanatalizia}. 
%%%%%%%%%%%%%%%%%%%%%%%%%%%%%%%%%%%%%%%%%%%%%%%%%%%%%%%%%%% 
\par
\subsection{How the paper is organized}
In the next section \ref{genteoria}, mainly utilizing  material from the  papers \cite{Fre:2015mla,Fr__2023}, I introduce the geometrical reformulation of Fluid Dynamics and within that contest I recall the notions and properties of Beltrami Flows. 
In section \ref{assialone}  I introduce axial symmetry into Hydrodynamics which is the simplest continuous one a hydroflow can have. Comparison between axial symmetric Beltrami Flows and discrete symmetric ones is one of my and my coauthor's \cite{matthewandme} future goals. 
Section \ref{cimentoarmonia} that I have quoted already several times is the theoretical core of the present paper. The analysis of harmonic $1$-forms with the appropriate boundary conditions for the topology of the periodic tube is the main instrument to find the appropriate functional basis for the axial symmetric functional space $L^2_{tube}$. That basis is established and analyzed in sections \ref{baseintubata} and \ref{algebraNS}.
Finally in section \ref{ricorrenzanatalizia} we briefly analyse the general structure of the problem to be solved by means of a recursive optimazation algorithm: my preferred version of a \textit{Physics Informed Neural Network}.
%%%%%%%%%%%%%%%%%%%%%%%%%%%%%%%%%%%%%%%%%%%%%%%%
\section{Geometric Reformulation of  Navier Stokes Equations}
\label{genteoria}
 Our primary object of study  is the fundamental
equation of classical hydrodynamics of an \textit{ideal,
incompressible, viscous fluid} subject to some external forces,
namely the Navier Stokes equation in three dimensional Euclidian
space $\mathbb{R}^3$, which, in our adopted notation, reads as
follows:
\begin{equation}\label{EulerusEqua}
    \frac{\partial}{\partial t}\mathbf{u}\, + \, \mathbf{u}\cdot \nabla \mathbf{u} \, =
     \, - \, \nabla p  \, + \, \nu \, \Delta \,\mathbf{u} \, + \, \mathbf{f} \quad ;
     \quad \nabla \cdot \mathbf{u} \, = \, 0
\end{equation}
In equation (\ref{EulerusEqua}), $\mathbf{u} \, = \,
\mathbf{u}\left( x \, , \,t\right)$ denotes the local velocity
field, $p(\mathbf{x},t)$ denotes the local pressure field, $\nu$ is
viscosity and $\mathbf{f}$ is the external force, if it is
introduced. The symbol $\Delta\, = \,\sum_{i=1}^3
\,\frac{\partial^2}{\partial x_i^2}$ stands for the standard
laplacian. In vector notation eq. (\ref{EulerusEqua}) takes the
following form:
\begin{equation}\label{EulerusEquaVec}
    \frac{\partial}{\partial t} u^i\, + \, u^j \, \partial_j \, u^i
    \, = \, - \, \partial^i \, p  \, + \, \nu \,\Delta u^i\, + \, f^i \quad ;
    \quad \partial^\ell  \, u_\ell\, = \, 0
\end{equation} and admits some straightforward rewriting that, notwithstanding
the kinder-garden arithmetic involved in its derivation, is at the
basis of several profound and momentous theoretical  developments
which have kept  the community of  dynamical system theorists busy
for already fifty years
\cite{arnoldus,Childress0,Childress,Henon,Dombre,ArnoldBook,Bogoyav0,Bogoyav,
arnoldorussopapero,FFMF,Dynamo,Gilbert,Etnyre2000,Ghrist2007}.
\par
Here we aim at extending to the case where $\nu \neq 0$ previous
results applying to the case of null viscosity, namely to Euler
equation. The scope, however, is more ample since, as we already
anticipated, we introduce a more direct reference to contact
structures.
\par
The core of the exposition in \cite{Fr__2023} is the group-theoretical approach, initiated
in \cite{Fre:2015mla} that brings into the classical field of
mathematical fluid-mechanics, when periodic boundary conditions are imposed, a brand new vision allowing for a more
systematic classification and  algorithmic construction of the so
named \textit{Beltrami flows}, providing new insight into their properties.
Combining the group theoretical classification of Beltrami
(anti-Beltrami) fields and their generalized relation with
\textit{contact structures}  is one of  the promising followup of the new group theoretical formulation of 
periodic Hydro Dynamics. The
other, as we already stressed, is the general scheme for the
construction of exact or approximate solutions of Navier Stokes
equations with prescribed hidden symmetries and calculable Beltrami
spectra.
\par
In the setup of periodic fluid-mechanics as addressed in \cite{Fr__2023} one dominant item
is the notion of \textit{Universal Classifying Group} (UCG) originally introduced in
\cite{Fre:2015mla}. UCG   is an
intrinsic property of the considered crystallographic lattice
$\Lambda$ and of its point group $\mathfrak{P}^{max}_\Lambda$,
which, by definition, is the maximal finite subgroup of
$\mathrm{SO(3)}$ leaving the lattice $\Lambda$ invariant.
\par
The reason why lattices and crystallography were brought into the
study of fluid dynamics is that  in \cite{Fre:2015mla,Fr__2023} the focus was on \textit{hydro-flows}
that are confined within some bounded domain, as it happens in a
large variety of cases of interest for technological applications
like industrial autoclavs, pipelines, thermal machines of various
kind, blood vessels in physiology, liquid helium micro-tubes in
superconducting magnets, chemical reactors with mechanical agitation
systems and so on. The argument goes as follows. Solutions of partial
differential equations (PDE.s) like the NS-equation in
(\ref{EulerusEqua}), that encode the characterizing feature of being
confined to finite regions of space can be obtained essentially by
means of two alternative strategies:
\begin{description}
  \item[A)] By brutally imposing boundary conditions that simulate the
  walls of the chamber, tube, box or whatever else contains the
  flowing fluid. This strategy is the most direct and suitable for
  numerical computer aided integration of the PDE.s but it is hardly
  viable in the search of exact analytic solutions of the same PDE.s with the
  ambition of establishing some rational taxonomy.
  \item[B)] The use of periodic boundary conditions which amounts to
  restricting one's attention to a compact space $\mathcal{M}_3$ without boundary ($\partial \mathcal{M}_3 \, = \, 0$) as
  a mathematical model of the finite volume region of interest.
\end{description}
The use of alternative B) amounts to developing in some suitable
Fourier series some functions (the velocity components) that are not
necessarily periodic but which, on a bounded support, coincide with
periodic functions admitting a Fourier series development.
\par
This being clarified a systematic way of imposing periodic boundary
conditions  is the  identification of the $\mathcal{M}_3$ manifold
with a $\mathrm{T^3}$ torus obtained by modding $\mathbb{R}^3$ with
respect to a three dimensional lattice $\Lambda \subset
\mathbb{R}^3$:
\begin{equation}\label{metricT3}
 \mathcal{M}_3 \, = \,    \mathrm{T}^3_g \, = \, \frac{\mathbb{R}^3}{\Lambda}
\end{equation}
Abstractly the  lattice $\Lambda$ is an abelian infinite group
isomorphic to $\mathbb{Z}\times \mathbb{Z} \times \mathbb{Z}$ which
is embedded in some way into $\mathbb{R}^3$. Using
eq.(\ref{metricT3}) the topological torus
\begin{equation}\label{tritorustop}
    \mathrm{T^3} \simeq \mathbb{S}^1 \times \mathbb{S}^1 \times \mathbb{S}^1
\end{equation}
comes out automatically equipped with a flat constant metric.
Indeed, according with (\ref{metricT3}) the flat Riemaniann space
$\mathrm{T}^3_g$ is defined as the set of equivalence classes with
respect to the following equivalence relation: $ {\mathbf{r}}^\prime
\, \sim \,  {\mathbf{r}} \quad \mbox{iff} \quad  {\mathbf{r}}^\prime
\, - \,  {\mathbf{r}} \, \in \, \Lambda $. The metric $g$ defined on
$\mathbb{R}^3$ is inherited by the quotient space and therefore it
endows  the topological torus (\ref{tritorustop}) with a flat
Riemaniann structure. Seen from another point of view the space of
flat metrics on  $\mathrm{T}^3$ is just the coset manifold
$\mathrm{SL(3,\mathbb{R})}/\mathrm{O(3)}$ encoding all possible
symmetric matrices, alternatively all possible space lattices, each
lattice being spanned by an arbitrary triplet of basis vectors.
\paragraph{\bf Lattices} To make the above statement
precise let us consider the standard $\mathbb{R}^3$ manifold and
introduce a basis of three linearly independent 3-vectors that are
not necessarily orthogonal to each other and of equal length:
\begin{equation}\label{sospirone}
 {\mathbf{w}}_\mu \, \in \, \mathbb{R}^3 \quad \mu \, = \, 1,
\dots ,\, 3
\end{equation}
Any vector in $\mathbb{R}$ can be decomposed along such a basis and
we have:
\begin{equation}\label{xvec}
 {\mathbf{r}} \, = \,  r^\mu {\mathbf{w}}_\mu
\end{equation}
The flat, constant metric on $\mathbb{R}^3$ is defined by:
\begin{equation}\label{gmunu}
g_{\mu\nu} \, = \, \langle   {\mathbf{w}}_\mu \, , \,
 {\mathbf{w}}_\nu \rangle
\end{equation}
where $\langle \, ,\, \rangle$ denotes the standard euclidian scalar
product. The space lattice $\Lambda$ consistent with the metric
(\ref{gmunu}) is the free abelian group (with respect to the sum)
generated by the three basis vectors (\ref{sospirone}), namely:
\begin{equation}\label{reticoloLa}
\mathbb{R}^3 \, \ni \,    {\mathbf{q}}  \, \in \, \Lambda \,
\Leftrightarrow \,  {\mathbf{q}} \, = \, q^\mu \,
 {\mathbf{w}}_\mu \quad \mbox{where} \quad q^\mu \, \in \,
\mathbb{Z}
\end{equation}
\paragraph{\bf Dual lattices} Any time we are given a lattice in the sense of the definition (\ref{reticoloLa}) we obtain
a dual lattice $\Lambda^\star$ defined by the property:
\begin{equation}\label{reticoloLastar}
\mathbb{R}^3 \, \ni \,    {\mathbf{p}}  \, \in \, \Lambda^\star \,
\Leftrightarrow \, \langle  {\mathbf{p}} \, , \,
 {\mathbf{q}}\rangle \, \in \, \mathbb{Z} \quad \forall \,
 {\mathbf{q}}\, \in \, \Lambda
\end{equation}
A basis for the dual lattice is provided by a set of three
\textit{dual vectors} $ {\mathbf{e}}^\mu$ defined by the
relations\footnote{In the sequel for the scalar product of two
vectors we utilize also the equivalent shorter notation $
{\mathbf{a}}\, \cdot  {\mathbf{b}} \, = \, \langle
 {\mathbf{a}}\, \cdot  {\mathbf{b}}\rangle $}:
\begin{equation}\label{dualvecti}
    \langle  {\mathbf{w}}_\mu \, , \,  {\mathbf{e}}^\nu \rangle \, = \, \delta^\nu_\mu
\end{equation}
so that
\begin{equation}\label{pcompi}
\forall \,  {\mathbf{p}} \, \in \, \Lambda^\star \quad
 {\mathbf{p}} \, = \, p_\mu \,  {\mathbf{e}}^\mu \quad
\mbox{where } \quad p_\mu \, \in \, \mathbb{Z}
\end{equation}
According with such a definition it immediately follows that the
original lattice is always a subgroup of the dual lattice and
necessarily a normal one, due to the abelian character of both the
larger and smaller group:
\begin{equation}\label{tarragonese}
    \Lambda \, \subset \, \Lambda^\star
\end{equation}
\subsection{Rewriting equations of  hydrodynamics in a
geometrical set up}
\label{riscrivitutto} Let us then begin by rewriting 
eq.(\ref{EulerusEquaVec}) which is the starting point of the entire
adventure. The first step to be taken in our raising conceptual
ladder is that of promoting the fluid trajectories, defined as the
solutions of the following first order differential
system\footnote{In mathematical hydrodynamics people distinguish two
notions, that of trajectories, which are the solutions of the
differential equations (\ref{streamlines}) and that of streamlines.
Streamlines are the instantaneous curves that at any time $t=t_0$
admit the velocity field $u^{i}(x,t_0)$ as tangent vector.
Introducing a new parameter $\tau$, streamlines at time $t_0$, are
the solutions of the differential system:
\begin{equation}\label{strimotti}
    \frac{d}{d\tau}x^i(\tau) \, = \, u^i(\mathbf{x}(\tau),t_0)
\end{equation}
In the case of steady flows where the velocity field is independent
from time, trajectories and streamlines coincide.}:
\begin{equation}\label{streamlines}
    \frac{d}{dt}x^i(t) \, = \, u^i(x(t),t)
\end{equation}
to smooth maps:
\begin{equation}\label{mappini}
    \mathcal{S} \, : \, \mathbb{R}_t \, \rightarrow \, \mathcal{M}_g
\end{equation}
from the time real line $\mathbb{R}_t $ to a smooth Riemaniann
manifold $\mathcal{M}_g$ endowed with a metric $g$. The classical
case corresponds to $\mathcal{M} \, = \, \mathbb{R}^3$, $g_{ij}(x)
\, = \, \delta_{ij}$, but any other Riemaniann three-manifold might
be used and there exist generalization also to higher dimensions.
Adopting this point of view, the velocity field $\mathbf{u}\left( x
\, , \,t\right)$ is turned into a time evolving vector field on
$\mathcal{M}$ namely into a smooth family of sections of the tangent
bundle $\mathcal{T}\mathcal{M}$:
\begin{equation}\label{Ufildo}
    \forall \, t \, \in \, \mathbb{R} \, : \, u^i(x,t) \, \partial_i \,
    \equiv \, {\mathrm{U}}(t) \, \in \, \Gamma\left(\mathcal{T}\mathcal{M}\, , \, \mathcal{M}\right)
\end{equation}
Next, using the Riemaniann metric, which allows to raise and lower
tensor indices, to any ${\mathrm{U}}(t)$ we can associate a family
of sections of the cotangent bundle $\mathcal{T}^\star\mathcal{M}$
defined by the following time evolving one-form:
\begin{equation}\label{Omfildo}
  \forall \, t \, \in \, \mathbb{R} \, : \,    \Omega^{[\mathrm{U}]}(t)
  \, \equiv \, g_{ij} \, u^i(x,t) \, dx^j \, \in \, \Gamma\left(\mathcal{T}^\star\mathcal{M}\, , \, \mathcal{M}\right)
\end{equation}
Utilizing the exterior differential and the contraction operator
acting on differential forms, we can evaluate the Lie-derivative of
the one-form $\Omega^{[\mathrm{U}]}(t)$ along the vector field
$\mathrm{U}$. Applying definitions (see for instance
\cite{pietrobook}, chapter five, page 120 of volume two) we obtain:
\begin{eqnarray}\label{trivialone}
\mathcal{L}_\mathrm{U} \Omega^{[\mathrm{U}]}(t) & \equiv & {\rm
i}_\mathrm{U}  \cdot {\rm d} \Omega^{[\mathrm{U}]} \, + \, {\rm d}
\left({\rm i}_\mathrm{U} \cdot \Omega^{[\mathrm{U}]} \right) \, = \,
\left ( u^\ell \partial_\ell \, u^i  \, + \, g^{i k}
\partial_{k} \, \underbrace{\parallel \mathrm{U} \parallel^2}_{g_{mn}
\, u^m \, u^n} \right)\, g_{ij} \, dx^j
\end{eqnarray}
and the Navier Stokes  equation can be rewritten in the the
following  index-free reformulation
\begin{eqnarray}
    - \, {\rm d } \left( p \, + \, \ft 12 \,\parallel \mathrm{U} \parallel^2\right ) & = &
    \partial_t \Omega^{[\mathrm{U}]} \, + \,  {\rm i}_\mathrm{U}
    \cdot {\rm d}  \Omega^{[\mathrm{U}]} \,  \, - \,\nu \,  \Delta_g \, \Omega^{[\mathrm{U}]}
    \, - \, \mathbf{f} \label{bernullini}
\end{eqnarray}
Where $\Delta_g$ is  the Laplace-Beltrami operator on 1-forms,
written in an index free notation as it follows:
\begin{equation}\label{laplacciobeltramo}
    \Delta_g \, = \, \delta\, \mathrm{d} \, + \, \mathrm{d} \, \delta \quad ;
    \quad\delta \equiv \star_g \, \mathrm{d} \star_g
\end{equation}
where with $\star_g$ we have denoted the Hodge duality operation in
the background of the metric $g$.
\par
Eq.(\ref{bernullini}) is one of the possible formulations of
classical Bernoulli theorem. To begin with, consider inviscid fluids
($\nu \, = \, 0$) with no external forces ($\mathbf{f} = 0$). Then
equation eq.(\ref{bernullini}) becomes:
\begin{equation}\label{bernullone}
- \, {\rm d } \left( p \, + \, \ft 12 \,\parallel \mathrm{U}
\parallel^2\right ) \, = \,
\partial_t \Omega^{[\mathrm{U}]} \, + \,  {\rm i}_\mathrm{U}
\cdot {\rm d}  \Omega^{[\mathrm{U}]} \,
\end{equation}
and from eq.(\ref{bernullini}) we immediately conclude that the Bernoulli function defined as follows:
\begin{equation}\label{finocchionabiscotta}
    H_B \, = \, p \, + \, \ft 12 \,\parallel \mathrm{U} \parallel^2
\end{equation}
is constant along the trajectories defined by
eq.(\ref{streamlines}). Turning matters around we can say that in
\textbf{steady flows} of inviscid free fluids, where
\begin{equation}\label{stedone}
\partial_t \,\Omega^{[\mathrm{U}]} \, = \,0
\end{equation}
the fluid trajectories necessarily lay on the level surfaces
$H_B(\mathrm{x})\, = \, h \, \in \, \mathbb{R}$ of the function:
\begin{equation}\label{ignobilia}
    H \, : \, \mathcal{M} \, \rightarrow \, \mathbb{R}
\end{equation}
defined by (\ref{finocchionabiscotta}) and hereafter named, as it is
traditional in Fluid Mechanics, the \textbf{Bernoulli function}.
\par
An identical conclusion can be reached in the case of non vanishing
viscosity if the steady flow condition (\ref{stedone}) is replaced
by:
\begin{equation}\label{genstedona}
    \partial_t \,\Omega^{[\mathrm{U}]} \, = \,\nu \, \Delta_g\, \Omega^{[\mathrm{U}]} \, + \, \mathbf{f}
\end{equation}
For instance if at time $t=t_0$, the $1$-form
$\Omega^{[\mathrm{U}]}$ is the superposition of a collection of $N$
eigenstates of the Laplace-Beltrami operator:
\begin{equation}\label{zeroide}
    \Omega^{[\mathrm{U}]}\mid_{t=t_0}\, = \, \sum_{i=1}^N \, \omega_{i} \quad ;
    \quad \Delta_g\,\omega_{i}\, = \, \lambda_i \, \omega_i
\end{equation}
choosing a subset of such forms, say those from $i=1$ to $i=M < N$,
one can solve the condition (\ref{genstedona}) by setting the
driving force as follows:
\begin{equation}\label{laterano}
    \mathbf{f} \, = \, -\, \nu \, \sum_{i=1}^{M} \lambda_i \, \omega_i
\end{equation}
and the 1-form flow as follows:
\begin{equation}\label{carnitina}
\Omega^{[\mathrm{U}]}\, = \, \sum_{i=1}^{M} \, \omega_i \, +
\sum_{i=M+1}^N \, \omega_i \, \exp\left[-\lambda_i \, t\right]
\end{equation}
For viscid fluids, flows satisfying eq.(\ref{genstedona}) will be
referred to as \textbf{generalized steady flows}. It follows that in
the case of  steady and generalized steady flows the fluid
trajectories necessarily lay on the level surfaces
$H_B(\mathrm{x})\, = \, h \, \in \, \mathbb{R}$ of the Bernoulli
function (\ref{ignobilia}) defined by (\ref{finocchionabiscotta}).
\vskip 0.2cm
\subsubsection{Foliations} Then if $H_B(\mathbf{x})$ has a non trivial
$x$-dependence, locally, in open charts $\mathcal{U}_n\subset
\mathcal{M}_n$ of the considered  $n$-dimensional manifold, it
defines a natural foliation of such charts $\mathcal{U}_n$  into a
smooth family of $(n-1)$-manifolds (all diffeomorphic among
themselves) corresponding to the level surfaces.
\par
The global topological  and analytic structure of level  surfaces of
the Bernoulli function is the object of interesting recent
mathematical studies (see for instance \cite{cardonebotto}) that we
avoid addressing since the focus of the present discussion is only
local and heuristic since the $3$-dimensional manifold eventually
considered in this paper is  the flat compact manifold 
$\boldsymbol{\mathfrak{T}}$ of fig. \ref{tubastro} and eq.(\ref{torone}) just as
in \cite{Fre:2015mla,Fr__2023} the considered manifolds  were the flat, non singular torii
$\mathbb{R}^3/\Lambda$ where $\Lambda$ is a lattice. Then in the
mentioned open charts, as already advocated, the trajectories,
\textit{i.e.} the solutions of eq.(\ref{streamlines}), lay on these
surfaces. In other words the dynamical system encoded in
eq.(\ref{streamlines}) is effectively $(n-1)$-dimensional admitting
$H$ as an additional conserved hamiltonian. In the classical  case
$n\, = \, 3$ this means that the differential system
(\ref{streamlines}) is actually two-dimensional, namely non chaotic
and in some instances even integrable\footnote{Here we rely on a
general result established by the theorem of Poincar\'e-Bendixson
\cite{PoincareNonchaos,bendix} on the limiting orbits of planar
differential systems  whose corollary is generally accepted to
establish  that two-dimensional continuous systems cannot be
chaotic.}. Consequently we reach the conclusion that no chaotic
trajectories (or streamlines) can exist in those domains where the
Bernoulli function $H_B(x)$ has a non trivial $x$-dependence: the
only window open for lagrangian chaos occurs in those domains where
$H_B$ is a constant function. Looking at
eq.s(\ref{bernullini}-\ref{bernullone}) we realize that the previous
argument implies that in steady and generalized steady flows,
 chaotic trajectories can occur only if   velocity
field satisfies the following condition:
\begin{equation}\label{cinesinotulipano}
    {\rm i}_\mathrm{U}  \cdot {\rm d}  \Omega^{[\mathrm{U}]} \, = \, 0
\end{equation}
This weak condition (\ref{cinesinotulipano}) is certainly satisfied
if the velocity field $\mathrm{U}$ satisfies the following strong
condition that is named \textbf{Beltrami equation}:
\begin{equation}
    {\rm d}  \Omega^{[\mathrm{U}]} \, = \, \lambda \, \star_g \, \Omega^{[\mathrm{U}]} \quad \Leftrightarrow \quad
   \star_g \, {\rm d}  \Omega^{[\mathrm{U}]} \, = \, \lambda \,  \Omega^{[\mathrm{U}]} \label{Belatramus}
\end{equation}
where $\star_g$, as already specified, denotes the Hodge duality
operator in the metric $g$:
\begin{eqnarray}\label{gongo}
    \star_g \, \Omega^{[\mathrm{U}]} &  = & \epsilon_{\ell m n} \, g^{\ell k}\, \Omega^{[\mathrm{U}]}_k \, dx^m \, \wedge \,dx^n
    \, =\,  u^\ell \, dx^m \, \wedge \,dx^n \, \epsilon_{\ell m n}\label{gruscia1}\\
    \star_g \, \mathrm{d} \, \Omega^{[\mathrm{U}]} &  = & \epsilon_{\ell m n} \, g^{m p}\,g^{n q}
    \partial_p \left( g_{qr} u^r\right)  \, dx^\ell \label{gruscia2}
\end{eqnarray}
\vskip 0.2cm
\subsubsection{Arnold theorem}
\label{arnaldoseczia} The heuristic argument
which leads to  consider velocity fields that satisfy
\textit{Beltrami condition} (\ref{Belatramus}) as the unique steady
candidates compatible with chaotic trajectories was transformed by
Arnol'd into a rigorous theorem \cite{ArnoldBook} which, under the
strong hypothesis that $(\mathcal{M},g)$ is a closed, compact
Riemaniann three-manifold, states the following:
%%%%%%%%%%%%%%%%%%%%%%%%%%%%%%%%%%%%%%%%%%%%%%%%%%%%%%%%%%%%%%%%%%%
\begin{teorema}
\label{fundarnoltheo} Assume that a region $D\subset \mathcal{M}$ of
the considered \textbf{three-dimensional} Riemannian manifold $\left
(\mathcal{M},g\right)$ is bounded by a compact analytic surface and
that the velocity field $\mathbf{U}$ does not satisfy Beltrami
equation everywhere in $D$, namely $\Omega^{[\mathbf{U}]}\, \neq \,
\lambda \star_g \, d\Omega^{[\mathbf{U}]}$, where $\lambda \in
\mathbb{R}$ is a real number. Then the region of the flow can be
partitioned by an analytic submanifold into a finite number of
cells, in each of which the flow is constructed in a standard way.
Namely the cells are of two types: those fibered into tori invariant
under the flow and those fibered into surfaces invariant under the
flow, diffeomorphic to the annulus $\mathbb{R}\times \mathbb{S}^1$.
On each of these tori the flow lines are either all closed or all
dense, and on each annulus all the flow lines are closed.
\end{teorema}
As one sees, in steady flows, when the velocity vector field of the
fluid  is not a Beltrami field then streamlines either lie on
surfaces that have the topology of torii or on surfaces that are
cylindrical. In both cases chaotic streamlines are excluded.
\par
%%%%%%%%%%%%%%%%%%%%%%%%%%%%%%%%%%%%%%%%%%%%%%%%%%%%%%%%%%%%%%%%%%%
\textit{Chaotic trajectories or streamlines} are of particular
interest, both from the point of view of  theory and of
applications, since, in many scenarios, chaotic flows are  desirable
in order either to homogenize the heath exchange between the fluid
and some external objects immersed in the flow, like it happens in
\textit{autoclavs}, or to promote the mixing of two different
fluids, like it happens in \textit{chemical reactors}. The examples
are multiple and the mentioned ones are just an illustration.
\par
On the other hand the chaotic trajectories are desirable in all
these applications at \textit{small scales} while on \textit{larger
scales} the fluid should appear as moving steadily in some given
direction. The intrinsic non linearity of the NS equation forbids
the linear combination of solutions as new solutions and the
superposition of different regimes at different scales is a very
difficult mathematical problem that requires specialized analysis.
\par
The desire to investigate the on-set of chaotic trajectories in
steady (or generalized steady) flows of incompressible fluids
motivated the interest of the  dynamical system community in
Beltrami vector fields defined by  the condition (\ref{Belatramus}).
Furthermore, in view of the above powerful theorem proved by Arnold,
the focus of attention concentrated on the mathematically very
interesting case  of compact three-manifolds. Within this class, the
most easily treatable case is that of flat compact manifolds without
boundary, so that the most popular playground turned out to be the
three torus $\mathrm{T^3}$, whose possible role in applications has
already been emphasized. 
\par
My interest into the relation between Beltrami flows and symmetry groups
which,  in the case of torii $T^3$, appeared as a correspondence with discrete space groups 
and their irreducible representations, induced me to consider instead the simplest
case of a continuous Lie group
symmetry, mainly the axial one: that meant to trade the torus $T^3$ for a
cylinder with periodic boundary condition in the longitudinal direction.
This topology is also attractive for application issues.
\par
Certainly many physical contexts for fluid
dynamics do not correspond to the idealized situation of a motion in
a compact manifold or, said differently, periodic boundary
conditions are not the most appropriate to be applied either in a
river, or in the atmosphere or in the charged plasmas environing a
compact star, yet the message conveyed by Arnold theorem that
Beltrami vector fields play a distinguished role in chaotic behavior
is to be taken seriously into account and gives an important hint.
 In view of what we are going to discus in section
\ref{viacontattosa} this hint is properly developed by considering
the one-to-one relation between Beltrami fields and contact
structures on three-manifolds.
\subsection{The path leading to contact geometry}
\label{viacontattosa} Beltrami vector fields are intimately related
with the mathematical conception of \textit{contact geometry and
contact topology}.   As we have seen from our sketch of Arnold
Theorem, the main obstacle to the onset of chaotic trajectories has
a distinctive geometrical flavor: trajectories are necessarily
ordered and non chaotic if the manifold where they take place has a
foliated structure $\Sigma_h \times \mathbb{R}_h$, the two
dimensional level sets $\Sigma_h$ being invariant under the action
of the velocity vector field $U$. In this case each streamline lays
on some surface $\Sigma_h$. Equally adverse to chaotic trajectories
is the case of \textit{gradient flows} where there is  a foliation
provided by the level sets of some function $H(x)$ and the velocity
field $\mathrm{U}=\nabla H$ is just the gradient of $H$. In this
case all trajectories are orthogonal to the leaves $\Sigma_h$ of the
foliation and their well aligned tangent vectors are parallel to its
normal vector.
\par
\begin{figure}[!hbt]
\begin{center}
%\iffigs
\includegraphics[height=40mm]{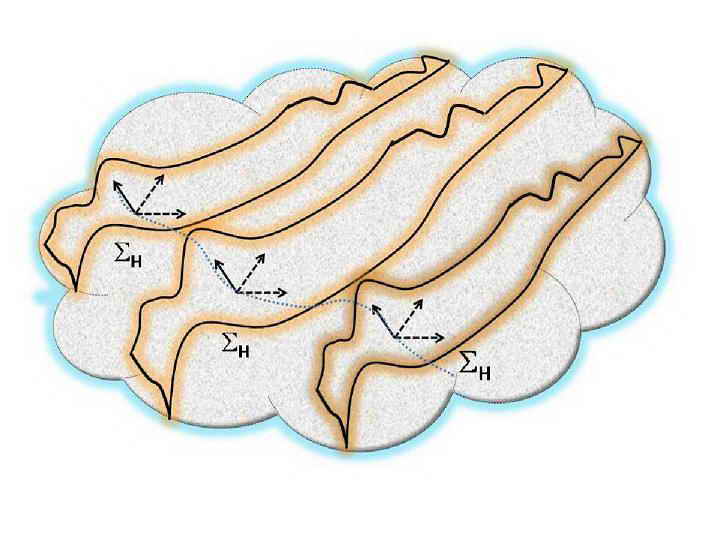}
%\else
\end{center}
 %\fi
\caption{\it  Schematic view of the foliation of a three dimensional
manifold $\mathcal{M}$. The family of two-dimensional surfaces
$\Sigma_h$ are typically the level sets $H(\mathbf{x}) =h$ of some
function $H\, : \, \mathcal{M} \, \rightarrow\, \mathbb{R}$. At each
point of $p\, \in \, \Sigma_h \subset \mathcal{M}$ the dashed
vectors span the tangent space $T_p^\parallel\Sigma_h$, while the
solid vector span the normal space to the surface
$T_p^\perp\Sigma_h$. Equally adverse to chaotic trajectories is the
case where the velocity field $\mathrm{U}$  lies in
$T_p^\perp\Sigma_h$ (gradient flow) or in $T_p^\parallel\Sigma_h$ }
\label{fogliattone}
 %\iffigs
 \hskip 1cm \unitlength=1.1mm
 %\end{center}
 % \fi
\end{figure}
In conclusion in presence of a foliation (or a local foliation) we
have the following decomposition of the tangent space to the
manifold $\mathcal{M}$ at any point $p \in \mathcal{M}$
 \begin{equation}\label{finchius}
    T_p\mathcal{M} \, = \, T_p^\perp\Sigma_h \, \oplus \, T_p^\parallel\Sigma_h
\end{equation}
and no chaotic trajectories are possible in a region $\mathfrak{S}
\subset \mathcal{M}$ where $\mathrm{U}(p) \, \in \,
T_p^\perp\Sigma_h $ or $\mathrm{U}(p) \, \in \,
T_p^\parallel\Sigma_h$ for $\forall p \, \in \, \mathfrak{S} $ (see
fig.\ref{fogliattone}).
\par
This matter of fact motivates an attempt to capture the geometry  of
the bundle of subspaces orthogonal to the lines of flow by
introducing  an intrinsic topological indicator that distinguishes
necessarily non chaotic flows from possibly chaotic ones. Let us
first consider the extreme case of a gradient flow where
$\Omega^{[\mathrm{U}]} \, = \, \mathrm{d} H$ is an exact form. For
such flows we have:
\begin{equation}\label{canedellascala}
\Omega^{[\mathrm{U}]} \, \wedge \, \mathrm{d}\Omega^{[\mathrm{U}]}
\, = \, \Omega^{[\mathrm{U}]} \, \wedge \, \underbrace{\mathrm{d} \,
\mathrm{d}H}_{= \, 0} \, = \, 0
\end{equation}
Secondly let us consider the opposite case where the velocity field
$\mathrm{U}$ is orthogonal to a gradient vector field $\nabla H$ so
that the integral curves of $\mathrm{U}$  lay on the level surfaces
$\Sigma_h$. Furthermore let us assume that $\mathrm{U}$ is self
similar on neighboring level surfaces. We can characterize this
situation in a Riemaniann manifold $(\mathcal{M},g)$ by the
following two conditions:
\begin{equation}\label{gattoascensore}
{\rm i}_{\nabla H}\Omega^{[\mathrm{U}]} \, \Leftrightarrow \,
g\left(\mathrm{U}\, ,\, \nabla H \right) \, = \, 0 \quad ; \quad
\left[ \mathrm{U}\, , \, \nabla H\right] \, = \, 0
\end{equation}
The first of eq.s(\ref{gattoascensore}) is obvious. To grasp the
second it is sufficient to introduce, in the neighborhood of any
point $p\, \in \, \mathcal{M}$, a local coordinate system composed
by $(h,x^\parallel )$ where $h$ is the value of the function $H$ and
$x^\parallel $ denotes some local coordinate system on the level set
$\Sigma_h$.  The situation we have described corresponds to assuming
that:
\begin{equation}\label{caruccio}
    \mathrm{U} \, \simeq \, U^{\parallel} (x^\parallel) \, \partial_\parallel  \quad ; \quad \partial_h
    \, U^{\parallel} (x^\parallel) \, = \, 0
\end{equation}
Under the conditions spelled out in eq.(\ref{gattoascensore}) we can
easily prove that:
\begin{equation}\label{suschione}
    {\rm i}_{\nabla H} \, \mathrm{d}\Omega^{[\mathrm{U}]} \, = \,0
\end{equation}
Indeed from the definition of the Lie derivative we obtain:
\begin{eqnarray}\label{cincischiando}
    {\rm i}_{\nabla H} \, \mathrm{d}\Omega^{[\mathrm{U}]} \, = \, \underbrace{\mathcal{L}_{\nabla H}
    \,\Omega^{[\mathrm{U}]}}_{= \, \Omega^{\left[[U\, ,\, \nabla H]\right]} \, = \, 0} \, - \,
    \mathrm{d}\left(\underbrace{{\rm i}_{\nabla H}\Omega^{[\mathrm{U}]}}_{= 0}\right)
\end{eqnarray}
Since we have both ${\rm i}_{\nabla H}\Omega^{[\mathrm{U}]} \, = \,
0$ and ${\rm i}_{\nabla H}\mathrm{d}\Omega^{[\mathrm{U}]} \, = \, 0$
it follows that also in this case:
\begin{equation}\label{franceschiello}
    \Omega^{[\mathrm{U}]} \, \wedge \, \mathrm{d}\Omega^{[\mathrm{U}]} \, = \, 0
\end{equation}
Therefore in order not to exclude chaotic trajectories one has to
assume that
\begin{equation}\label{rebomba}
    \Omega^{[\mathrm{U}]} \, \wedge \, \mathrm{d}\Omega^{[\mathrm{U}]} \, \neq \, 0
\end{equation}
and the above condition is what leads us to \textit{contact
geometry}.
\subsection{Contact structures in $D=3$ and hydro-flows}
Let us now consider the case relevant for fluid dynamics, namely
that of three dimensional contact manifolds $\left(\mathcal{M}_3\,
,\,\xi_\alpha\right)$, where, in the notation $\xi_\alpha$, we
mention the contact form $\alpha$ defining the contact structure.
In such contact
manifolds, the Legendrian submanifolds (see \cite{Fr__2023} and Appendix A in \cite{geotermico}) are all 1-dimensional, namely
they are \textit{curves} or, as it is customary to name them in the
present context, \textit{knots}.
\par
Hence in three dimensions there are two kind of knots, the
\textbf{Legendrian knots} whose tangent vector belongs to
$\text{ker}\, \alpha$ and the \textbf{transverse knots} whose
tangent vector is parallel to the Reeb field at each point of the
trajectory.
\par
\begin{figure}[!hbt]
\begin{center}
\includegraphics[height=150mm]{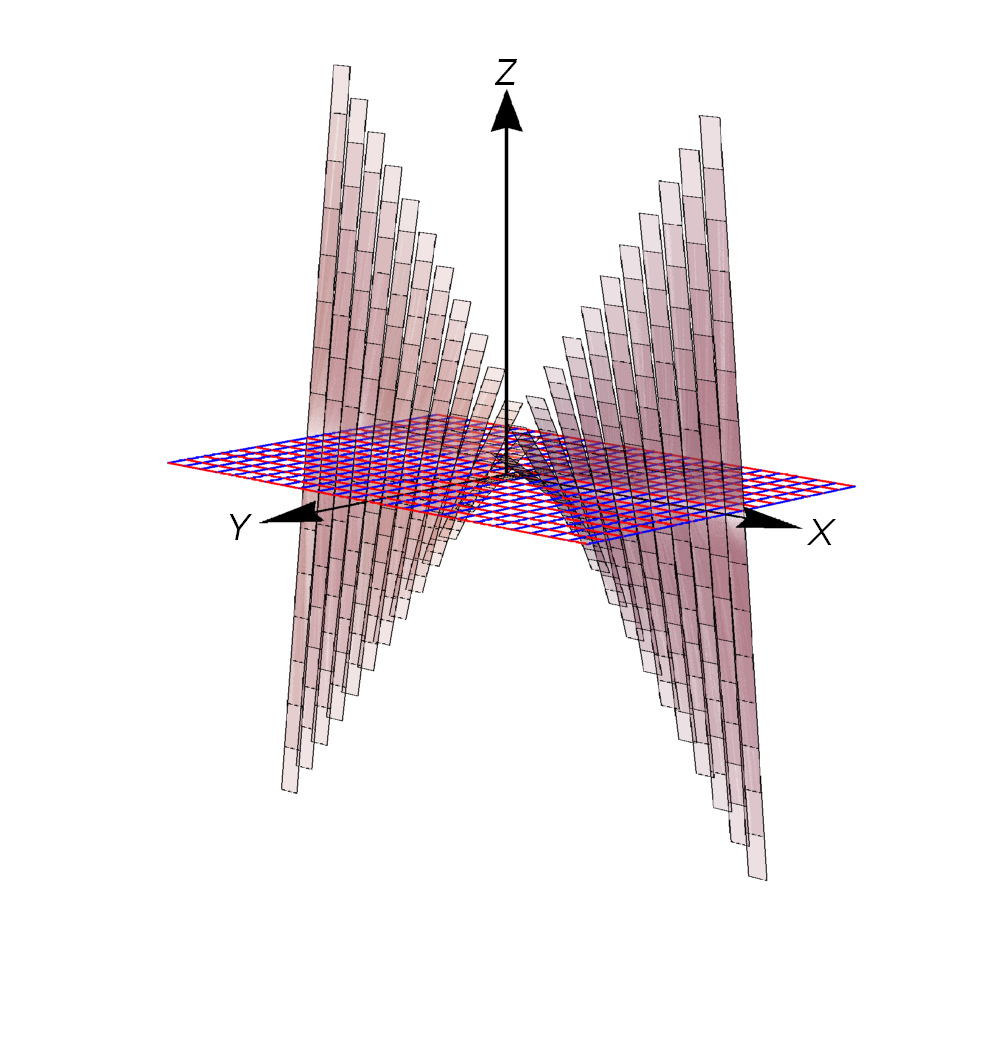}
\end{center}
\caption{\it Schematic vision of the standard contact structure
in $\mathbb{R}^3$.} \label{berlucchino}
 \hskip 1cm \unitlength=1.1mm
\end{figure}

Furthermore in D=3 the condition on the Reeb field (see eq.(A.27) of \cite{geotermico}):
\begin{equation}\label{gridolino}
    \epsilon^{\lambda \mu_1\nu_1\mu_2\nu_2\dots\mu_n\nu_m} \,
    R_\lambda \, \partial_{\mu_1}\,R_{\nu_1}
    \,\partial_{\mu_2}\,R_{\nu_2}\, \dots\,
    \partial_{\mu_n}\,R_{\nu_n}\, \neq \, 0 \quad \text{nowhere vanishes}
\end{equation}
becomes
\begin{equation}\label{urlonotturno}
    \epsilon^{\lambda\mu\nu}  \,R_\lambda \, \partial_\mu \, R_\nu
    \, \neq \, 0
\end{equation}
The flat Euclidian space in three dimensions whose coordinates we denote
as $x,y,z$ is endowed with a standard contact structure that admits
the  contact form displayed below
\begin{equation}\label{romildino}
    \alpha_s \, = \, \mathrm{d}z+ x \,\mathrm{d}y
\end{equation}
and pictorially shown in fig. \ref{berlucchino}.
\subsubsection{Relation with Beltrami vector fields}
\label{belatramocontatto}
As wee see the main reason to introduce the contact form  conception
is that, so doing one liberates the notion of a vector field capable
to generate chaotic trajectories from the use of any  metric
structure. A vector field $U$ is potentially interesting for chaotic
regimes if it is a Reeb  field for at least one contact form
$\alpha$. In this way the mathematical theorems about the
classification of contact structures modulo diffeomorphisms
(theorems that are metric-free and of topological nature) provide
new global methods to capture the topology of hydro-flows.
\par
Instead if we work in a Riemaniann manifold endowed with a metric
$(\mathcal{M},g)$ we can always invert the procedure and define the
contact form $\alpha$ that can admit $\mathrm{U}$ as a Reeb vector
field by identifying
\begin{equation}\label{gominato}
    \alpha \, = \, \Omega^{[\mathrm{U}]}
\end{equation}
In this way the first of the two conditions (see \cite{geotermico}) defining the Reeb field $\mathbf{R}_\alpha$ of a contact $1$-form
$\alpha$:
\begin{eqnarray}\label{Ribbo}
&&\alpha\left(\mathbf{R}_\alpha\right) \, = \, \lambda(\mathbf{x})
\quad =
 \quad
 \text{nowhere vanishing  function on $\mathcal{M}_{2n+1}$}\nonumber\\
 &&\forall \mathbf{X}\in \Gamma\left[\mathcal{TM}_{2n+1},\mathcal{M}_{2n+1}\right] \quad :
 \quad
 \mathrm{d}\alpha\left(\mathbf{R}_\alpha,\mathbf{X}\right)\, = \,
 0
\end{eqnarray} is
automatically satisfied: ${\rm i}_\mathrm{U}\Omega^{[\mathrm{U}]} \,
= \, \parallel \mathrm{U}\parallel^2 \, > 0$. It remains to be seen
whether $\Omega^{[\mathrm{U}]}$ is indeed a contact form, namely
whether $\Omega^{[\mathrm{U}]} \, \wedge \, \mathrm{d}
\Omega^{[\mathrm{U}]} \, \ne \,0$ and whether the second condition
${\rm i}_\mathrm{U} \, \mathrm{d} \Omega^{[\mathrm{U}]} \, = \, 0$
is also satisfied. Both conditions are automatically fulfilled if
$\mathrm{U}$ is a Beltrami field, namely if it is an eigenstate of
the operator $\star_g \, \mathrm{d}$ as advocated in
eq.(\ref{Belatramus}). Indeed the implication ${\rm i}_\mathrm{U} \,
\mathrm{d} \Omega^{[\mathrm{U}]} \, = \, 0$ of Beltrami equation was
shown  in eq. (\ref{cinesinotulipano}), while from the Beltrami
condition it also follows:
\begin{eqnarray}\label{cirimellaG}
 \Omega^{[\mathrm{U}]} \, \wedge \, \mathrm{d} \Omega^{[\mathrm{U}]}
 & = &\Omega^{[\mathrm{U}]} \, \wedge \, \star_g \Omega^{[\mathrm{U}]}
 \, = \, \parallel \mathrm{U}\parallel^2 \, \mathrm{Vol}  \, \ne \, 0
 \quad ; \quad
 \mathrm{Vol} \,  \equiv \, \frac{1}{3!} \,
 \times \, \epsilon_{ijk} \, dx^i \, \wedge \, dx^j \, \wedge \, dx^k
\end{eqnarray}
In this way the conceptual circle closes and we see that all
Beltrami vector fields can be regarded as Reeb  fields for a
bona-fide contact form. Since the same contact structure (in the
topological sense) can be described by different contact forms, once
Beltrami fields have been classified it remains the task to discover
how many inequivalent contact structures they actually describe. Yet
it is reasonable to assume that every contact structure has a
contact form representative that is derived from a Beltrami Reeb
field. Indeed a precise correspondence is established by a theorem
proved in \cite{Ghrist2007}:
\begin{teorema}
\label{ghristheo} Any rotational Beltrami vector field on a
Riemaniann $3$-manifold is a Reeb  field for some contact form.
Conversely any Reeb field associated to a contact form on a
$3$-manifold is a rotational Beltrami field for some Riemaniann
metric. Rotational Beltrami field means an eigenfunction of the
$\star_g \mathrm{d}$ operator corresponding to a non vanishing
eigenvalue $\lambda$.
\end{teorema}
\section{Axial Symmetry in Fluid Dynamics and Beltrami Flows}
\label{assialone}
Having recalled the general principles that promote Beltrami (or anti Beltrami) fields to a distinguished role in fluid dynamics we turn to the manifold of interest to us in which we will establish a functional basis made of Beltrami/anti-Beltrami fields plus closed forms (gradient flows) all endowed with axial symmetry. The natural path
leading to our result consists of exchanging the cartesian coordinate  $(x,y,z)$ with the cylindrical ones $(r,\phi,z)$.
\subsection{Flat geometry in cylindrical coordinates}
Given the standard flat metric of \(\mathbb{R}^3\) in orthonormal cartesian coordinates $\{$x,y,z$\}$
\begin{equation}\label{flatmetric}
 \text{ds}^2=\text{dx}^2+\text{dy}^2+\text{dz}^2
\end{equation}
the transformation
\begin{equation}\label{cylindframe}
 \{x\to r \cos (\phi ),y\to r \sin (\phi ),z\to z\}
\end{equation}
provides the new cylindrical coordinate system $\{$r,$\phi$,z$\}$ in which the metric becomes
\begin{equation}\label{gordone}
  \text{ds}^2=\text{dr}^2+\text{d$\phi $}^2 r^2+\text{dz}^2
\end{equation}
which implies :
\begin{equation}\label{truzzone}
  g_{\text{ij}}=\left(
\begin{array}{ccc}
 1 & 0 & 0 \\
 0 & r^2 & 0 \\
 0 & 0 & 1 \\
\end{array}
\right) \quad ; \quad g^{\text{ij}} =\left(
\begin{array}{ccc}
 1 & 0 & 0 \\
 0 & \frac{1}{r^2} & 0 \\
 0 & 0 & 1 \\
\end{array}
\right)
\end{equation}
Obviously the above metric is flat as the original one, yet in cylindrical coordinates \(y^i\)= $\{$r,$\phi $,z$\}$,
the Levi-Civita connection is not zero and we have:
\begin{equation}\label{crinierarossa}
 \pmb{\pmb{\Gamma }_{\text{  }b}^a}\equiv \Gamma _{\text{cb}}^a\text{dy}^c\, = \,\left(
\begin{array}{ccc}
 0 & -\text{d$\phi $}\, r & 0 \\
 \frac{\text{d$\phi $}}{r} & \frac{\text{dr}}{r} & 0 \\
 0 & 0 & 0 \\
\end{array}
\right)
\end{equation}
The Jacobian of the coordinate transformation is the following one:
\begin{align}\label{crisallo}
 J \, &\equiv \,  \left(
\begin{array}{ccc}
 \text{Cos}[\phi ] & \text{Sin}[\phi ] & 0 \\
 -r \,\text{Sin}[\phi ] & r \,\text{Cos}[\phi ] & 0 \\
 0 & 0 & 1 \\
\end{array}
\right) \quad ; \quad 
 J^{-1} \, = \, \frac{\partial (r,\phi ,z)}{\partial (x,y,z)}\, = \, \left(
\begin{array}{ccc}
 \text{Cos}[\phi ] & -\frac{\text{Sin}[\phi ]}{r} & 0 \\
 \text{Sin}[\phi ] & \frac{\text{Cos}[\phi ]}{r} & 0 \\
 0 & 0 & 1 \\
\end{array}
\right)
\end{align}
In view of what follows we need to recall a few more geometrical relations applying to flat geometry in cylindrical coordinates.
\par
The dreibein description of flat metric in these coordinates is the following:
\begin{eqnarray}\label{dreigamba}
  \mathrm{ds}^2 & = & \mathbf{e}^r \times \mathbf{e}^r \, + \, \mathbf{e}^\phi \times \mathbf{e}^\phi \, + \, \mathbf{e}^z \times \mathbf{e}^z \nonumber\\
  && \mathbf{e}^r\, = \, \mathrm{dr}\quad ; \quad \mathbf{e}^\phi\, = \, r \,\text{d$\phi$}\quad ; \quad \mathbf{e}^z \, = \, \mathrm{dz}
\end{eqnarray}
The volume $3$-form is defined as usual:
\begin{equation}\label{frandunjo}
  \mathrm{Vol} \, = \, \mathbf{e}^r\wedge \mathbf{e}^\phi \wedge \mathbf{e}^z \, = \, r \times \mathrm{dr}\wedge \text{d$\phi$} \wedge \mathrm{dz} \, = \, \sqrt{\text{det} g} \times \mathrm{dr}\wedge \text{d$\phi$} \wedge \mathrm{dz}
\end{equation}
\subsection{The velocity field \pmb{ U} and the associated 1-form \pmb{ $\Omega$}}
The velocity field of Hydrodynamics is a vector field of the form
\begin{eqnarray}\label{uffildo}
 {\mathbf{U}}&=&g(r,z,\phi )\frac{\overset{\rightharpoonup }{\partial }}{\partial r}+h(r,z,\phi )\frac{\overset{\rightharpoonup }{\partial }}{\partial
\phi }+w(r,z,\phi )\frac{\overset{\rightharpoonup }{\partial }}{\partial z}
\end{eqnarray}
where $g(r, z,\phi )$, $h(r,z,\phi )$, $w(r,z,\phi) $ are the above mentioned  three functions of the three coordinates. 
By means of the inverse Jacobian, the same vector field can be written in the standard Cartesian basis as follows:
\begin{eqnarray}\label{carolingio}
\mathbf{U}&=&\left(g(r,z,\phi )\,\cos[\phi]-\frac{h(r,z,\phi )}{r}\,\sin[\phi]\right)\,\frac{\overset{\rightharpoonup }{\partial }}{\partial
x}\nonumber \\ 
&&+\left(g(r,z,\phi )\,\sin[\phi ] +\frac{h(r,z,\phi )}{r}\,\cos[\phi ]\right)\frac{\overset{\rightharpoonup }{\partial }}{\partial y}
+w(r,z,\phi)\,\frac{\overset{\rightharpoonup }{\partial }}{\partial z}
\end{eqnarray}
The norm of the velocity field is:
\begin{equation}\label{duracchio}
  \mid\mathbf{U}\mid^2 \, \equiv \, g_{\text{ij}}\,U^iU^j=\frac{1}{2}\left(g(r,\phi ,z)^2+r^2h(r,\phi ,z)^2+w(r,\phi ,z)^2\right)
\end{equation}
so that the Bernoulli function is as follows:
\begin{equation}\label{valona}
  H_B=\frac{1}{2} \left(g(r,\phi ,z)^2+r^2 h(r,\phi ,z)^2+w(r,\phi ,z)^2\right)+p(r,\phi ,z)
\end{equation}
where \textit{ p (r, $\phi $, z)} is the pressure field.
\par
On the other hand the 1-form description of the velocity field is provided by the following expression:
\begin{eqnarray}\label{omegassiale}
  \Omega^{[\mathbf{U}]} &=&\text{dy}^j \, g_{\text{ij}}\,  U^i \, =\,\text{dr} \, {g}(r,\phi ,z)\, 
  +\,\text{d$\phi $}\, r^2 {h}(r,\phi ,z)\, + \, \text{dz} \,  {w}(r,\phi ,z)
\end{eqnarray}
\paragraph{\sc Invariant exact solution}
Before performing CFD simulations one can ask oneself the question if exact solutions do exist with the specified symmetry. Such solutions are obviously ideal cases, since no symmetry is completely exact in practical contexts, yet the consideration of the idealized case of exact symmetry can provide a global viewpoint to be compared with the results of detailed numerical simulations of the particular case under consideration. 
\subsection{Axial symmetry}
When the container of the fluid is cylindrically shaped, we are in presence of an axial symmetry, namely we deal with a rotation group SO(2) around
the z-axis. The general question to be formulated to begin with is the following one. What is the general form of an ansatz for the velocity vector
field \(\boldsymbol{\mathrm{U}}(t,\pmb{y}) =\mathrm{U}_i(t,\pmb{y})\) which is invariant under the $\mathrm{SO(2)}\subset \mathrm{SO(3)}$ transformations. Let us pose the invariance statement
in general terms . If $\Gamma \subset \mathrm{SO(3)}$  is any subgroup we must have:
\begin{equation}\label{godrone}
  \forall \gamma  \in \Gamma \quad :\quad \gamma ^{-1} \cdot  \pmb{\mathrm{U}}(t, \gamma  \cdot \pmb{x}) = \pmb{\mathrm{U}}(t,\pmb{x})
\end{equation}
\subsubsection{Imposing axial symmetry}
An element of the rotation group in the xy-plane is represented by the following matrix
\begin{equation}\label{gammoide}
  \gamma =\left(
\begin{array}{ccc}
%\hline
 \cos[\phi] & -\cos[\phi] & 0 \\
%\hline
 \sin[\phi ] & \cos[\phi ] & 0 \\
%\hline
 0 & 0 & 1 \\
%\hline
\end{array}
\right)
\end{equation}
If we use cylindrical coordinates as defined above and we rely on the standard trigonometric identities:
\begin{equation}\label{idatrigo}
  \cos[\phi -\theta ]= \cos[\phi ) \cos[\theta ]+ \sin[\phi )\sin[\theta ] \quad;\quad\sin[\phi -\theta
]= - \cos[\phi] \, \sin[\theta ]+ \sin[\phi ]\cos[\theta ]
\end{equation}
observing that in cylindrical coordinates a rotation of an angle theta around the z-axis correspond to the shift
\begin{equation}\label{contadino}
  \phi  \longrightarrow  \phi  + \theta
\end{equation}
we conclude that the general form of an axial symmetric vector field is as follows 
\begin{eqnarray}\label{fabbro}
  \pmb{\mathrm{U}}_{axial}(\pmb{\text{\textit{$x$}}}) & = &\left\{ \, \mathrm{G}(r,z)\cos[\phi]-\frac{\mathrm{H}(r,z)}{r}\,\sin[\phi ], \, \, G(r,z)\sin[\phi ]+\frac{\mathrm{H}(r,z)}{r}\cos[\phi],\,\, \mathrm{W}(r,z)\,\right\} \nonumber\\
   {\Omega }^{[\pmb{\mathrm{U}}]}_{axial} & = & \text{dr} \,\text{G}(r,z)+\text{d$\phi $} \, r^2 \text{H}(r,z)+\text{dz}\, \text{W}(r,z)
\end{eqnarray}
where $\mathrm{G}(r,z), \mathrm{H}(r,z)$ and $\mathrm{W}(r, z)$ are three functions of two variables, namely the components of the vector field in cylindrical coordinates are independent from the angle $\phi $. We are interested in considering the NS equations reduced to these two variables in the case
of a steady flow, namely posing $\partial _t\pmb{U}(\pmb{\textit{x}})=0$.
\subsection{General form of the NS equations in cylindrical coordinates}
\label{tarquinioprisco}
The most convenient way of writing the NS equations is the geometrical one discussed in  section \ref{fluidostuff}:
\begin{equation}\label{gongolone}
  \pmb{d}H_B+i_{\pmb{\mathrm{U}}}\cdot \pmb{d}\Omega ^{\pmb{\mathrm{U}}}+\frac{\partial \Omega ^{[\pmb{\mathrm{U}}]}}{\partial t}\, -\, \nu  \Delta {\Omega }^{[\pmb{\mathrm{U}}]}\, = \,0
\end{equation}
where $\mathbf{d}$ is the exterior derivative operator, $ i_{\pmb{X}} $ is the contraction operator along a vector field $\pmb{X}$ and $\Delta $ is the Laplace-Beltrami operator:
\begin{equation}\label{pignattasbilenca}
 \pmb{\Delta}\, = \, \pmb{d}\,\delta +\text{$\delta \, \pmb{d} $} =-\pmb{d}*\pmb{d}* -*\pmb{d}*\pmb{d}
\end{equation}
where $\delta $ is the coadjoint exterior derivative, $ *$ denoting the Hodge duality operator:
\begin{equation}\label{cocacchio}
  \delta =-*\pmb{d}*
\end{equation}
\subsubsection{Explicit  cylindrical coordinate expression of the Laplacian on the hydro $1$-form $\Omega^{[\boldsymbol{\mathrm{U}}]}$}
With reference to the generic  $1$-form description of the fluid velocity vector field, as displayed in eq.(\ref{omegassiale}), we have:
\begin{eqnarray}
\label{forbaccio}
\pmb{\Delta}{\Omega }^{[\pmb{\mathrm{U}}]} &=& \frac{1}{r^2}\times\left[\text{dr}\left(r^2 \partial^2_z g(r,\phi ,z)+r^2 \partial^2_r g(r,\phi ,z)+r
   \partial_r g(r,\phi ,z)+\partial^2_\phi g(r,\phi ,z)-g(r,\phi ,z)-2 r \partial_\phi h(r,\phi
   ,z)\right)\right.\nonumber\\
   &&\left.+\text{dz} \left(r^2 \partial^2_z w(r,\phi ,z)+r \left(\partial_r w(r,\phi
   ,z)+r \partial^2_r w(r,\phi ,z)\right)+\partial^2_\phi w(r,\phi ,z)\right)\right.\nonumber\\
   &&\left. +\text{d$\phi $} \, r
   \left(2 \partial_\phi g(r,\phi ,z)+r^3 \partial^2_z h(r,\phi ,z)+r \left(r^2
   \partial^2_r h(r,\phi ,z)+3 r \partial_r h(r,\phi ,z)+\partial^2_\phi h(r,\phi
   ,z)\right)\right)\right]
\end{eqnarray}
In the case of axial symmetric flows as described in eq.(\ref{fabbro}) the expression of the Laplacian simplifies
and we have:
\begin{eqnarray}
\label{axsymbaccio}
 \pmb{\Delta}{\Omega }^{[\pmb{\mathrm{U}}]}_{axial} &=& -\frac{\left(r^2 \partial^2_z \text{G}(r,z)+r^2 \partial^2_r \text{G}(r,z)+r
   \partial_r \text{G}(r,z)-\text{G}(r,z)\right)}{r^2}\, \text{dr}\nonumber\\
   &&-r \left(r \partial^2_z \text{H}(r,z)+3 \partial_r \text{H}(r,z)+r \partial^2_r \text{H}(r,z)\right)\,\text{d$\phi
   $}\nonumber \\ 
    &&-\frac{ \left(r
   \partial^2_z \text{W}(r,z)+\partial_r \text{W}(r,z)+r \partial^2_r \text{W}(r,z)\right)}{r}\, \text{dz}\nonumber\\
\end{eqnarray}
The other ingredient of the steady NS equation (\ref{gongolone}) is the object:
\begin{equation}\label{cordaro}
  \boldsymbol{\mathrm{T}}^{[\mathbf{U}]} \, \equiv \, \pmb{d}H_B+i_{\pmb{\mathrm{U}}}\cdot \pmb{d}\Omega ^{\pmb{\mathrm{U}}}
\end{equation}
Utilizing the definition (\ref{finocchionabiscotta}) of the Bernoulli function one finds:
\begin{equation}\label{sganarello}
\boldsymbol{\mathrm{T}}^{[\mathbf{U}]}  \, = \,   \text{dy}^k g_{\text{rk}} \left(g^{\text{sr}} \nabla _sp+U^i \nabla _iU^r\right)
\end{equation}
and in cylindrical coordinates we have:
\begin{eqnarray}
\label{ganzone}
 \boldsymbol{\mathrm{T}}^{[\mathbf{U}]} &=&  \left(\partial_\phi g(r,\phi ,z) h(r,\phi ,z)+\partial_z g(r,\phi ,z) w(r,\phi ,z)+g(r,\phi ,z)
   \partial_r g(r,\phi ,z)-r h(r,\phi ,z)^2+\partial_r p(r,\phi ,z)\right)\,\text{dr}\nonumber\\  
   &&+ \left(r^2 g(r,\phi ,z) \partial_r h(r,\phi ,z)+2 r
   g(r,\phi ,z) h(r,\phi ,z)+r^2 \partial_z h(r,\phi ,z) w(r,\phi ,z)+r^2 h(r,\phi ,z) \partial_\phi h(r,\phi
   ,z)\right.\nonumber\\
  &&\left. +\partial_\phi p(r,\phi ,z)\right) \, \text{d$\phi $} \nonumber\\
  &&+ \left(g(r,\phi ,z)
   \partial_r w(r,\phi ,z)+h(r,\phi ,z) \partial_\phi w(r,\phi ,z)+\partial_z p(r,\phi ,z)+w(r,\phi ,z)
   \partial_z w(r,\phi ,z)\right)\, \text{dz}
\end{eqnarray}
If we impose axial symmetry we obtain:
\begin{eqnarray}
\label{paranco}
  \boldsymbol{\mathrm{T}}^{[\mathbf{U}]}_{axial} &=& \left(\partial_z\mathrm{G}(r,z) \mathrm{W}(r,z)+\mathrm{G}(r,z) \partial_r\mathrm{G}(r,z)-r
   \mathrm{H}(r,z)^2+\partial_r\mathrm{P}(r,z)\right)\,\text{dr} \nonumber\\ 
   &&+\left(r^2
   \mathrm{G}(r,z) \partial_r\mathrm{H}(r,z)+2 r \mathrm{G}(r,z) \mathrm{H}(r,z)+r^2 \partial_z\mathrm{H}(r,z) \mathrm{W}(r,z)\right)\,\text{d$\phi $} \nonumber\\
    &&+ \left(\mathrm{G}(r,z)
   \partial_r\mathrm{W}(r,z)+\partial_z\mathrm{P}(r,z)+\mathrm{W}(r,z) \partial_z\mathrm{W}(r,z)\right)\, \text{dz} 
\end{eqnarray}
Combining eq.s (\ref{axsymbaccio}) and (\ref{paranco}) one obtains the  Navier Stokes equations with axial symmetry:
\begin{equation}\label{carlingacorta}
  \boldsymbol{\mathrm{T}}^{[\mathbf{U}]}_{axial}\, -\, \nu  \pmb{\Delta}{\Omega }^{[\pmb{\mathrm{U}}]}_{axial}\, = \,-\,\partial_t\,{\Omega }^{[\pmb{\mathrm{U}}]}_{axial}
\end{equation}
\section{Cylindrical harmonic $1$-forms}
\label{cimentoarmonia}
As  building blocks for solutions of eq.(\ref{carlingacorta}), it is interesting to consider axial $1$-forms
$\boldsymbol{\mathfrak{harm}}[\Lambda]$ that are harmonic, namely eigenstates of the Laplacian operator:
\begin{equation}\label{harminio}
  \pmb{\Delta}\boldsymbol{\mathfrak{harm}}[\Lambda] \, = \, \Lambda \,\boldsymbol{\mathfrak{harm}}[\Lambda] \quad ; \quad \Lambda \in \mathbb{R}
\end{equation}
recalling also that all Beltrami/anti Beltrami flows are necessarily harmonic. Indeed, by definition Beltrami flows are eigenstates of the Beltrami operator $\pmb{\Game}$:
\begin{equation}\label{belatrus}
  \pmb{\Game} \, \Omega_{\pm \varpi}^{[\mathbf{U}]}\, = \, \pm \varpi\, \Omega_{\pm \varpi}^{[\mathbf{U}]}
  \quad \text{where} \quad \pmb{\Game} \, \equiv \, \star \,\pmb{\mathrm{d}} \quad ; \quad \varpi > 0
\end{equation}
and referring to equation (\ref{pignattasbilenca}) we see that on Beltrami flows the Laplacian operator becomes minus the square of the Beltrami operator:
\begin{equation}\label{pulcinella}
  \pmb{\Delta} \Omega_{\pm \varpi}^{[\mathbf{U}]} \, = \, \left(-\pmb{\mathrm{d}}\,\star \,\pmb{\mathrm{d}}\,\star\, - \,
  \star \,\pmb{\mathrm{d}}\,\star \,\pmb{\mathrm{d}} \right) \, \Omega_{\pm \varpi}^{[\mathbf{U}]} \, = \, - \, \pmb{\Game}^2 \,\Omega_{\pm \varpi}^{[\mathbf{U}]} \, = \, - \varpi^2 \, \Omega_{\pm \varpi}^{[\mathbf{U}]}
\end{equation}
In the above equation the operator $-\pmb{\mathrm{d}}\,\star \,\pmb{\mathrm{d}}\,\star\,$ annihilates a Beltrami flow, since from the Beltrami condition and the square property of the Hodge operator $\star^2 \, = \, - \mathrm{Id}$ it follows that $\star \Omega_{\pm \varpi}^{[\mathbf{U}]} \ltimes \mathbf{d}\Omega_{\pm \varpi}^{[\mathbf{U}]}$. 
\par
Hence if we determine the space of harmonic $1$-forms with certain boundary behavior in the same space we find, as particular cases the Beltrami/anti Beltrami $1$-forms  with the same boundary behavior. 
\par 
\subsubsection{Periodic boundary conditions in the $z$-direction}
\begin{figure}[!hbt]
\begin{center}
\includegraphics[height=80mm]{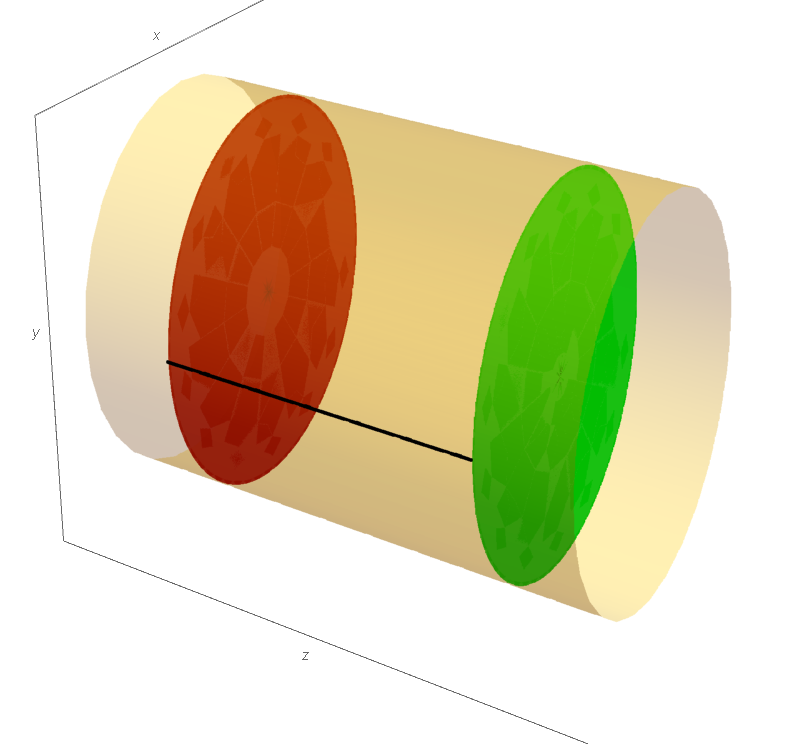}
\end{center}
\caption{\it Schematic vision of a cylinder hosting an axial symmetric flow. The cylinder is infinite in the $z$-direction but we are interested in a flow that is confined to the finite region of the cylinder comprised between the two disks, the red and the green one, that are separated by a horizontal distance $\ell$, evidenced by the black line. The most convenient way to describe such a flow is to impose periodic boundary conditions on the flow in the $z$-direction} \label{tubastro}
 \hskip 1cm \unitlength=1.1mm
\end{figure}
As we schematically display in fig.\ref{tubastro} the situation of interest to us is that of an axial symmetric flow 
confined in a finite portion of an infinite cylinder that we delimit with two  disks. The appropriate way of modeling such a flow is by requiring:
\begin{enumerate}
  \item Vanishing of the velocity field as $r\to \infty$ which simulates the confinement in the finite radius cylinder.
  \item Periodicity in the horizontal direction $z$ which simulates the confinement in the finite portion of the cylinder.    
\end{enumerate}
Referring to the general form (\ref{fabbro}) of the axial symmetric hydro $1$-form, the second of the above requests is satisfied by setting:
\begin{eqnarray}\label{castagnaccio}
  G(r,z)&=& G(r)\, \left( a_k \cos (2 \pi  k z)+b_k \sin (2 \pi  k z)\right) \nonumber\\
  H(r,z)&=& H(r) \, \left( c_k \cos (2 \pi  k z)+d_k \sin (2 \pi  k z)\right) \nonumber\\
   W(r,z)&=& W(r)\, \left( f_k \cos (2 \pi  k z)+g_k \sin (2 \pi  k z)\right) 
\end{eqnarray}
where $z$ is measured in units of the distance $\ell$ between the two disks of fig.\ref{tubastro} and $k\in \mathbb{Z}$. Inserting the ansatz (\ref{castagnaccio}) in the Laplacian eigenvalue equation (\ref{harminio}) one finds three differential equations for the three functions of the radial variable $r$ whose general integral can be expressed in term of Bessel functions. In this way we find the following set of harmonic hydro $1$-forms with the
prescribed periodicity:
\begin{eqnarray}
\label{armoniosa}
\boldsymbol{\mathfrak{harm}}[k,\lambda] &=& \left(a_k \cos (2 \pi  k z)+b_k \sin (2 \pi  k z)\right) \left(\text{$\gamma_1 $}
   J_1\left(2 k \pi  r \sqrt{\lambda ^2-1}\right)+  \text{$\gamma_2 $} Y_1\left(2 k \pi 
   r \sqrt{\lambda ^2-1}\right)\right)\, \text{dr}  \nonumber\\
 &+& r  \left(c_k \cos (2 \pi  k z)+d_k \sin (2 \pi  k z)\right) \left(\text{$\beta_1
   $} J_1\left(2 k \pi  r \sqrt{\lambda ^2-1}\right)+  \text{$\beta_2 $} Y_1\left(2 k
   \pi  r \sqrt{\lambda ^2-1}\right)\right)\, \text{d$\phi$}  \nonumber\\
&+&  \left(f_k \cos (2 \pi  k z)+g_k \sin (2 \pi  k z)\right) \left(\text{$\alpha_1 $}
   J_0\left(2 k \pi  r \sqrt{\lambda ^2-1}\right)+  \text{$\alpha_2 $} Y_0\left(2 k \pi 
   r \sqrt{\lambda ^2-1}\right)\right)\, \text{dz} \nonumber\\
\end{eqnarray}
that satisfy eq.(\ref{harminio}) with eigenvalue:
\begin{equation}\label{cromotasso}
 \Lambda \, = \, - \,  4 \pi ^2 k^2 \lambda ^2
\end{equation}
\begin{figure}[!hbt]
\begin{center}
\includegraphics[width=75mm]{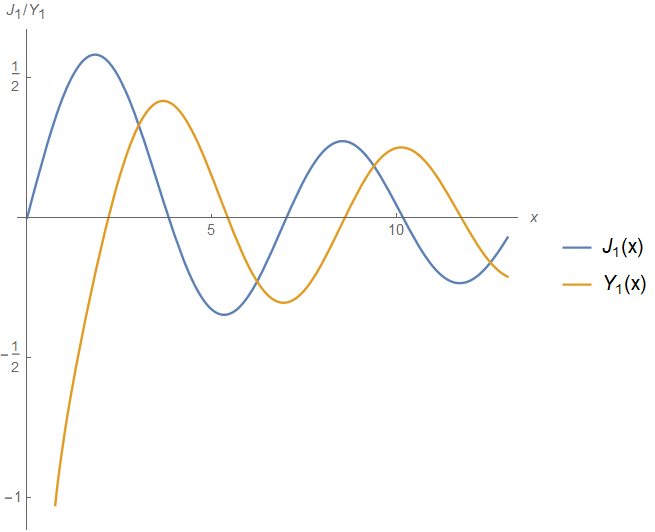}
\includegraphics[width=75mm]{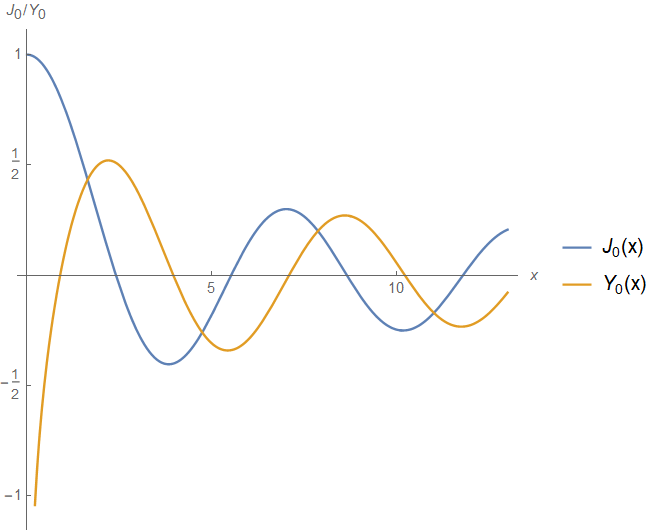}
\end{center}
\caption{\it Behavior of the Bessel functions appearing in the expression of the harmonic $1$-form displayed in
eq.(\ref{armoniosa}). As we see the second solution $Y_{1,0}\left(2 k \pi 
   r \sqrt{\lambda ^2-1}\right)$ has to be excluded since it has a singularity for $r=0$ namely at the very center of the tube. Instead the first solution $J_{1,0}\left(2 k \pi 
   r \sqrt{\lambda ^2-1}\right)$  has the perfect behavior to satisfy the desired boundary conditions. }  \label{besselloni}
 \hskip 1cm \unitlength=1.1mm
\end{figure}
Let us next consider the behavior of the Bessel functions appearing in eq.(\ref{armoniosa}) (see fig.\ref{besselloni}). As we remark also in the figure caption, the second solutions of the Bessel differential equation $Y_{1,0}\left(2 k \pi r \sqrt{\lambda ^2-1}\right)$ have to be excluded since they diverge for $r\to 0$. On the other hand, the initial zero of $J_{1}\left(2 k \pi r \sqrt{\lambda ^2-1}\right)$ at  $r=0$ is just welcome. It means that the components of the velocity field in the radial and winding directions are zero on the axis of the tube: instead the constant value of $J_{0}\left(2 k \pi r \sqrt{\lambda ^2-1}\right)$ at $r=0$ is just coherent with the previous fact. Indeed at the central axis of the tube the fluid velocity  is only in the horizontal $z$-direction and such component does not vanish. 
\par
These observations imply that the relevant harmonic $1$-forms are those with 
\begin{equation}\label{curtabullone}
  \alpha_2\, = \, \beta_2 \, = \, \gamma_2 \, = \, 0 
\end{equation}
The free parameters at fixed $\lambda$ and $k$ are just six since after imposing eq.(\ref{curtabullone}) the parameters $\alpha_1,\,\beta_1, \,\gamma_1$ become pleonastic as they can be reabsorbed in the other six
$a_k,b_k,c_k,d_k,f_k,g_k$. For reasons that will become evident soon, it is also convenient to redefine  $\lambda$ by setting:
\begin{equation}\label{cartolinanera}
 \lambda \, = \,  \frac{\sqrt{\mu ^2+1}}{\mu } \quad ; \quad \mu >0
\end{equation}
and we obtain the following six parameter harmonic $1$-form for every spectral pair $\{\mu,k\}$:
\begin{eqnarray}
\label{baltimora}
\boldsymbol{\mathfrak{har}}\left[\mu,k|a_1,\dots,a_6\right] &=& \, J_1\left(\frac{2 k \pi  r}{\mu }\right)\,\left(a_1 \cos (2 \pi  k z)\, + \, a_2 \sin (2
   \pi  k z)\right)\, \text{dr}\nonumber\\
&+&  \, r \,   J_1\left(\frac{2 k \pi  r}{\mu }\right) \left(a_3 \cos
   (2 \pi  k z) \,+\, a_4 \sin (2 \pi  k z)\right)\, \text{d$\phi$}\nonumber\\
   &+& \, J_0\left(\frac{2 k \pi  r}{\mu }\right)
   \left(a_5 \cos (2 \pi  k z)\, + \, a_6 \sin (2 \pi  k z)\right)\, \text{dz} 
\end{eqnarray}
\subsection{Beltrami/anti Beltrami and closed flows}
\label{beltrachiude}
As we already remarked above, all Beltrami flows, namely all the hydro $1$-forms satisfying eq.(\ref{belatrus}) are also harmonic $1$-forms, yet the opposite is not true. Since we determined the $6$-parameter family of harmonic $1$-forms (\ref{baltimora}) one can try to specialize the $6$-parameters in such a way that the $1$-form 
$\boldsymbol{ \mathfrak{har}}\left[\mu,k|\hat{a}_1,\dots,\hat{a}_6\right]$ satisfies also Beltrami equation 
(\ref{belatrus}) with one of the two available signs for the eigenvalue:
\begin{equation}\label{baraccone}
  \pm\varpi \, = \, \pm \sqrt{-\Lambda}\quad ; \quad \varpi \, = \,  \, 2\, k\, \pi \, \frac{\sqrt{\mu ^2+1}}{\mu } 
\end{equation}
The $1$-forms for which the plus sign applies are \textbf{Beltrami flows}, those for which the minus sign
applies are \textbf{anti Beltrami flows}.\
\par
The result of our analysis is that the $6$-parameter space of harmonic $1$-form splits in three two-dimensional subaspaces: the space of Beltrami flows, the space of anti-Beltrami flows and the space of \textbf{closed $1$-form flows}.
\par
Explicitly we have, the following Beltrami flows
\begin{eqnarray}\label{Beltramiflows}
  \boldsymbol{ \mathfrak{Belt}}_{+}\left[\mu,k|\alpha_+,\beta_+\right] &=&  J_1\left(\frac{2 k \pi  r}{\mu }\right) \left(\alpha _+ \cos (2 \pi  k z)+\beta _+
   \sin (2 \pi  k z)\right)\, \text{dr}\nonumber\\
  &+& \frac{\sqrt{\mu ^2+1}\, r \, J_1\left(\frac{2 k \pi  r}{\mu }\right) \left(\beta _+
   \cos (2 \pi  k z)-\alpha _+ \sin (2 \pi  k z)\right)}{\mu }\, \text{d$\phi$}\nonumber\\
  &+& \frac{ J_0\left(\frac{2 k \pi  r}{\mu }\right) \left(\beta _+ \cos (2 \pi  k
   z)-\alpha _+ \sin (2 \pi  k z)\right)}{\mu } \, \text{dz} \nonumber\\
   \pmb{\Game} \, \boldsymbol{ \mathfrak{Belt}}_{+}\left[\mu,k|\alpha_+,\beta_+\right] & = & \varpi \, 
   \boldsymbol{ \mathfrak{Belt}}_{+}\left[\mu,k|\alpha_+,\beta_+\right]
\end{eqnarray}
the following anti-Beltrami flows:
\begin{eqnarray}\label{Beltramiflows}
  \boldsymbol{ \mathfrak{Belt}}_{-}\left[\mu,k|\alpha_-,\beta_-\right] &=& -J_1\left(\frac{2 k \pi  r}{\mu }\right) \left(\alpha _- \cos (2 \pi  k z)-\beta _-
   \sin (2 \pi  k z)\right)\,  \text{dr}\nonumber\\
  &-& \frac{\sqrt{\mu ^2+1} r J_1\left(\frac{2 k \pi  r}{\mu }\right) \left(\alpha _- \sin
   (2 \pi  k z)+\beta _- \cos (2 \pi  k z)\right)}{\mu } \text{d$\phi$}\nonumber\\
  &+& \frac{J_0\left(\frac{2 k \pi  r}{\mu }\right) \left(\alpha _- \sin (2 \pi  k z)+\beta
   _- \cos (2 \pi  k z)\right)}{\mu } \, \text{dz} \nonumber\\
   \pmb{\Game} \, \boldsymbol{ \mathfrak{Belt}}_{-}\left[\mu,k|\alpha_-,\beta_-\right] & = & - \,\varpi \, 
   \boldsymbol{ \mathfrak{Belt}}_{-}\left[\mu,k|\alpha_-,\beta_-\right]
\end{eqnarray}
Finally we have the following closed form flows:
\begin{eqnarray}\label{Beltramiflows}
  \boldsymbol{ \mathfrak{clos}}\left[\mu,k|\alpha_0,\beta_0\right] &=&J_1\left(\frac{2 k \pi  r}{\mu }\right) \left(\alpha _0 \cos (2 \pi  k z)+\beta _0 \sin
   (2 \pi  k z)\right) \,  \text{dr}\nonumber\\
  &+& J_0\left(\frac{2 k \pi  r}{\mu }\right) \left(\alpha _0 \mu  \sin (2 \pi  k z)-\beta _0
   \mu  \cos (2 \pi  k z)\right) \text{dz} \nonumber\\
   \pmb{d} \, \boldsymbol{ \mathfrak{clos}}\left[\mu,k|\alpha_0,\beta_0\right] & = & 0
\end{eqnarray}
The above announced decomposition is provided by the following identity:
\begin{equation}\label{frippone}
  \boldsymbol{ \mathfrak{har}}\left[\mu,k|a_1,\dots,a_6\right] \, =\, \boldsymbol{ \mathfrak{Belt}}_{+}\left[\mu,k|\alpha_+,\beta_+\right]+\boldsymbol{ \mathfrak{Belt}}_{-}\left[\mu,k|\alpha_-,\beta_-\right]+\boldsymbol{ \mathfrak{clos}}\left[\mu,k|\alpha_0,\beta_0\right]
\end{equation}
where the relation between the parameters of the harmonic form and the parameters of the other three forms is the following:
\begin{eqnarray}\label{crana}
  &&a_1= -\alpha _-+\alpha _++\alpha _0,\,a_2= \beta _-+\beta _++\beta _0,\,a_3=
   \frac{\beta _+ \sqrt{\mu ^2+1}-\beta _- \sqrt{\mu ^2+1}}{\mu },\,\nonumber\\
   &&a_4= \frac{\alpha
   _- \left(-\sqrt{\mu ^2+1}\right)-\alpha _+ \sqrt{\mu ^2+1}}{\mu },\,a_5=
   \frac{-\beta _0 \mu ^2+\beta _-+\beta _+}{\mu },\,a_6= \frac{\alpha _0 \mu ^2+\alpha
   _--\alpha _+}{\mu }
\end{eqnarray}
The relation between the two sets of parameters $\text{par}_1 \, = \,\left\{a_1,a_2,a_3,a_4,a_5,a_6\right\}$ and
$\text{par}_2\, = \, \left\{\alpha _+,\beta _+,\alpha _-,\beta _-,\alpha _0,\beta _0\right\}$ that is encoded in eq.(\ref{crana}) can be expressed in matrix form:
\begin{equation}\label{leonorstop}
  \text{par}_1 \, = \, \mathcal{Q} \, \text{par}_2 \quad ; \quad  \mathcal{Q} \, = \, \left(
\begin{array}{cccccc}
 1 & 0 & -1 & 0 & 1 & 0 \\
 0 & 1 & 0 & 1 & 0 & 1 \\
 0 & \frac{\sqrt{\mu ^2+1}}{\mu } & 0 & -\frac{\sqrt{\mu ^2+1}}{\mu } & 0 & 0 \\
 -\frac{\sqrt{\mu ^2+1}}{\mu } & 0 & -\frac{\sqrt{\mu ^2+1}}{\mu } & 0 & 0 & 0 \\
 0 & \frac{1}{\mu } & 0 & \frac{1}{\mu } & 0 & -\mu  \\
 -\frac{1}{\mu } & 0 & \frac{1}{\mu } & 0 & \mu  & 0 \\
\end{array}
\right)
\end{equation}
The matrix $\mathcal{Q} $ is non singular and can be inverted so that we can write:
\begin{equation}\label{leonorgoon}
  \text{par}_2 \, = \, \mathcal{Q}^{-1} \, \text{par}_1 \quad ; \quad  \mathcal{Q}^{-1} \, = \, \left(
\begin{array}{cccccc}
 \frac{\mu ^2}{2 \mu ^2+2} & 0 & 0 & -\frac{\mu }{2 \sqrt{\mu ^2+1}} & 0 & -\frac{\mu
   }{2 \mu ^2+2} \\
 0 & \frac{\mu ^2}{2 \mu ^2+2} & \frac{\mu }{2 \sqrt{\mu ^2+1}} & 0 & \frac{\mu }{2 \mu
   ^2+2} & 0 \\
 -\frac{\mu ^2}{2 \mu ^2+2} & 0 & 0 & -\frac{\mu }{2 \sqrt{\mu ^2+1}} & 0 & \frac{\mu
   }{2 \mu ^2+2} \\
 0 & \frac{\mu ^2}{2 \mu ^2+2} & -\frac{\mu }{2 \sqrt{\mu ^2+1}} & 0 & \frac{\mu }{2
   \mu ^2+2} & 0 \\
 \frac{1}{\mu ^2+1} & 0 & 0 & 0 & 0 & \frac{\mu }{\mu ^2+1} \\
 0 & \frac{1}{\mu ^2+1} & 0 & 0 & -\frac{\mu }{\mu ^2+1} & 0 \\
\end{array}
\right)
\end{equation}
Eq.(\ref{leonorgoon}) allows to extract from any given harmonic $1$-form its Beltrami, anti-Beltrami and closed component.
\section{A basis of Beltrami/anti Beltrami plus closed axial symmetric flows in the compact
tube}\label{baseintubata}
Eventually we are interested in hydro-flows confined in the finite tube of fig.\ref{tubastro} which is a compact space with boundary. We have already taken care of the finite extension in the horizontal direction by imposing periodicity in the $z$-coordinate but we have also to confine the flow  in the radial direction in order to take the finite size of the tube into account. To this effect the existence of the infinite set of almost equally spaced zeros of the Bessel functions is very useful. In Wolfram Mathematica the zeros $j_{\ell,n}$ of $J_\ell(x)$ are an available built-in object named \textit{Besselzero$[\ell,n]$}, where $n\in \mathbb{N}$ is their enumeration in increasing order, namely
$j_{\ell,n}<j_{\ell,m}$ if $n<m$. Utilizing such an ingredient we can introduce the following two infinite set of functions:
\begin{eqnarray}
\label{saldatore}
  \mathcal{G}_1^{[n]}(r) &=& \frac{2 J_1\left(r j_{1,n}\right)}{\left|
   J_0\left(j_{1,n}\right)\right| }\nonumber\\
  \mathcal{G}_0^{[n]}(r) &=& \frac{j_{1,n} J_0\left(r j_{1,n}\right)}{\left|
   J_0\left(j_{1,n}\right)\right| } 
\end{eqnarray}
that are well defined in the interval $r\in[0,1]$ and have the following properties:
\begin{eqnarray}
\label{concistoro}
  \mathcal{G}_1^{[n]}(r) &=& -\frac{2}{\left(j_{1,n}\right){}^2} \partial_r \mathcal{G}_0^{[n]}(r) \nonumber\\
  \int_{0}^{1}\,r\, \mathcal{G}_1^{[n]}(r)\, \mathcal{G}_1^{[m]}(r) \, dr   &=& 2 \, \delta_{n,m} \nonumber\\
 \int_{0}^{1}\,r\, \mathcal{G}_0^{[n]}(r)\, \mathcal{G}_0^{[m]}(r) \, dr  &=& \frac{\left(j_{1,n}\right){}^2}{2} \, \delta_{n,m} 
\end{eqnarray}
In terms of these functions we can write the Beltrami anti/Beltrami $1$-forms discussed in the previous section as follows:
\begin{eqnarray}
\label{beltrifinali}
  \Omega^{(n,k)}_\pm\left[\alpha_\pm,\beta_\pm\right] &=& \mathcal{G}_1^{[n]}(r)
   \left(\alpha _\pm^{(k,n)}\,\cos (2 \pi  k z) +\beta _\pm^{(k,n)}\,\sin (2 \pi 
   k z) \right)\,\text{dr} \nonumber\\
  &\pm &\frac{ \varpi(k,n)}{2 \pi  k} \,  r \,  \mathcal{G}_1^{[n]}(r)\,
   \left(\beta _\pm^{(k,n)}\,\cos (2 \pi  k z) -\alpha _\pm^{(k,n)}\,\sin (2 \pi 
   k z) \right)\, \text{d$\phi$} \nonumber\\
   &+&\frac{1}{\pi  k}\,
   \mathcal{G}_0^{[n]}(r) \left(\beta_\pm^{(k,n)}\,\cos (2 \pi  k z) -\alpha_\pm^{(k,n)}\,\sin (2 \pi  k z) \right)\,\text{dz}    
\end{eqnarray}
where:
\begin{equation}\label{eigencoso}
  \varpi(n,k) \, = \, \sqrt{4 \pi ^2 k^2+\left(j_{1,n}\right){}^2}
\end{equation}
is the eigenvalue of the Beltrami operator:
\begin{equation}\label{beltramini}
  \boldsymbol{\Game }\,\Omega^{(n,k)}_\pm\left[\alpha_\pm,\beta_\pm\right]\,  = 
  \, \pm \varpi(n,k) \, \Omega^{(n,k)}_\pm\left[\alpha_\pm,\beta_\pm\right]\,
\end{equation}
and $\alpha_\pm,\beta_\pm$ are the two free parameters at fixed
spectral type $(n,k)$. Indeed as $k$ is the quantized momentum in the horizontal direction $z$ in the same way the number $n$ enumerating the zeros of the Bessel function is the analogue of the quantized momentum in the radial direction. 
\par
Altogether, including also the closed $1$-forms discussed in section \ref{beltrachiude}, at fixed spectral type $(n,k)$ we can utilize a basis of six hydro $1$-forms as follows:
\begin{eqnarray}
\label{cunegonda}
  {\Omega}A_\pm[n,k] &=& \cos (2 \pi  k z)
   \mathcal{G}_1^{[n]}(r)\,\text{dr} \mp\frac{\varpi (n,k)}{2 \pi 
   k} \, \sin (2 \pi  k z)\, r    \mathcal{G}_1^{[n]}(r)\,\text{d$\phi$}-\frac{ \sin (2 \pi  k z) \mathcal{G}_0^{[n]}(r)}{\pi  k}\,\text{dz}   \nonumber\\
  {\Omega}B_\pm[n,k] &=& \sin (2 \pi  k z)
   \mathcal{G}_1^{[n]}(r)\,\text{dr}\pm\frac{\varpi (n,k)}{2 \pi 
   k} \, \cos (2 \pi  k z)\, r    \mathcal{G}_1^{[n]}(r)\,\text{d$\phi$}+\frac{ \cos (2 \pi  k z) \mathcal{G}_0^{[n]}(r)}{\pi  k}\,\text{dz}\nonumber\\
    {\Omega}A_0[n,k] &=& \cos
   (2 \pi  k z) \mathcal{G}_1^{[n]}(r)\, \text{dr}\, +\, \frac{4 \pi  k  \sin (2 \pi  k z) \mathcal{G}_0^{[n]}(r)}{\mathfrak{j}_n^2}\, \text{dz} \nonumber\\
   {\Omega}B_0[n,k] &=& \sin
   (2 \pi  k z) \mathcal{G}_1^{[n]}(r)\, \text{dr}\, -\, \frac{4 \pi  k  \cos (2 \pi  k z) \mathcal{G}_0^{[n]}(r)}{\mathfrak{j}_n^2}\, \text{dz}
\end{eqnarray}
that satisfy Beltrami equation in the form:
\begin{alignat}{5}\label{gambellino}
  \boldsymbol{\Game} \,{\Omega}A_\pm[n,k] &\quad =\quad & \pm \, \varpi(k,n) \,{\Omega}A_\pm[n,k] & \quad ; \quad &
  \boldsymbol{\Game} \,{\Omega}B_\pm[n,k] & \quad =\quad & \pm \, \varpi(k,n) \,{\Omega}B_\pm[n,k] &\null \nonumber\\
  \boldsymbol{\Game} \,{\Omega}A_0[n,k] & \quad =\quad & 0 & \quad ; \quad &
  \boldsymbol{\Game} \,{\Omega}B_0[n,k] &\quad =\quad & 0 & \null
\end{alignat}
The second line of the above equation (\ref{gambellino}) is certainly true since the harmonic $1$-forms $\boldsymbol{\Game} \,{\Omega}A_0[n,k]$,$\boldsymbol{\Game} \,{\Omega}B_0[n,k]$ are closed.
\par 
Hence we can write the general ansatz for a steady axial symmetric flow as follows:
\begin{equation}
\label{cassiodoro}
\Omega^{[\mathbf{U}]}\, = \, \sum_{n=1}^\infty \left(\sum_{k=1}^\infty \, \left( a_\pm[n,k] \, {\Omega}A_\pm[n,k] + 
b_\pm[n,k] \, {\Omega}B_\pm[n,k]\right)\, + \, \sum_{k=0}^\infty \, \left( a_0[n,k] \, {\Omega}A_0[n,k] + 
b_0[n,k] \, {\Omega}B_0[n,k]\right)\right)
\end{equation}
where the coefficients:
\begin{equation}\label{comencini}
  c[s] \, \equiv \, \left\{a_\pm[n,k], b_\pm[n,k], a_0[n,k], b_0[n,k] \right\}
\end{equation}
are the cylindrical analogue of the Fourier coefficients in a conventional Fourier expansion and are those
anticipated in the introduction in the schematic formula \ref{dilatonato}. Furthermore similarly to the case of the standard Fourier expansion, but with a significant difference,  one can split each spectral
cell $(n,k)$ in three parts, the \textbf{Beltrami part}  the \textbf{anti-Beltrami part} and the \textbf{irrotational part}, namely
the contribution from the closed $1$-forms. Indeed at fixed spectral numbers $(n,k)$ and at fixed type $A$ or $B$,
the difference between Beltrami, anti Beltrami and closed is the sign of the angular velocity $\mathbf{U}_\phi$. 
Beltrami and anti-Beltrami components of the fluid rotate around the central axis in opposite direction, while 
closed $1$-form components do not rotate at all. The physical interpretation of Beltrami flows is therefore
very simple and intuitive in the cylindrical setup. Such interpretation also clarifies the importance of the inclusion of the closed $1$-forms in the general ansatz. Indeed when a Beltrami and an anti-Beltrami component with the same spectral numbers meet in the non-linear interaction term provided by the diamond product (see below section
\ref{gibuti}), they coalesce into an irrotational component, namely into a closed $1$-form contribution. 
\par
In view of this simple and beautiful physical interpretation,  the construction of the Beltrami index of a flow advocated in \cite{beltraspectra} and reviewed in \cite{Fr__2023} has to be reconsidered and slightly modified when dealing with flows in a cylindrical set up, mathematically modeled by the tube (\ref{torone}): the Beltrami index in this case is tripartite: \textbf{the laevo-rotatory, dextro-rotatory and irrotational} percentage of the flow.  
\par
One might ask why the summations on $n$ and $k$ in (\ref{cassiodoro}) start from $1$ for the Beltrami $1$-forms and not from $0$. The answer arises from a careful consideration of the zero modes. 
\subsection{The zero-modes}
Some special care is needed with the zero-modes namely when either $k$ or $n$, or both go the value $0$. The Beltrami flows have to be defined by means of a limiting procedure in these cases. Let us first consider the case 
$k=0$, namely the trigonometric zero modes.
\subsubsection{Trigonometric zero modes}
The correct starting point is the consideration of the eigenvalue of  Beltrami equation as given in eq.(\ref{eigencoso}). We see that $\varpi(0,n) \, = \, j_{1,n}$ hence we have to perform a limit $k\to 0$ on the formulae (\ref{cunegonda}) in such a way as to obtain four $1$-forms 
${\Omega}A_\pm[n,0],{\Omega}B_\pm[n,0]$ that satisfy Beltrami equations as in eq.(\ref{gambellino}) with eigenvalue $\pm\varpi(0,n) \, = \, \pm j_{1,n}$. For the case of the ${\Omega}A_\pm[n,0]$ hydro $1$-forms it is just sufficient to expand the object ${\Omega}A_\pm[n,k]$ in power series of $k$ and isolate the first constant term. We obtain
\begin{equation}\label{funiculi}
  {\Omega}A_\pm[n,0] \, = \, \mathcal{G}_1^{[n]}(r)\,\text{dr} \, \mp \,  j_{1,n} \,z \, r   \,  \mathcal{G}_1^{[n]}(r)\,\text{d$\phi $} \, - \, 2 \, z \,
   \mathcal{G}_0^{[n]}(r)\, \text{dz}
\end{equation}
For the case of the ${\Omega}B_\pm[n,0]$ hydro $1$-forms we have to multiply it first by $k$ (a rescaling of the corresponding coefficients in the expansion) and then expand in power series of $k$ as in the previous case. So doing we obtain: 
\begin{equation}\label{funicula}
  {\Omega}B_\pm[n,0] \, = \, \frac{1}{2 \pi } \left( \pm \, j_{1,n}\, r \,\mathcal{G}_1^{[n]}(r)\,\text{d$\phi $} \, +\, 2  \, \mathcal{G}_0^{[n]}(r) \, \text{dz}\right)
\end{equation}
Both ${\Omega}A_\pm[n,0]$ and ${\Omega}B_\pm[n,0]$ satisfy Beltrami   eq.(\ref{gambellino}) with the correct eigenvalue:
\begin{equation}\label{porcellino}
  \boldsymbol{\Game} \,{\Omega}A_\pm[n,0] \, = \, \pm \, j_{1,n} \,{\Omega}A_\pm[n,0] \quad  ; \quad
  \boldsymbol{\Game} \,{\Omega}B_\pm[n,0] \, = \, \pm \, j_{1,n} \,{\Omega}B_\pm[n,0]
\end{equation}
It must be observed that the flows (\ref{funiculi}) are not periodic in $z$, due to the explicit linear $z$-dependence of some components. The flows (\ref{funicula}) are instead periodic in $z$ for the simple reason that they do not depend on $z$.
\subsubsection{Bessel zero modes}
Next we consider the Bessel zero modes, namely those where we put $n=0$. Also in this case we need to provide an interpretation of the definition (\ref{saldatore}). The object $j_{1,0}$ is the zero of $J_1(x)$ of position $0th$ 
which does not mean anything. We interpret $j_{1,0}=1$ and we obtain the two functions
\begin{equation}\label{definoni}
  \mathcal{G}_0^{[0]}(r) \, = \, \frac{J_0(r)}{J_0(1)} \quad ; \quad \mathcal{G}_1^{[0]}(r) \, = \, 2\frac{J_1(r)}{J_0(1)}
\end{equation}
with such a definition we see that the ${\Omega}A_\pm[0,k]$ and ${\Omega}B_\pm[0,k]$ satisfy Beltrami equation with the correct eigenvalue that now is $\varpi(0,k) \,= \,  \sqrt{4 \pi ^2 k^2+1}$.
\par
\begin{figure}[!hbt]
\begin{center}
\includegraphics[width=80mm]{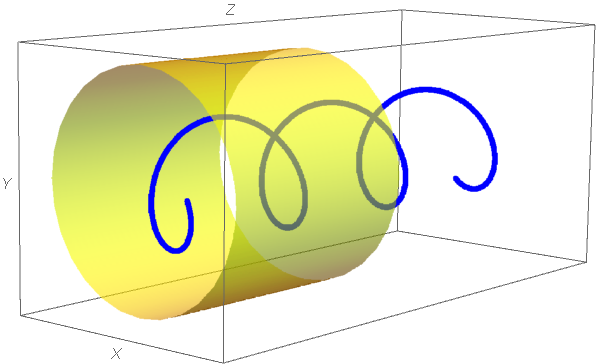}
\caption{\it  In this figure we show one streamline starting at $r=1/2,\phi=0,z=-1$ of the Beltrami flow
${\Omega}B_0[5]$. As one sees the radius never changes during the time evolution, while the longitudinal coordinate $z$ advances linearly in time as the angle $\phi$. The result is the helix shown in the picture.}\label{helica}
\end{center}
\end{figure}
\begin{figure}[!hbt]
\begin{center}
\includegraphics[width=80mm]{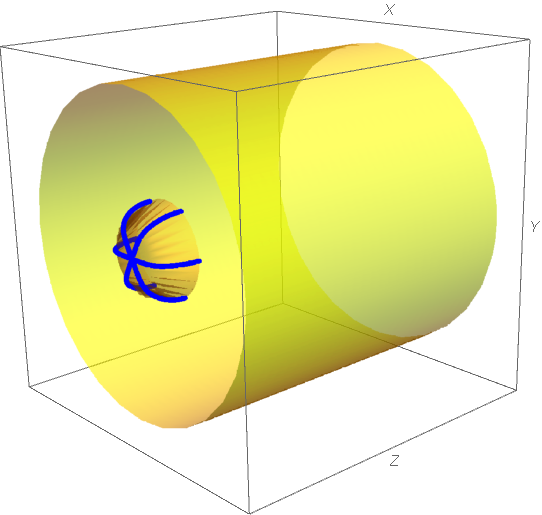}
\caption{\it  In this figure we show a few streamlines  of the Beltrami flow
${\Omega}B_+[1,1]$. The initial point is at $r_0 = 0.3,\, z_0=-0.75$ for all the displayed stremalines that differ among themselves only for the initial angle $\phi_0$. In the evolution of the streamlines associated with closed $1$-form flows the radius always goes to zero and the streamline stops there. The coordinate $z$ oscillates a little bit and then reaches  an asymptotic value. The angle $\phi$ remains constant. Instead of enveloping a torus surface as the Beltrami, anti/Beltrami streamline do, the closed $1$-form streamlines envelop a cone whose vertex is the stop point.}\label{gustocono}
\end{center}
\end{figure}

However as we just observed in the case of the trigonometric zero-modes, also the Bessel zero-modes are unacceptable
since they violate the tube-boundary conditions. Indeed at $r=1$ the velocity field in the radial direction should vanish, but this is not true since $\mathcal{G}_1(0,1)  \neq 0$.
\paragraph{The only admitted zero-modes}
In conclusion the only $0$-modes that might eventually be included in the summation, if necessary for closure of the
function set, are ${\Omega}B_\pm[n,0]$, displayed in eq.(\ref{funicula}). The corresponding flows have zero velocity in the radial direction and a constant angular velocity and longitudinal velocity at every value of the radius. 
The streamlines of such a flow are helicoidal longitudinal push-forward lines as shown in picture \ref{helica}
Hence the expansion (\ref{cassiodoro}) can be improved by adding an extra term as follows:
\begin{eqnarray}
\label{cassiodargento}
\Omega^{[\mathbf{U}]}& = & \sum_{n=1}^\infty \left(\sum_{k=1}^\infty \, \left( a_\pm[n,k] \, {\Omega}A_\pm[n,k] + 
b_\pm[n,k] \, {\Omega}B_\pm[n,k]\right)\,\right.\nonumber\\
&&\left. + \, \sum_{k=0}^\infty \, \left( a_0[n,k] \, {\Omega}A_0[n,k] + 
b_0[n,k] \, {\Omega}B_0[n,k]\right)+c_\pm[n]{\Omega}B_\pm[n,0]\right)\nonumber\\
\end{eqnarray}
In fig.\ref{gustocono} we display an example of streamlines for a closed $1$-form flow.
\subsection{Functional space, scalar product and norms}
That those in eq.(\ref{cunegonda}) constitute a basis of orthogonal functions for the development in the double series of the hydro $1$-form as specified by eq.(\ref{cassiodargento}) follows by appropriately defining the functional space $L^2_{tube}$ of $\mathbf{U}$ hydro vector fields (or dual hydro $1$-forms) on the \text{tube} $\mathfrak{T}$ of fig. \ref{tubastro}. Topologically the space $\mathfrak{T}$  is the product of a circle (spanned by the angle variable $\phi$) with  a parallelogram spanned by the coordinates $z\in [-1,1]$ $\times$ $r\in[0,1]$. Imposing perdiodic boundary conditions, actually we  have identified the boundary disks (the red and the green one of fig. \ref{tubastro} ) so that the effective topology of $\mathfrak{T}$ is the interior of a genus 1 torus. Mathematically we can state that:
\begin{equation}\label{torone}
  \mathfrak{T} \, \simeq \, \mathbb{S}^1 \times \mathbb{S}^1 \times [0,1]
\end{equation}
where the first circle is spanned by the angle $\phi$, the second circle by the angle $\psi \, \equiv \, \pi (z+1)$ and the interval $[0,1]$ by the radial variable $r$. The boundary of the tube $\mathfrak{T}$, which is indeed a $2$-torus corresponds to the locus $r=1$:
\begin{equation}\label{bendaggio}
  \partial \mathfrak{T} \, = \, \{1,\phi,z\} \, \simeq \, \mathbb{S}^1 \times \mathbb{S}^1 
\end{equation}
On the boundary the velocity field $\mathbf{U}$ must be tangent to the boundary, namely its first (radial) component must vanish:
\begin{equation}\label{vontruppen}
  \mathbf{U}^1\left(1,\phi,z\right) \, = \, 0
\end{equation}
which is guaranteed by the universal zero (by construction) at $r=1$  of the functions $\mathcal{G}_1^{[n]}(r)$. In force of these observations the vector fields dual to the hydro $1$-forms (\ref{cunegonda}), namely
\begin{eqnarray}
\label{Adelchi}
  {\mathbf{U}}A_\pm[n,k] &=& \left\{ \cos (2 \pi  k z)
   \mathcal{G}_1^{[n]}(r)\, ,\, \mp\frac{\varpi (k,n)}{2 \pi 
   k} \, \sin (2 \pi  k z)\, \frac{1}{r}    \mathcal{G}_1^{[n]}(r)\,,\,-\frac{ \sin (2 \pi  k z) \mathcal{G}_0^{[n]}(r)}{\pi  k} \right\}  \nonumber\\
  {\mathbf{U}}B_\pm[n,k] &=& \left\{\sin (2 \pi  k z)
   \mathcal{G}_1^{[n]}(r)\,,\, \pm\frac{\varpi (k,n)}{2 \pi 
   k} \, \cos (2 \pi  k z)\, \frac{1}{r}    \mathcal{G}_1^{[n]}(r)\, ,\,\frac{ \cos (2 \pi  k z) \mathcal{G}_0^{[n]}(r)}{\pi  k}\,\right\} \nonumber\\
    {\mathbf{U}}A_0[n,k] &=& \left\{ \cos (2 \pi  k z)
   \mathcal{G}_1^{[n]}(r)\, , \, 0\, , \,\frac{4\pi\,k\, \sin (2 \pi  k z) \mathcal{G}_0^{[n]}(r)}{j^2_{1,n}}\right\}
   \nonumber\\
    {\mathbf{U}}B_0[n,k] &=&\left\{ \sin (2 \pi  k z)
   \mathcal{G}_1^{[n]}(r)\, , \, 0\, , \,-\frac{4\pi\,k\, \cos (2 \pi  k z) \mathcal{G}_0^{[n]}(r)}{j^2_{1,n}} \right\}\nonumber\\
   {\mathbf{U}}B_\pm[n,0] &=& \left\{\,0\, ,\,\pm\frac{j_{1,n}}{2 \pi } \, \, \frac{\mathcal{G}_1^{[n]}(r)}{r}    \, ,\,\frac{ \mathcal{G}_0^{[n]}(r)}{\pi}\,\right\} \nonumber\\
\end{eqnarray}
 are all well defined on the tube $\mathfrak{T}$ and satisfy the boundary condition (\ref{vontruppen}). 
 Identically it happens for the closed $1$-forms ${\Omega}A_0[n,k]$,$\,\,{\Omega}B_0[n,k]$ (see eq.(\ref{cunegonda})). In the case of the zero modes ${\Omega}B_\pm[n]$ the radial velocity vanishes everywhere so the condition is a fortiori realized.   
 \par
For the vector fields ${\mathbf{U}}$ well defined on  $\mathfrak{T}$ and satisfying the boundary condition 
(\ref{vontruppen}) we calculate the squared norm as:
\begin{equation}\label{norm}
\|\mathbf{U}\|^2 \, \equiv \,  \int_{\mathfrak{T}} \, \mathbf{U}^i\mathbf{U}^j \, g_{ij} \boldsymbol{\text{d$\mu$}}[r,\phi,z]
\, = \, \int_{0}^{2\pi} \, \text{d$\phi$} \, \int_{0}^{1} \, \text{dr} \, \int_{-1}^{1} \, \text{dz}
\,\left[ (\mathbf{U}^1)^2  \, + \, r^2 \, (\mathbf{U}^2)^2 \, + \, (\mathbf{U}^3)^2 \right]
\end{equation}
The vector field $\mathbf{U}$ belongs to the functional space $L^2_{tube}$ if its norm is finite:
\begin{equation}\label{finitelenorme}
 \mathbf{U}\in L^2_{tube} \quad \text{iff} \quad   \|\mathbf{U}\|^2 \, < \, \infty \quad \text{and} \quad \mathbf{U}^1\left(1,\phi,z\right) \, = \, 0
\end{equation}
It is important to stress that in the definition (\ref{finitelenorme}) it is nowhere assumed that the components of the vector field should depend only on $r$ and $z$. Namely the velocity fields $\mathbf{U}$ well defined and meaningful inside the tube form a functional space much larger than that formed by those that are axial symmetric. Then we have a proper functional subspace:
\begin{equation}\label{axialsymL2}
  L^2_{tube} \, \supset \, L^2_{axial} \quad = \quad \left\{ \mathbf{U} \in L^2_{tube}  \, \mid \, \partial_\phi \mathbf{U}_i \, = \,0, i=1,2,3\right\}
\end{equation}
Within the whole functional space $L^2_{tube}$ equation (\ref{norm}) is generalized to define a hermitian scalar product of two vector fields $\mathbf{U},\mathbf{V}\in L^2_{tube}$ as follows:
\begin{equation}\label{frugiferentis}
  \langle \mathbf{U}\, \mid \, \mathbf{V}\rangle \, =\, \int_{\mathfrak{T}} \, \mathbf{U}^i\mathbf{V}^j \, g_{ij} \boldsymbol{\text{d$\mu$}}[r,\phi,z]\, = \, \int_{0}^{2\pi} \, \text{d$\phi$} \, \int_{0}^{1} \, \text{dr} \, \int_{-1}^{1} \, \text{dz}
\,\left[ \mathbf{U}^1\mathbf{V}^1  \, + \, r^2 \, \mathbf{U}^2\mathbf{V}^2 \, + \, \mathbf{U}^3\mathbf{V}^3 \right]
\end{equation}
The same scalar product (\ref{frugiferentis}) and hence  the squared norm (\ref{norm}) $\|\mathbf{U}\|^2 \, = \, 
\langle \mathbf{U}\, \mid \, \mathbf{U}\rangle$ can also be rewritten in terms of the corresponding hydro $1$-forms in a way that is very useful when dealing with Beltrami flows:
 \begin{equation}\label{frostasapore}
  \langle \mathbf{U}\, \mid \, \mathbf{V}\rangle \, = \, \int_{\mathfrak{T}} \Omega^{\mathbf{U}} \wedge \star_g \, \Omega^{\mathbf{V}} 
\end{equation}
The above being established we can calculate the scalar products of the vector fields listed in (\ref{Adelchi}):
\begin{eqnarray}
\label{conglomerato}
 \langle {\mathbf{U}}A_\pm[n_1,k_1] \, \mid \, {\mathbf{U}}A_\pm[n_2,k_2]\rangle  &=& \delta_{n1,n2} \, \times \,\delta_{k1,k2} \times \frac{\varpi^2(n_1,k_1)}{k_1^2 \, \pi^2} \nonumber\\
 \langle {\mathbf{U}}B_\pm[n_1,k_1] \, \mid \, {\mathbf{U}}B_\pm[n_2,k_2]\rangle  &=& \delta_{n1,n2} \, \times \,\delta_{k1,k2} \times \frac{\varpi^2(n_1,k_1)}{k_1^2 \, \pi^2} \nonumber\\
 \langle {\mathbf{U}}A_0[n_1,k_1] \, \mid \, {\mathbf{U}}A_0[n_2,k_2]\rangle  &=& \delta_{n1,n2} \, \times \,\delta_{k1,k2} \times \frac{2\,\varpi^2(n_1,k_1)}{j^2_{1,n_1}} \nonumber\\
 \langle {\mathbf{U}}B_0[n_1,k_1] \, \mid \, {\mathbf{U}}B_0[n_2,k_2]\rangle  &=& \delta_{n1,n2} \, \times \,\delta_{k1,k2} \times \frac{2\,\varpi^2(n_1,k_1)}{j^2_{1,n_1}} \nonumber\\
 \langle {\mathbf{U}}A_0[n_1,k_1] \, \mid \, {\mathbf{U}}B_0[n_2,k_2]\rangle  &=& 0 \nonumber\\
 \langle {\mathbf{U}}A_0[n_1,k_1] \, \mid \, {\mathbf{U}}B_\pm[n_2,k_2]\rangle &=& 0 \nonumber\\
 \langle {\mathbf{U}}B_0[n_1,k_1] \, \mid \, {\mathbf{U}}A_\pm[n_2,k_2]\rangle  &=& 0 \nonumber\\
 \langle {\mathbf{U}}B_0[n_1,k_1] \, \mid \, {\mathbf{U}}B_\pm[n_2,k_2]\rangle &=& 0 \nonumber\\
\langle {\mathbf{U}}A_\pm[n_1,k_1] \, \mid \, {\mathbf{U}}A_\mp[n_2,k_2]\rangle&=& 0 \nonumber \\
\langle {\mathbf{U}}B_\pm[n_1,k_1] \, \mid \, {\mathbf{U}}B_\mp[n_2,k_2]\rangle & = & 0 \nonumber\\
\langle {\mathbf{U}}A_\pm[n_1,k_1] \, \mid \, {\mathbf{U}}B_\pm[n_2,k_2]\rangle & = & 0 \nonumber\\
\langle {\mathbf{U}}A_\pm[n_1,k_1] \, \mid \, {\mathbf{U}}B_\mp[n_2,k_2]\rangle & = & 0 \nonumber\\
\end{eqnarray}
The vector fields (\ref{Adelchi}) constitute an orthogonal  basis for the functional space $L^2_{axial}$ of axial symmetric velocity fields in the tube. This was anticipated by means of the expansion (\ref{cassiodoro}) namely:
\begin{eqnarray}
\label{apuleio}
\mathbf{U}& = & \sum_{n=1}^\infty \left(\sum_{k=1}^\infty \, \left( a_\pm[n,k] \, {\mathbf{U}}A_\pm[n,k] + 
b_\pm[n,k] \, {\mathbf{U}}B_\pm[n,k]\right)\,\right.\nonumber\\
&&\left. + \,\sum_{k=0}^\infty \, \left( a_0[n,k] \, {\mathbf{U}}A_0[n,k] + 
b_0[n,k] \, {\mathbf{U}}B_0[n,k]\right)+c_\pm[n]{\mathbf{U}}B_\pm[n,0] \right)
\end{eqnarray}
In the above series, if we have $\mathbf{U}$ given by any other source of information, 
the coefficients $a_\pm[n,k]$,$b_\pm[n,k]$ are obtained by the following scalar products:
\begin{equation}\label{franziskus}
  a_\pm[n,k] \, = \, \frac{k^2 \, \pi^2}{\varpi^2(n,k)} \, \langle\mathbf{U} \, \mid \, {\mathbf{U}}A_\pm[n,k] \rangle
  \quad ; \quad b_\pm[n,k] \, = \, \frac{k^2 \, \pi^2}{\varpi^2(n,k)} \, \langle\mathbf{U} \, \mid \, {\mathbf{U}}B_\pm[n,k]\rangle
\end{equation}
and similarly for all the other elements of the basis.
\subsubsection{Stream-lines}
\begin{figure}[!hbt]
\begin{center}
\includegraphics[width=80mm]{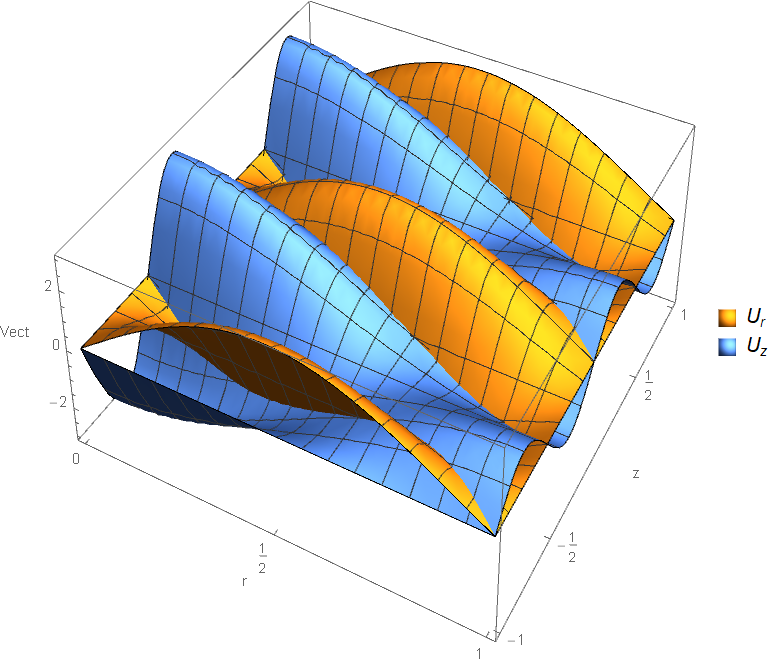}
\includegraphics[width=70mm]{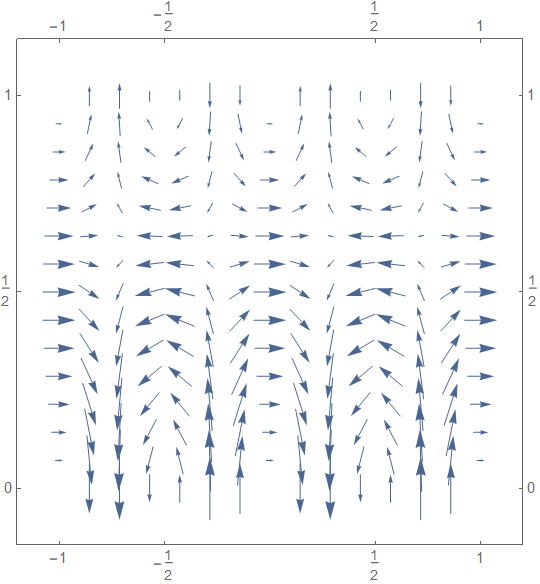}
\end{center}
\caption{\it Visualization of the Beltrami vector field in eq.(\ref{Ermengarda}). In the picture on the left we have made a plot of the two components of the velocity $\mathbf{U}_z$ and $\mathbf{U}_r$. In the picture on the right we have instead displayed the same information as a vector plot in two dimension. One notes the circulation structure of the flow, that is demonstrated in the fig.\ref{circulari1} where a set of stream-lines generated by this vector field is displayed. }  \label{lowestbeltra}
 \hskip 1cm \unitlength=1.1mm
\end{figure}

\begin{figure}[!hbt]
\begin{center}
\includegraphics[width=80mm]{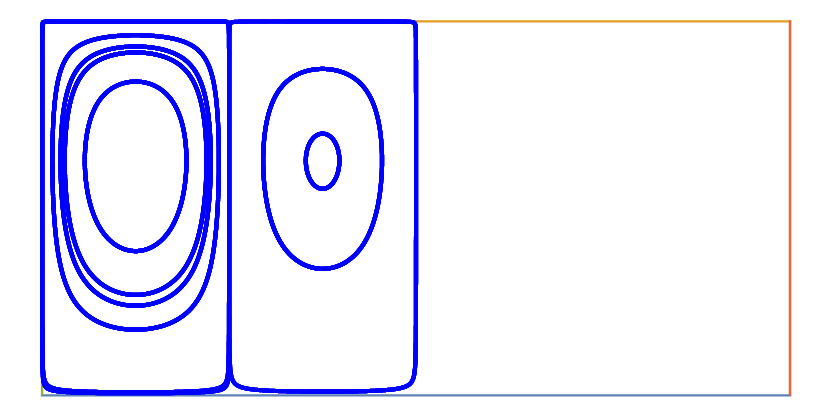}
\includegraphics[width=70mm]{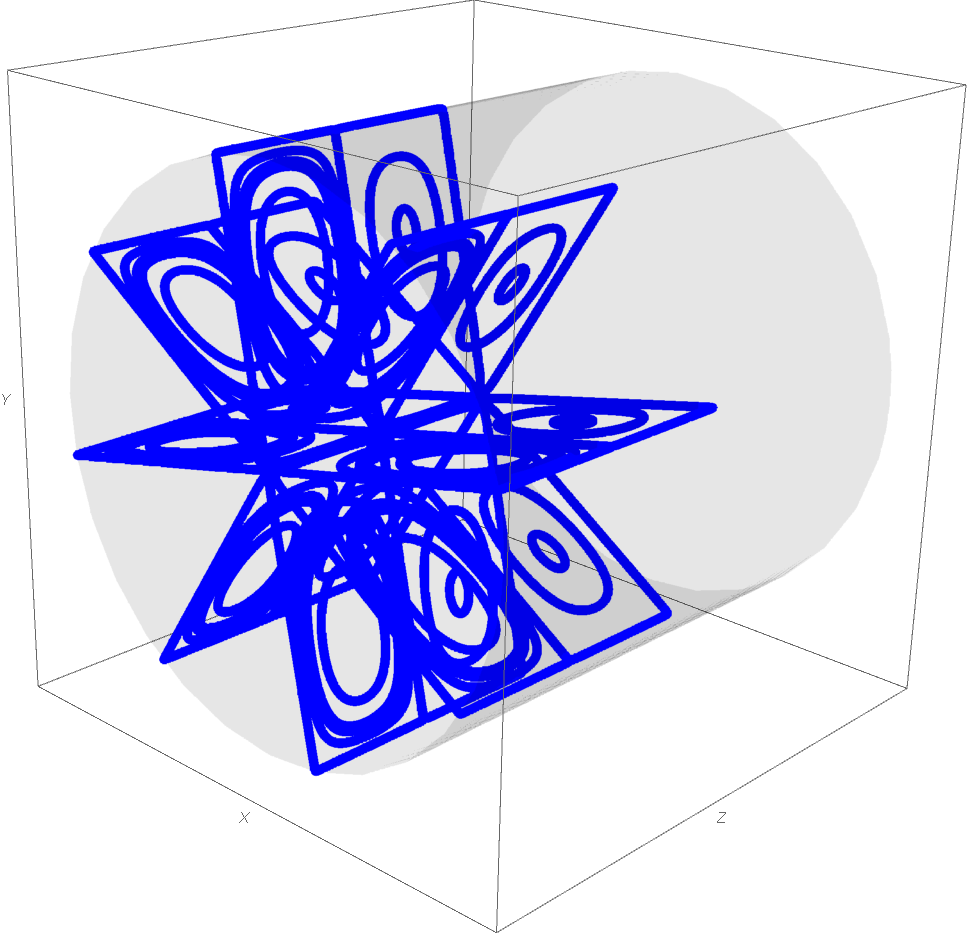}
\includegraphics[width=70mm]{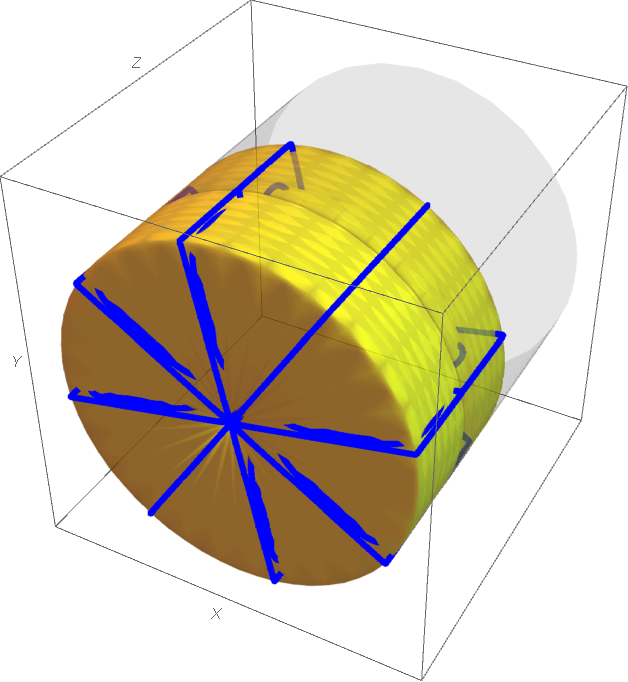}
\includegraphics[width=70mm]{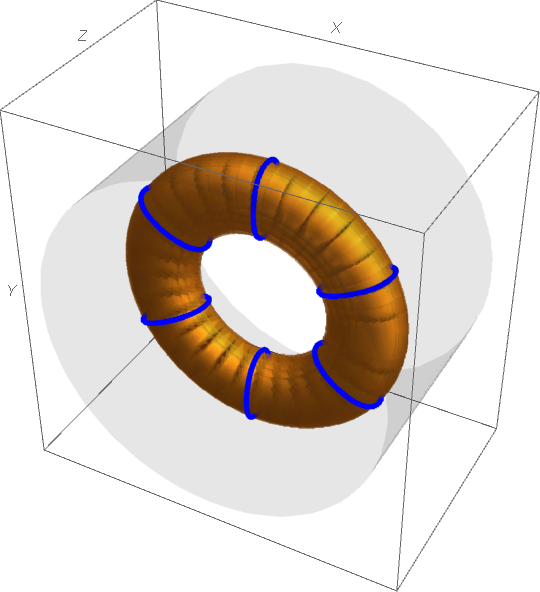}
\end{center}
\caption{\it Visualization of nine stream-lines of the vector field in eq.(\ref{Ermengarda}). In the first picture on the left we display the nine streamlines in the $z,r$ plane. As one sees, they are all closed loops in the plane.
In the second picture of the first line we show the same loops in the three-dimensional space. They are internal to the cylinder. Each loop is repeated eight times by taking equally spaced initial angles $\theta$. Taking all values of $\theta$ from $0$ to $2\pi$ from each loop one generates a torus, as displayed in the last picture on the right of
the second line. The stream-lines wind around such a torus as $A$-cycles. Therefore there is an infinity of such 
torii and the stream-lines wrap around them. In the first figure on the left in the second line we display the nine torii associated with the nine considered $(z,r)$-stream lines. }  \label{circulari1}
\hskip 1cm \unitlength=1.1mm
\end{figure}

\begin{figure}[!hbt]
\begin{center}
\includegraphics[width=80mm]{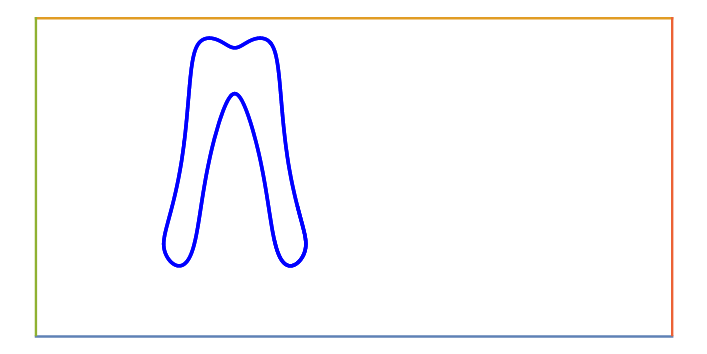}
\includegraphics[width=70mm]{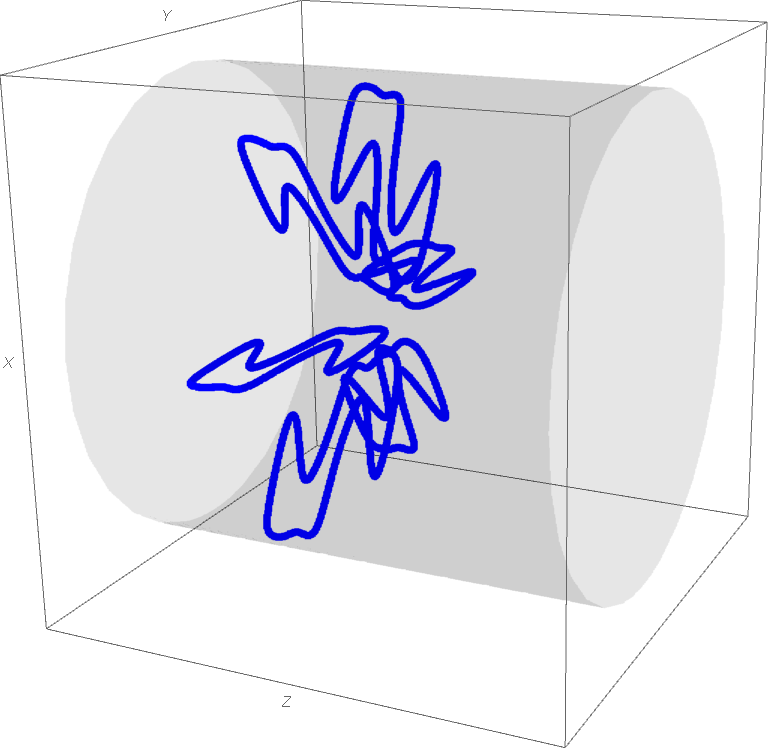}
\includegraphics[width=70mm]{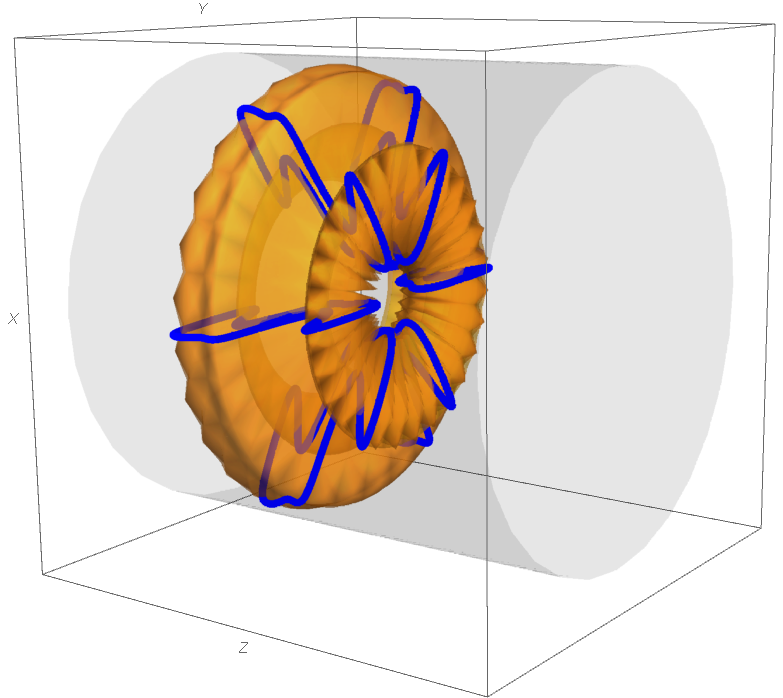}
\end{center}
\caption{\it Visualization of one stream-line of the vector field in eq.(\ref{comboU}). In the first picture on the left we display the streamlines in the $z,r$ plane. In the second picture of the first line we show the same loop in the three-dimensional space. The loop is repeated six times by taking equally spaced initial angles $\theta$.  Taking all values of $\theta$ from $0$ to $2\pi$  one generates a torus, as displayed in the last picture. }  \label{braghefruste}
\hskip 1cm \unitlength=1.1mm
\end{figure}

In order to understand the structure of the axial symmetric flows it is useful to consider the integration of the first order equations that provides the trajectory or stream-lines of a fluid infinitesimal element given an initial position and fixed the form of the velocity-field. We choose one $\mathbf{U}$, for instance an elementary Beltrami flow that solves the inviscid Navier Stokes stationary equation, namely equation (\ref{gongolone}) with constant Bernoulli function $H_B$ and zero viscosity $\nu$. The form of the first order differential equations for the stream-lines is the following:
\begin{eqnarray}
\label{baltica}
  \frac{\mathrm{d}}{\mathrm{d}t}\,r(t) &=& \mathbf{U}_1\left(r(t),z(t)\right) \nonumber\\
\frac{\mathrm{d}}{\mathrm{d}t}\,z(t) &=& \mathbf{U}_3\left(r(t),z(t)\right) \nonumber\\
\frac{\mathrm{d}}{\mathrm{d}t}\,\phi(t) &=& \mathbf{U}_2\left(r(t),z(t)\right) 
\end{eqnarray}
Because of the axial symmetry, the non-trivial differential system to be solved is provided only by the first two equations in (\ref{baltica}), for the unknown functions $r(t),z(t)$. In any given vector field $\mathbf{U}(r,z)$ such a differential system is highly non-linear and can be solved only numerically by fixing first the initial conditions $r_0,z_0$. Given the solution $r(t,r_0,z_0), z(t,r_0,z_0)$ by replacing it into the last equation we find
\begin{equation}\label{garittula}
  \frac{\mathrm{d}}{\mathrm{d}t}\,\phi(t) \, = \, \mathbf{U}_2\left(r(t,r_0,z_0),z(t,r_0,z_0)\right)
\end{equation}
which is already reduced to quadratures. Indeed the solution for $\phi(t)$ is:
\begin{equation}\label{finocchiona}
  \phi(t) \, = \, \int_0^t \, \mathbf{U}_2\left(r(\tau,r_0,z_0),z(\tau,r_0,z_0)\right) \, \mathrm{d}\tau \, + \, \theta \quad ; \quad \theta \in [0,2\pi]
\end{equation}
This means that each time we construct a solution $r(t,r_0,z_0), z(t,r_0,z_0)$ of the differential system:
\begin{eqnarray}
\label{prebaltica}
  \frac{\mathrm{d}}{\mathrm{d}t}\,r(t) &=& \mathbf{U}_1\left(r(t),z(t)\right) \nonumber\\
\frac{\mathrm{d}}{\mathrm{d}t}\,z(t) &=& \mathbf{U}_3\left(r(t),z(t)\right) 
\end{eqnarray}
by means of eq.(\ref{finocchiona}) we have not only just one  stream line, rather an entire surface of stream lines 
letting $\theta$ take all its possible values. 
\paragraph{An example with the Beltrami flow $\mathbf{U}A_+[1,1]$}
Referring to eq.(\ref{Adelchi}) consider:
\begin{eqnarray}
\label{Ermengarda}
  {\mathbf{U}}A_+[1,1] &=& \left\{ \cos (2 \pi   z)
   \mathcal{G}_1^{[1]}(r)\, ,\, \mp\frac{\varpi (1,1)}{2 \pi 
   } \, \sin (2 \pi   z)\, \frac{1}{r}    \mathcal{G}_1^{[1]}(r)\,,\,-\frac{ \sin (2 \pi   z) \mathcal{G}_0^{[1]}(r)}{\pi  } \right\} 
\end{eqnarray}
The best way to visualize the vector field is to consider the plot of the two relevant components of the reduced
differential system (\ref{prebaltica}) as functions over the rectangle $(z,r)$ and at the same time perform a vector plot in two-dimension. This is what we do in fig.\ref{lowestbeltra} for the Beltrami field (\ref{Ermengarda}).
Observing fig.\ref{lowestbeltra} one sees the periodicity of the field in the $z$-direction. Indeed the incoming arrows at $z=-1$ are just into one-to-one correspondence with the outgoing arrows at $z=1$. In the $r$ direction the flow goes up and down between $r=0$ and $r=1$ while it oscillates between two values of $z$. It means that we should expect stream lines that are closed loops in the $z,r$ plane. Indeed this is what is revealed by the explicit numerical integration of the differential system (\ref{prebaltica}). In fig.\ref{circulari1} we display nine
streamlines in the $z,r$ plane obtained taking equally spaced  initial boundary conditions
$r^i_0,z^i_0$, ($i=1,\dots,9$).
\paragraph{An example with a linear combination of two Beltrami flows} 
In order to appreciate the complexity of flows that can be generated by the superposition
of elementary Beltrami/anti-Beltrami flows we just consider one example where we use the following linear 
combination:
\begin{equation}\label{comboU}
  \mathbf{U}\, = \, \frac{1}{5} \, {\boldsymbol{U}}A_+[2,2]\, + \, \frac{1}{3}\, {\boldsymbol{U}}B_-[1,5]\,
\end{equation}
that involves a Beltrami and an anti-Beltrami flow with different spectral type. The structure of this composed velocity field  is displayed in fig.\ref{mistoconpanna}, to be compared with that of the pure Beltrami vector field (\ref{Ermengarda}), previously displayed in fig. \ref{lowestbeltra}.
\begin{figure}[!hbt]
\begin{center}
\includegraphics[width=80mm]{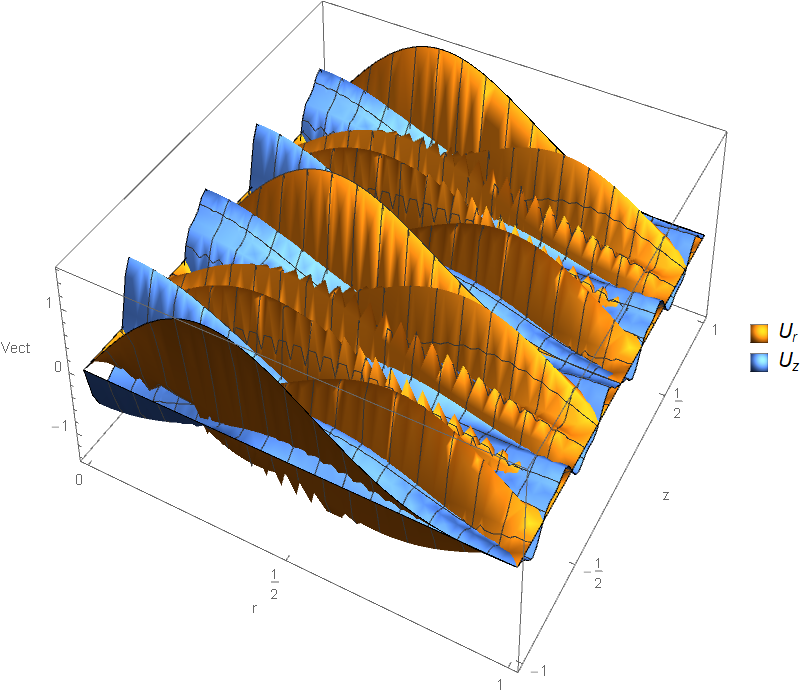}
\includegraphics[width=70mm]{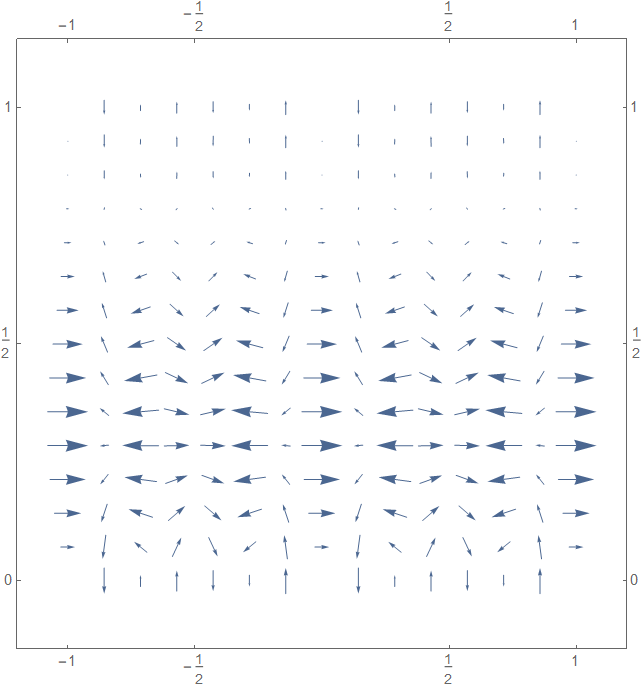}
\end{center}
\caption{\it Visualization of the two-componenta vector field of eq.(\ref{comboU}). In the picture on the left one sees plot of the two components of the velocity $\mathbf{U}_z$ and $\mathbf{U}_r$. In the picture on the right we have instead displayed the same information as a vector plot in two dimension.}  \label{mistoconpanna}
 \hskip 1cm \unitlength=1.1mm
\end{figure}
As a result of this structure the stream-lines are still closed loops in $z,r$ space, yet highly deformed in their shape: an example is shown in fig.\ref{braghefruste}. The rotation around the central axis of the tube still creates
revolution surfaces with torus-topology yet deformed in a rather capricious way as also shown in fig. \ref{braghefruste}. 
\section{Algebraic structure of axial symmetric NS equation in $L^2_{tube}$}
\label{algebraNS}
We come now to the setup for the search of steady solutions of the Navier-Stokes with axial symmetry and constant
Bernoulli function. Our starting point is eq.(\ref{gongolone}). If we put $H_B = h =\text{const}$ and $\partial_t \Omega^{[\mathbf{U}]} \, = \,0$ we get:
\begin{equation}\label{trombolone}
  i_{\pmb{\mathrm{U}}}\cdot \pmb{d}\Omega ^{\pmb{\mathrm{U}}}\, -\, \nu  \Delta {\Omega }^{[\pmb{\mathrm{U}}]}\, = \,0
\end{equation}
\paragraph{Constant Bernoulli function}
It is interesting to recall what was said in section 7.1 of \cite{Fr__2023} about the interpretation of the condition of constant Bernoulli function. We report the few lines of that paper verbatim.
\par
The condition of constant Bernoulli function is easily implemented by setting the \textit{pressure field}
equal to a constant $h$ minus the squared norm of velocity field:
\begin{equation}\label{tarielka}
    p\left(\mathbf{x},t\right) \, = \, h \, -\, \ft 12 \parallel
    U(\mathbf{x},t)\parallel^2 \, = \,\, h \, - \, \text{const}
    \,\times\,\frac{
    \Omega^{[\mathrm{U}]} \wedge \star_{g} \Omega^{[\mathrm{U}]}}{\text{Vol}}
\end{equation}
where
\begin{equation}\label{volume3forma}
    \text{Vol} \, \equiv \, \ft{1}{3!} \, \mathrm{det}\left(g\right)
    \, \mathrm{d}x\wedge \mathrm{d}y \wedge \mathrm{d}z
\end{equation}
is the volume $3$-form. If the velocity field satisfies Beltrami
equation with eigenvalue $\mu$
\begin{equation}\label{carnenonvale}
    \star_{g} \,\mathbf{d}\Omega^{[\mathrm{U}]} \, = \,  \mu \, \Omega^{[\mathrm{U}]}
\end{equation}
then $\Omega^{[\mathrm{U}]}$ is a \textbf{contact form} and the velocity field $\mathbf{U}$ is its Reeb field. Indeed we get:
\begin{equation}\label{pagnuflone}
    \Omega^{[\mathrm{U}]} \wedge \star_{g} \Omega^{[\mathrm{U}]} \,
    = \, \frac{1}{\mu} \, \Omega^{[\mathrm{U}]} \wedge
    \mathbf{d}\Omega^{[\mathrm{U}]} \, = \, \lambda(\mathbf{x},t) \,
    \text{Vol}
\end{equation}
So the physical pressure field (apart from the additive
constant $h$) obtains an inspiring geometrical interpretation:
 it is the nowhere vanishing function $\lambda(\mathbf{x},t)$
mentioned in the definition of the Reeb field that we recall from paper \cite{Fr__2023}. In that paper the same definition is numebered 3.8.
\par
\begin{definizione}\label{ribatriestina} Associated with a contact form $\alpha$ one has
the so called \textbf{Reeb vector field}  $\mathbf{R}_\alpha$,
defined by the two conditions:
\begin{eqnarray}\label{Ribbo}
&&\alpha\left(\mathbf{R}_\alpha\right) \, = \, \lambda(\mathbf{x})
\quad =
 \quad
 \text{nowhere vanishing function on $\mathcal{M}_{2n+1}$}\nonumber\\
 &&\forall \mathbf{X}\in \Gamma\left[\mathcal{TM}_{2n+1},\mathcal{M}_{2n+1}\right] \quad :
 \quad
 \mathrm{d}\alpha\left(\mathbf{R}_\alpha,\mathbf{X}\right)\, = \,
 0
\end{eqnarray}
\end{definizione}
The functional space $L^2_{tube}$ introduced in eq.(\ref{finitelenorme}) is a subspace of a larger one
$\widetilde{L}^2_{tube}$ defined as in eq. (\ref{finitelenorme}) but with the removal of the boundary condition
at $r=1$.
\begin{eqnarray}\label{carpettone}
  \widetilde{L}^2_{tube} &\ni& \mathbf{U} \, = \,  \left\{\mathbf{U}^1(r,\phi,z), \, \mathbf{U}^2(r,\phi,z), \, \mathbf{U}^3(r,\phi,z)\right\} \quad \text{iff} \quad \|\mathbf{U}\|^2 \, < \, \infty  \\
 \|\mathbf{U}\|^2 & \equiv &  \int_{\mathfrak{T}} \, \mathbf{U}^i\mathbf{U}^j \, g_{ij} \boldsymbol{\text{d$\mu$}}[r,\phi,z]
\, = \, \int_{0}^{2\pi} \, \text{d$\phi$} \, \int_{0}^{1} \, \text{dr} \, \int_{-1}^{1} \, \text{dz}
\,\left[ (\mathbf{U}^1)^2  \, + \, r^2 \, (\mathbf{U}^2)^2 \, + \, (\mathbf{U}^3)^2 \right] 
\end{eqnarray}
Why is it important to consider the nested sequence $\widetilde{L}^2_{tube}\supset{L}^2_{tube}\supset{L}^2_{axial}$?
The answer is easily given. The tube (\ref{torone}) contains the fundamental cell of the quotient $\mathbb{R}^3/\Lambda_{hexag}$ where $\Lambda_{hexag}$ is the hexagonal space lattice of three-dimensional space
\paragraph{The hexagonal lattice}
The basis vectors of $\Lambda_{Hex}$ can be taken as the following ones:
\begin{equation}\label{Hexagbase}
    {\mathbf{w}}_1 \, = \, \left\{\sqrt{2},0,0\right\} \quad ; \quad {\mathbf{w}}_2 \, = \,
    \left\{-\frac{1}{\sqrt{2}},{\frac{\sqrt{3}}{\sqrt{2}}},0\right\} \quad ; \quad {\mathbf{w}}_3 \, = \, \left\{0,0,\sqrt{2}\right\}
\end{equation}
which implies that the metric is the following non diagonal one:
\begin{equation}\label{nonkronecca}
    g_{\mu\nu} \, = \, \left(
\begin{array}{ccc}
 2 & -1  & 0 \\
 -1   & 2 & 0 \\
 0 & 0 & 2 \\
\end{array}
\right)
\end{equation}
The basis vectors $ {\mathbf{e}}^\mu$ of the dual momentum lattice
$\Lambda_{Hex}^\star$ do not coincide with those of the lattice
$\Lambda_{Hex}$. They are the following ones:
\begin{equation}\label{Hexagdualbas}
    {\mathbf{e}}^1 \, = \, \left\{\frac{1}{\sqrt{2}},\frac{1}{\sqrt{6}},0\right\} \quad ;
    \quad {\mathbf{e}}^2 \, = \,\left\{0,\sqrt{\frac{2}{3}},0\right\} \quad ;
    \quad {\mathbf{e}}^3 \, = \,\left\{0,0,\frac{1}{\sqrt{2}}\right\}
\end{equation}
so that the space lattice is now a proper subgroup of its dual
$\Lambda_{Hex}^\star$, named also the \textit{momentum-lattice}. In
order to understand the structure of the hexagonal lattice one ought
to consider first the hexagonal tessellation of a plane that is
generated by the first two basis vectors ${\mathbf{w}}_{1,2}$.
\par
\begin{figure}[!hbt]
\begin{center}
\includegraphics[width=150mm]{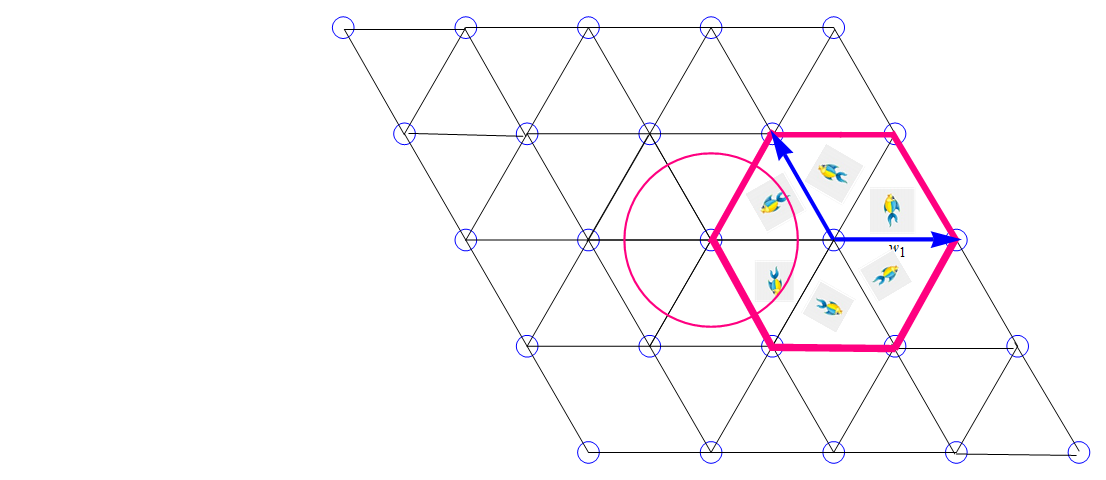}
\caption{{\it  A view of the hexagonal tesselation of the plane. The
hexagonal two dimensional lattice coincides with the $A_2$ root
lattice. Indeed the projection on the plane of the two basis vectors
$\mathbf{w}_1$ and $\mathbf{w}_2$ (the two blue vectors) are the two
simple roots of the $A_2$ Lie algebra. Each point of the lattice can
be regarded as the center of a regular hexagon whose vertices are
the first nearest neighbors. These hexagons provide a tesselation of
the infinite plane. } \label{Hexagtessere}}
\end{center}
\end{figure}

To this effect it is convenient to look at fig.\ref{Hexagtessere}
The space lattice which provides a tiling of the plane by means of
regular hexagons coincides with the root lattice of the  $A_2$ Lie
algebra, its generators being the two simple roots $\alpha_{1,2}$.
\par
The plane projection of the dual lattice $\Lambda^\star_{Hex}$ is
just the weight lattice of $A_2$, the plane projection of the basis
vectors $\mathbf{e}_{1,2}$ being just the fundamental weights
$\lambda_{1,2}$. This is illustrated in the next fig.\ref{pesorete}.
\begin{figure}[!hbt]
\begin{center}
\includegraphics[width=90mm]{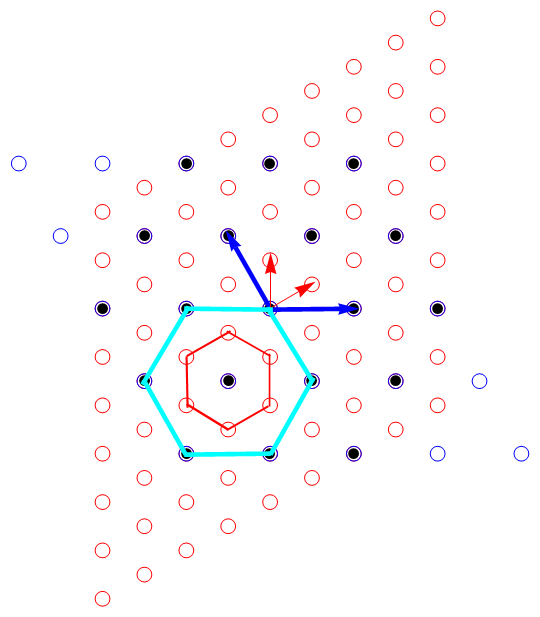}
\caption{{\it  Illustration of the dual momentum lattice of the
hexagonal lattice in the plane. The red circles are the points of
the momentum lattice, while the blue ones are the points of the
space lattice. In the finite portions of the two lattices that we
show in this picture the black points are the common ones. As we see
each point of the space--lattice is surrounded by two hexagons; the
vertices of the smaller hexagon are moment-lattice points that do
not belong to space-lattice, while the vertices of the bigger
hexagon are the space-lattice nearest neighbors, as already remarked
in the caption of fig.\ref{Hexagtessere}.} \label{pesorete}}
\end{center}
\end{figure}
There it is clearly shown that the space lattice is a sublattice of
the dual momentum lattice.
\par
The three-dimensional hexagonal lattice is obtained by adjoining an
infinite number of equally spaced planes each tiled in the way shown
in fig.s \ref{Hexagtessere} and \ref{pesorete}.
%%%%%%%%%%%%%%%%%
\begin{figure}[!hbt]
\begin{center}
\includegraphics[height=70mm]{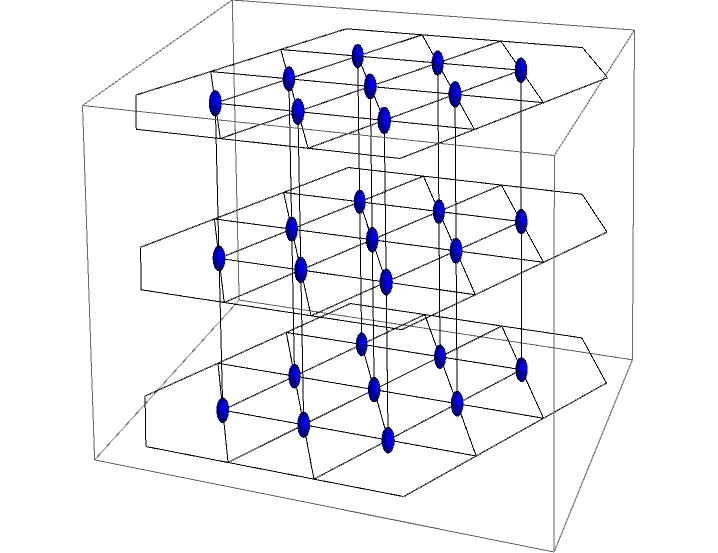}
\includegraphics[height=70mm]{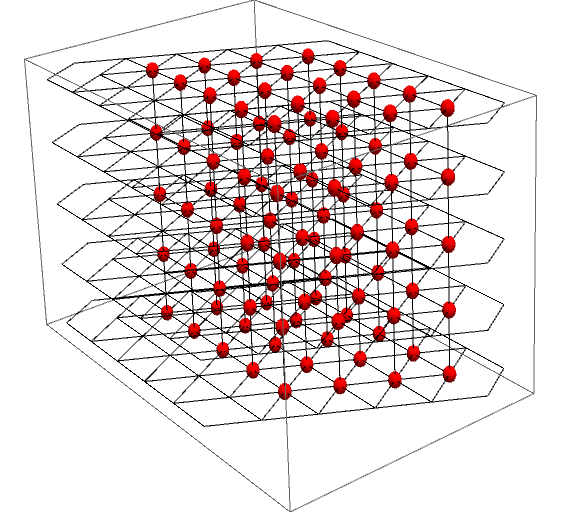}
\caption{\it  A view of the hexagonal space lattice $\Lambda_{Hex}$
(blue points on the left) and momentum momentum lattice
$\Lambda_{Hex}^\star$ (red points on the right)} \label{HexagLatPS}
\end{center}
\end{figure}
A view of the resulting three dimensional lattices is provided in
fig.\ref{HexagLatPS}.
\par
\begin{figure}[!hbt]
\begin{center}
\includegraphics[height=70mm]{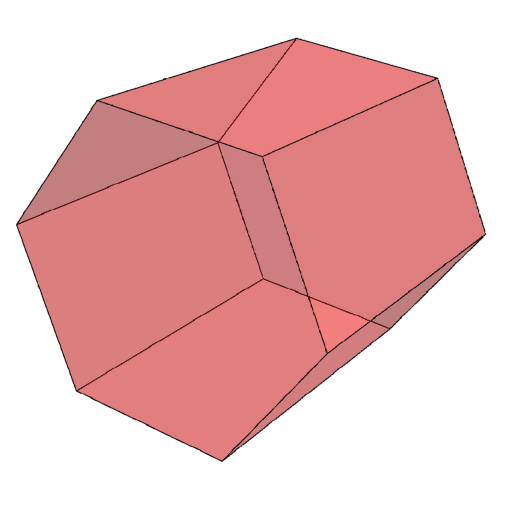}
\includegraphics[height=70mm]{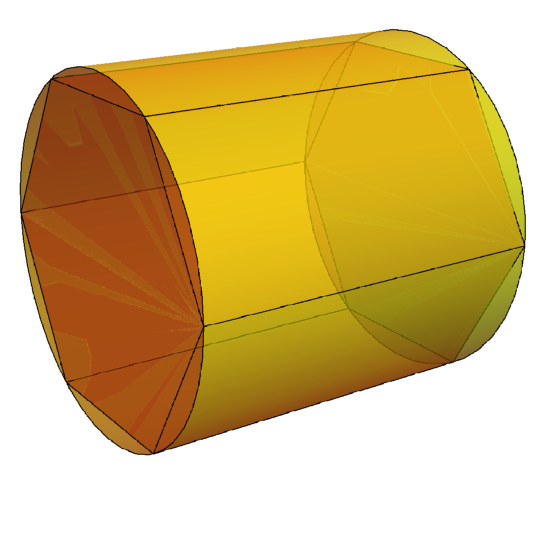}
\caption{\it  A view of the fundamental hexagonal box hosting the hydro-flows with periodic boundary conditions 
with respect to the Hexagonal Point Group (see \cite{Fr__2023}). The picture on the left shows the hexagonal box, while that on the right demonstrates that it is fully contained inside the tube of fig. \ref{tubastro}.
} \label{casuccia}
\end{center}
\end{figure}

In this way one can consider a fundamental hexagonal box like that in figure \ref{casuccia}. In \cite{Fr__2023}
the basis of Beltrami flows that can be used to expand a generic flow periodic with respect to the transformations of the hexagonal pointgroup $\mathrm{Dih_6}$ was constructed and shown to arrange, energy shell by energy shell into irreducible representations of the Universal Classifying Group $\mathfrak{U}_{72}$. Each linear combination $\mathbf{U}$ of such basic Beltrami / anti-Beltrami flows is a well defined function inside the hexagonal box and hence inside the tube. 
The norm  $\| \mathbf{\mathbf{U}}\|^2$ can be calculated with the definition (\ref{norm}) and it will be finite
if the field is finite in the hexagonal box. Therefore the axial symmetric flows and the flows with hexagonal periodcity leave inside the same functional space and can be compared and even combined. 
\par
This is quite relevant for the further developments of this project, once steady axial symmetric flows are determined. Axial symmetry can be broken by perturbations that will naturally arrange into representations of the one-dimensional rotation group $\mathrm{SO(2)}$ and, consequently, also of the Point Group of the Hexagonal lattice, whose intersection with $\mathrm{SO(2)}$ is the $\mathbb{Z}_6$ cyclic group.   
\subsection{The diamond product and its decomposition}
\label{gibuti}
Considering now both the expansion of the flow in modes (\ref{apuleio}) that is repeated identically in terms of the dual $1$-forms in eq.(\ref{cassiodargento})
and the non-linear term of the steady Navier Stokes equation at constant Bernoulli function (\ref{trombolone}), we
see that the latter is bilinear in the expansion coefficients $a_{\pm,0}[n,k],a_{\pm,0}[n,k],c_\pm[n,0]$. Indeed simplifying the notation by naming the coefficient collectively as:
\begin{eqnarray}\label{lescontamines}
 && c[n,k,i] \, = \, \{a_{\pm,0}[n,k],a_{\pm,0}[n,k]\} \quad ; \quad i=1,\dots , 6 \nonumber\\
  && i=1 \Leftrightarrow a_+ \quad , \quad i=2 \Leftrightarrow a_- \quad , \quad i=3 \Leftrightarrow b_+ \quad , \quad i=4 \Leftrightarrow b_- , \quad , \quad  i=5 \Leftrightarrow a_0 , \quad , \quad i=6 \Leftrightarrow b_0 \nonumber\\
\end{eqnarray}
we see that we can write
\begin{equation}
\label{pifferaio}
  \boldsymbol{\mathrm{i}}_{\pmb{\mathrm{U}}}\cdot \pmb{d}\Omega ^{\pmb{\mathrm{U}}}\, = \, 
  \sum^{4}_{i_1=1}\, \sum^{4}_{i_2=1} \, \sum^{\infty}_{k_1=1} \, \sum^{\infty}_{k_2=1} 
   \, \sum^{\infty}_{n_1=1} \, \sum^{\infty}_{n_2=1} \,  c[n_1,k_1,i_1]\,\,  c[n_2,k_2,i_2] \,\,\left(
   \boldsymbol{\mathrm{i}}_{\pmb{\mathrm{U}}[n_1,k_1,i_1]}\cdot \pmb{d}\Omega ^{\pmb{\mathrm{U}}[n_2,k_2,i_2]}\right)
\end{equation}
For $i_2 \leq 4$ the $1$-form $\Omega ^{\pmb{\mathrm{U}}[n_2,k_2,i_2]} $ is Beltrami or anti/Beltrami so that the
interaction term
\begin{equation}\label{interaturo}
  \boldsymbol{\mathrm{i}}_{\pmb{\mathrm{U}}[n_1,k_1,i_1]}\cdot \pmb{d}\Omega ^{\pmb{\mathrm{U}}[n_2,k_2,i_2]} 
\end{equation}
can be rewritten in another useful way utilizing the identity:
\begin{equation}\label{consolato}
  \pmb{d}\Omega ^{\pmb{\mathrm{U}}[n_2,k_2,i_2]} \, = \, (-1)^{1+i_2} \, \varpi(n_2,k_2) \, \underbrace{\star_{g} \Omega ^{\pmb{\mathrm{U}}[n_2,k_2,i_2]}}_{\text{Hodge dual}}
\end{equation}
which leads to:
\begin{equation}\label{ferrovecchio}
  \boldsymbol{\mathrm{i}}_{\pmb{\mathrm{U}}[n_1,k_1,i_1]}\cdot \pmb{d}\Omega ^{\pmb{\mathrm{U}}[n_2,k_2,i_2]} \, = \, (-1)^{1+i_2} \, \varpi(n_2,k_2) \left(\boldsymbol{\mathrm{i}}_{\pmb{\mathrm{U}}[n_1,k_1,i_1]}\cdot\star_{g} \Omega ^{\pmb{\mathrm{U}}[n_2,k_2,i_2]}\right)
\end{equation}
By explicit and immediate calculation we find:
\begin{eqnarray}\label{regredo}
  \boldsymbol{\mathrm{i}}_{\pmb{\mathrm{U}}[n_1,k_1,i_1]}\cdot\star_{g} \Omega ^{\pmb{\mathrm{U}}[n_2,k_2,i_2]}
  & = &\Omega^{\mathbf{V}} \quad \text{where} \quad\nonumber\\
  \mathbf{V}& = & \pmb{\mathrm{U}}[n_1,k_1,i_1] \diamond \pmb{\mathrm{U}}[n_2,k_2,i_2] \, \equiv \, 
  r \, g^{i\ell} \,\epsilon_{\ell\, j \, k} \pmb{\mathrm{U}}^j[n_1,k_1,i_1]\,\pmb{\mathrm{U}}^k[n_2,k_2,i_2]
\end{eqnarray}
We name \textbf{diamond product} the simple antisymmetric operation introduced  in eq.(\ref{regredo}) whose existence is a consequence of the use of a Beltrami/anti-Beltrami basis of functions in the considered functional space $L^2_{axial}$ of axial symmetric flows.
\par
Using the above compact notation we can decompose the diamond product of two elementary Beltrami/anti-Beltrami flows
in the same basis, using the scalar product (\ref{frostasapore},\ref{conglomerato}). We obtain the following result
\begin{eqnarray}\label{koblinka}
  \pmb{\mathrm{U}}[n_1,k_1,i_1] \diamond \pmb{\mathrm{U}}[n_2,k_2,i_2] & =&
  \sum_{n=1}^{\infty}\sum_{4}^{i=1} \left(\left\langle\,n_1,k_1,i_1,n_2,k_2,i_2\|n, k_1+k_2,i\right\rangle \, \pmb{\mathrm{U}}[n,k_1+k_2,i] \right.\nonumber\\
  & & \left.\, + \, \left\langle\,n_1,k_1,i_1,n_2,k_2,i_2\|n, k_1-k_2,i\right\rangle \, \pmb{\mathrm{U}}[n,k_1-k_2,i]\right)
\end{eqnarray}
Note that no request is made on the velocity field $\pmb{\mathrm{U}}[n_1,k_1,i_1]$ that can be also the dual of a closed form, namely the index $i_1$ can take also the values $5$ and $6$. On the contrary if $i_2=5$ of $i_2=6$ the interaction term vanishes and no other fields arise from the interaction.
\par
The coefficients:
\begin{equation}\label{comeco}
  \left\langle\,n_1,k_1,i_1,n_2,k_2,i_2\|n, k_1+k_2,i\right\rangle  \quad ; \quad \left\langle\,n_1,k_1,i_1,n_2,k_2,i_2\|n, k_1-k_2,i\right\rangle 
\end{equation}
have a rather general form. In tables (\ref{table1}-\ref{table16}) presented in appendix
\ref{carpaccio} I have shown the explicit structure of the coefficients when all the three indices $i_1,i_2,i$
are in the range $1,2,3,4$. Namely I have calculated the coefficients for the $4\times 4=16$ cases of
the diamond product of Beltrami, anti-Beltrami vector fields. One should still analyse the non symmetric case of closed $1$-forms with Beltrami, anti Beltrami cofactors. Furthermore one should calculate the coalescence coefficient of a pair of Beltrami (anti-Beltrami) fields into a closed $1$-form. The analytic structure of these cases is just similar to that of the 16 calculated coefficients and nothing more arises qualitatively from that analysis. It is just a long exercise to go through all the cases. I have skipped such an exercise since at the end
all diamond products will be evaluated numerically in the computer code to be constructed for the next paper \cite{matthewandme}. The analytic structure is completely clear from the exercise performed in the 16 cases displayed in appendix \ref{carpaccio}. 
Apart from the already introduced algebraic objects, the coefficients depend on the following two triple integrals of Bessel functions (see definitions (\ref{saldatore}-\ref{concistoro})):
\begin{eqnarray}
\label{calendula}
  N_{0,1}(n_1,n_2,n) &\equiv& \int^{1}_{0} \, r\, \mathcal{G}_0^{[n_1]}(r)\, \mathcal{G}_1^{[n_2]}(r)\, \mathcal{G}_1^{[n]}(r)\, \mathrm{dr} \nonumber\\
  N_{1,1}(n_1,n_2,n) &\equiv & \int^{1}_{0} \, r\, \mathcal{G}_1^{[n_1]}(r)\, \mathcal{G}_1^{[n_2]}(r)\, \mathcal{G}_0^{[n]}(r)\, \mathrm{dr}  
\end{eqnarray}
that are always well defined and convergent. Only in some special cases they can be evaluated analytically, yet, because of the nice behavior of their integrand, they can be easily evaluated numerically and even tabulated. 
\par
Before presenting a schematic structure of the neural network optimization procedure for the solution of the problem let me illustrate the complexity of streamlines that can be obtained by the superposition of our basis functions.
As a mixture of all the possible ingredients I have chosen:
\begin{equation}\label{cartolinarussa}
  \Omega^{[\mathbf{U}_{mix}]} \, = \, \ft 43 \, {\Omega}A_+[1,1]+ \ft 52\, \,{\Omega}A_-[3,2]+2 \, {\Omega}B_-[3,0]
\end{equation}
In fig.\ref{mistoconpanna} I show a single streamline of the vector field defined in (\ref{cartolinarussa}). By means of three instant pictures shot at times $t=1,t=3,t=7$ of its developing, I show how such streamline progressively invades and covers an entire region of space internal to the tube.
\begin{figure}[!hbt]
\begin{center}
\includegraphics[height=70mm]{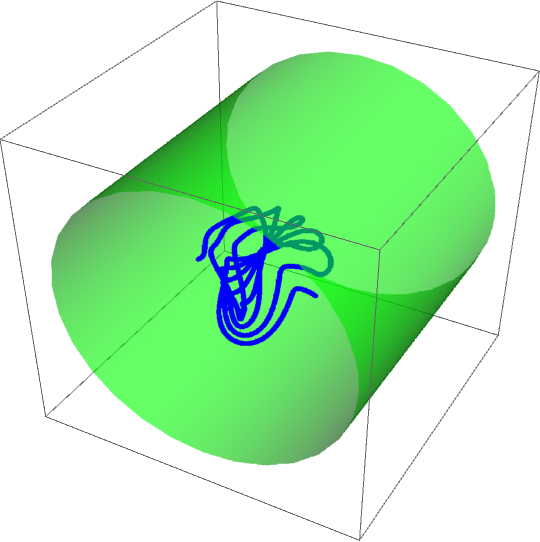}
\includegraphics[height=70mm]{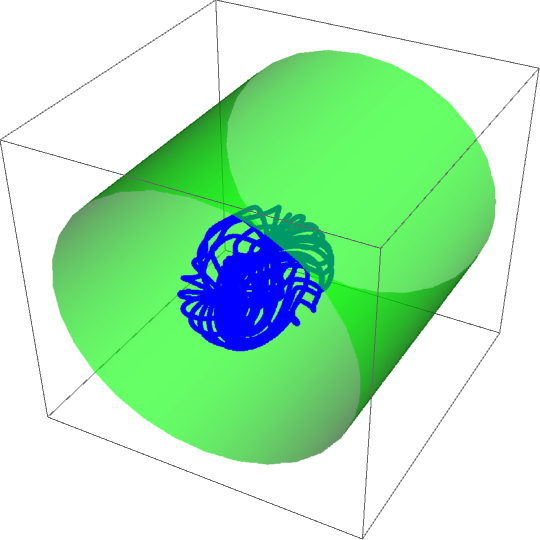}
\includegraphics[height=70mm]{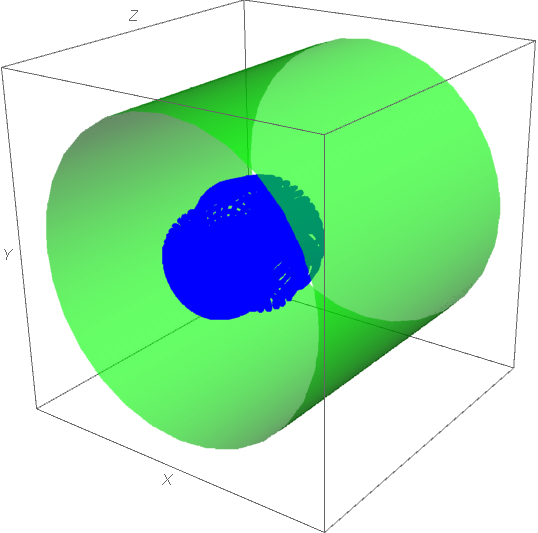}
\includegraphics[height=70mm]{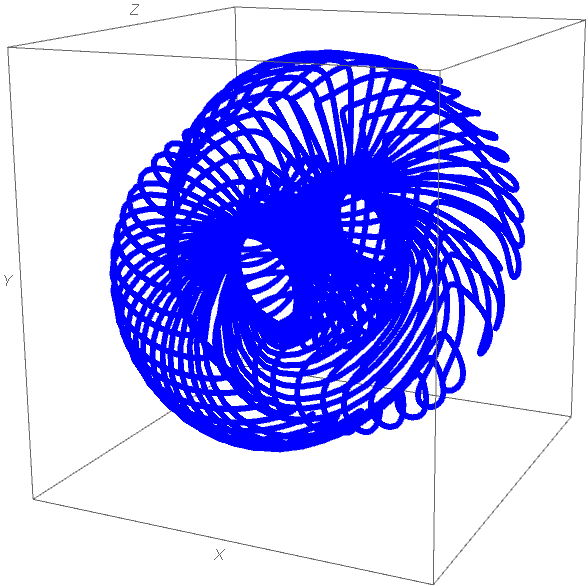}
\caption{\it In this figure I display just one streamline for the vector field (\ref{cartolinarussa}).
In the first row  I show on the left the streamline at $t=1$, on the right the same streamline at time $t=3$. In the second line I show the streamline at $t=7$: on the left as it appears insiede the tube, on the right I make a zoom on it looking also from a rotated viewpoint. 
} \label{mistoconpanna}
\end{center}
\end{figure}
\section{Quadratic recurrence relations and Neural Networks.}
\label{ricorrenzanatalizia}
Using the expansion presented in previous sections in eq.(\ref{cassiodargento}) and inserting it into the Navier-Stokes equation (\ref{trombolone}) we can define the new vector field $\mathbf{V}_{NS}$:
\begin{equation}\label{VNS}
  \Omega^{[\mathbf{V}_{NS}]}\, \equiv \, i_{\pmb{\mathrm{U}}}\cdot \pmb{d}\Omega ^{\pmb{\mathrm{U}}}\, -\, \nu  \Delta {\Omega }^{[\pmb{\mathrm{U}}]}
\end{equation}
If $\mathbf{V}_{NS}\, = \,0$ we have, in $\mathbf{U}$, an exact solution of the non linear equation which is the hard task of no easy solution without some additional inspiration from some additional symmetry or educated guess. On the other hand if 
$\|\mathbf{V}_{NS}\|^2 \, = \, 0$ the vector field $\mathbf{U}$ is not necessarily an exact  solution, yet it  some function that differs from an exact solution only on a subset of vanishing measure of the hosting compact manifold $\boldsymbol{\mathfrak{T}}$. This would just be  great! Next to such unattainable triumph, if we could make the norm $\|\mathbf{V}_{NS}\|^2 $ as small as possible, we would have a very good approximation to a solution of the non linear differential equation, the better, the smaller  $\|\mathbf{V}_{NS}\|^2 $ is. Smaller respect to what? This is the relevant question. The rather obvious answer is: \textit{with respect to the norm of the vector field $\pmb{\mathrm{U}}$ we want construct}. This provides an important hint. The functional of the unknown coefficients 
$c[s]$ defined below
\begin{equation}\label{chifun}
  \chi(c[s]) \, = \, \frac{\|\mathbf{V}_{NS}\|^2 }{\|\mathbf{U}\|^2 }
\end{equation}
is the best candidate to be utilized in the algorithm of optimization.
Indeed $\chi(c[s])$ defined in (\ref{chifun}) is the object we should try to make as close to zero as possible.
Obviously since the norm of $\mathbf{U}$ should be finite, the expansion coefficients should be such as to make the 
series convergent. The norm of $\mathbf{U}$ is:
\begin{equation}\label{cartollo}
  \|\mathbf{U}\|^2 \, = \, \sum^{\infty}_{n=1}\sum^{\infty}_{k=0}\sum^{6}_{i=1} c[n,k,i]^2 \, \|\mathbf{U}[n,k,i]\|^2
\end{equation}
Since the norms of the basic functions have a simple universal form displayed in eq.(\ref{conglomerato}) and grow quadratically in $n$, since the zeros of the Bessel functions grow approximately linearly in $n$, it follows that the coefficients $c[n,k,i]$ should behave as:
\begin{equation}\label{treporcelli}
  c[n,k,i] \, \stackrel{(n,k)\to \infty }{\approx} \frac{1}{\|\mathbf{U}[n,k,i]\|^\alpha} \quad ; \quad \alpha > 1
\end{equation}
where $\alpha$ is some power to be optimally adjusted by means of numerical experiments. Hence
a resetting of the unknown coefficients $c[s]$ such as the following one  is  to be made as the first step:
\begin{equation}\label{treporcelli}
  c[n,k,i] \, = \,  \frac{{\tilde{c}[n,k,i]}}{\|\mathbf{U}[n,k,i]\|^\alpha} \quad ; \quad \alpha > 1
\end{equation}
The idea is that the rescaled coefficients $\tilde{c}[s]$ should be of the order of $1$ or of at most of the \textit{tens}. With
this replacement eq.(\ref{cartollo}) becomes:
\begin{equation}\label{plasticone}
  \|\mathbf{U}\|^2 \, = \, \sum^{\infty}_{n=1}\sum^{\infty}_{k=0}\sum^{6}_{i=1} \tilde{c}[n,k,i]^2 \, \frac{1}{\|\mathbf{U}[n,k,i]\|^{2(\alpha-1)}}
\end{equation}
\par
The rescaling (\ref{treporcelli}) is just a change of variables that might be useful in the implementation of the optimization algorithm, yet it does not change anything from the theoretical viewpoint. Hence for notation simplicity I continue to use the original coefficients in the further development of the argument. 
\par
The second essential step is to consider the coefficients appearing in the norm of $\mathbf{V}_{NS}$. These are obtained by considering the scalar product of $\mathbf{V}_{NS}$ with each $\mathbf{U}[n,k,i]$ of the function basis. Explicitly we have:
\begin{eqnarray}\label{gerarchia}
 &d[s] = \, &\, \frac{1}{\|\mathbf{U}[n,k,i]\|^2} \,\langle \mathbf{U}[n,k,i] \, | \, \mathbf{V}_{NS}\rangle \, = \, - \, \nu\,  \varpi^2(n,k) \, c[n,k,i] \, + \nonumber\\
 &&\,\sum_{i_1=1}^6 \, \sum_{i_2=1}^4 \, 
 \sum_{n_1=1}^{\infty}\sum_{n_2=1}^{\infty}\, (-1)^{1+i_2} \left(\sum_{\kappa=1}^{k-1}\,\left(c[n_1,k-\kappa,i_1] \, c[n_2,\kappa,i_2] \left\langle\,n_1,k-\kappa,i_1,n_2,\kappa,i_2\|n, k,i\right\rangle \right)     \right. \nonumber\\
 && \left.  + \,\sum_{\kappa=1}^{\infty} \left( c[n_1,k+\kappa,i_1] \, c[n_2,\kappa,i_2] \left\langle\,n_1,k+\kappa,i_1,n_2,\kappa,i_2\|n, k,i\right\rangle \right)
 \right)  
\end{eqnarray}
As we see from eq.(\ref{gerarchia}), the coefficients of the vector field in the numerator of the functional (\ref{chifun}) are quadratic and linear expressions in the coefficients of the denominators and we have
\begin{equation}\label{cagnettogiallo}
  \chi \, = \, \frac{\sum^{\infty}_{n=1}\sum^{\infty}_{k=0}\sum^{6}_{i=1} d[n,k,i]^2 \, \|\mathbf{U}[n,k,i]\|^2}{\sum^{\infty}_{n=1}\sum^{\infty}_{k=0}\sum^{6}_{i=1} c[n,k,i]^2 \, \|\mathbf{U}[n,k,i]\|^2}
\end{equation}
Hence the optimization algorithm should try to reduce the numerator without reducing the denominator, a fact that   would be against the reduction of $\chi$. The precise strategy to reach the goal is postponed to the next paper \cite{matthewandme}. I can just make a general observation. Clearly the exact solution of Navier Stokes equation in the tube is not unique, rather we expect to have a large number of such solutions. At vanishing viscosity $\nu\,=\,0$ all the elements of our functional basis
are exact solutions so that the number of inviscid flows is just infinite. As the $\nu$ parameter is turned on, each solution of Euler equation (the Navier Stokes equation at $\nu=0$ ) violates the equation by means of a vector field $\mathbf{V}_{NS}$ whose norm is exactly calculable:
\begin{equation}\label{nortallo}
  \|\mathbf{V}_{NS}\|^2 \, = \, \nu^2 \,\varpi(n,k)^2 \, \|\mathbf{U}\|^2
\end{equation}
which means that the $\chi$ functional in this case is:
\begin{equation}\label{balordo}
  \chi \, = \, \nu^2 \,\varpi(n,k)^2 
\end{equation}
Adopting a perturbative approach in the viscosity parameter $\nu$, leads us  to introduce, with coefficients of order $\nu$, other elements of the functional basis, whose diamond product with $\mathbf{U}$ should have a non vanishing projection on the original $\mathbf{U}$, so as to be able to compensate the contribution (\ref{balordo}) to the functional $\chi$. This comment suggests that the various coefficients might be represented as polynomials in the viscosity  parameter $\nu$ and
what decides the type of solution is the choice of a dominant Beltrami (or anti-Beltrami) flow whose coefficient survives the limit $\nu\to 0$.  
\section{Conclusions}
The motivations and the perspective of the present paper have been amply illustrated in the introductory section and also in the main body of the article: I will not repeat myself. I just stress once again that the goal is not to derive NS solutions for immediate use in any practical problem. The goal is that of studying the approach to exact solutions of the non linear equation from a different viewpoint based on the properties of the utilized functional basis that already incorporates the chosen symmetry (in this case the axial one) trying to monitor the mechanism of non linearity through the product law (the diamond product) of the building blocks one with the other. The hope is that the approximate solutions hopefully retrieved by the neural network optimization algorithms might reveal the hidden rules by means of which one should choose the combination of the ingredients in order to get an exact solution of the differential equation. Imposing axial symmetry reduces the game to a manageable basis of functions that still keeps the main features of the inspected phenomena, in particular the Beltrami anti-Beltrami structure of the spectrum. In any case axial symmetric solutions of the Navier Stokes equations are interesting for their own sake and, as I showed with a few examples, the richness of structure of such flows is impressive.
\par
From the mathematical technical viewpoint, the main result of my paper is the construction of a complete basis of harmonic one-forms well-defined in the tube topology and with correct boundary condition on the cylindrical surface using trigonometric functions in the longitudinal direction and Bessel $J_{0,1}$ functions in the radial direction. The rescaling of the Bessel function variable by means of the Bessel zeros $j_{\nu,n}$ is a trick, that is quite simple, yet, up to my knowledge used here for the first time in order to introduce a function basis on the interval $[0,1]$. The re-expansion in the same functional basis of any two products of the same basis works very well and a finite number of terms reproduces any product of two with an impressive precision (see Appendix
\ref{impressiveprod}). 
\vskip 0.3cm
\section*{Acknowledgements} It is my pleasure to acknowledge extremely important and clarifying conversations with my frequent collaborator and great friend Mario Trigiante in summer 2023 when this investigation started to come soon to an early stop because of the priority given to the other project on Machine Learning mathematical foundations \cite{pgtstheory,TSnaviga,naviga,tassellandum,geotermico}. At this stage the reactivation of this Fluid Dynamics project was quite to the point in order to combine it with the exploration of Neural Networks of Physics Informed type. I am also pleased to mention more recent illuminating conversations with my other most dear collaborator and friend Alexander Sorin. Finally I would like to thank the Ph.D. student and my coauthor Matteo Santoro for our recent constructive conversations and for his positive answer to my invitation to develop together the optimization algorithm needed to progress further on the here opened road. 
\newpage
\appendix
\begin{landscape}
\section{Tables of the diamond product decomposition coefficients}
\label{carpaccio}
\begin{eqnarray}\label{table1}
  && \text{$\Omega $A}_+[n_1,k_1]\diamond \text{$\Omega $A}_+[n_2,k_2] \, = \, \nonumber\\
  &&\begin{array}{rl}
 +\frac{\left(k_1+k_2\right) \left(N_{0,1}\left(n_1,n_2,n\right)
   \left(\left(k_1+k_2\right) \varpi _{k_2,n_2}+k_2 \varpi
   _{k_1+k_2,n}\right)-N_{0,1}\left(n_2,n_1,n\right) \left(\left(k_1+k_2\right) \varpi
   _{k_1,n_1}+k_1 \varpi _{k_1+k_2,n}\right)-N_{1,1}\left(n_1,n_2,n\right) \left(k_2 \varpi
   _{k_1,n_1}-k_1 \varpi _{k_2,n_2}\right)\right)}{8 k_1 k_2 \varpi _{k_1+k_2,n}^2} &
   \text{$\Omega $A}_+\left(n,k_1+k_2\right) \\
 +\frac{\left(k_1-k_2\right) \left(N_{0,1}\left(n_1,n_2,n\right) \left(k_2 \varpi
   _{k_1-k_2,n}+\left(k_2-k_1\right) \varpi
   _{k_2,n_2}\right)+N_{0,1}\left(n_2,n_1,n\right) \left(\left(k_1-k_2\right) \varpi
   _{k_1,n_1}+k_1 \varpi _{k_1-k_2,n}\right)-N_{1,1}\left(n_1,n_2,n\right) \left(k_2 \varpi
   _{k_1,n_1}+k_1 \varpi _{k_2,n_2}\right)\right)}{8 k_1 k_2 \varpi _{k_1-k_2,n}^2} &
   \text{$\Omega $A}_+\left(n,k_1-k_2\right) \\
 +0 & \text{$\Omega $B}_+\left(n,k_1+k_2\right) \\
 +0 & \text{$\Omega $B}_+\left(n,k_1-k_2\right) \\
 +\frac{\left(k_1+k_2\right) \left(N_{0,1}\left(n_1,n_2,n\right)
   \left(\left(k_1+k_2\right) \varpi _{k_2,n_2}-k_2 \varpi
   _{k_1+k_2,n}\right)-N_{0,1}\left(n_2,n_1,n\right) \left(\left(k_1+k_2\right) \varpi
   _{k_1,n_1}-k_1 \varpi _{k_1+k_2,n}\right)-N_{1,1}\left(n_1,n_2,n\right) \left(k_2 \varpi
   _{k_1,n_1}-k_1 \varpi _{k_2,n_2}\right)\right)}{8 k_1 k_2 \varpi _{k_1+k_2,n}^2} &
   \text{$\Omega $A}_-\left(n,k_1+k_2\right) \\
 -\frac{\left(k_1-k_2\right) \left(\varpi _{k_1-k_2,n} \left(k_2
   N_{0,1}\left(n_1,n_2,n\right)+k_1 N_{0,1}\left(n_2,n_1,n\right)\right)+\varpi
   _{k_2,n_2} \left(\left(k_1-k_2\right) N_{0,1}\left(n_1,n_2,n\right)+k_1
   N_{1,1}\left(n_1,n_2,n\right)\right)+\varpi _{k_1,n_1} \left(\left(k_2-k_1\right)
   N_{0,1}\left(n_2,n_1,n\right)+k_2 N_{1,1}\left(n_1,n_2,n\right)\right)\right)}{8 k_1
   k_2 \varpi _{k_1-k_2,n}^2} & \text{$\Omega $A}_-\left(n,k_1-k_2\right) \\
 +0 & \text{$\Omega $B}_-\left(n,k_1+k_2\right) \\
 +0 & \text{$\Omega $B}_-\left(n,k_1-k_2\right) \\
\end{array}\nonumber\\
\end{eqnarray}
\begin{eqnarray}
\label{table2}
  && \text{$\Omega $A}_+[n_1,k_1]\diamond \text{$\Omega $B}_+[n_2,k_2] \, = \, \nonumber\\
  &&\begin{array}{rl}
 +0 & \text{$\Omega $A}_+\left(n,k_1+k_2\right) \\
 +0 & \text{$\Omega $A}_+\left(n,k_1-k_2\right) \\
 +\frac{\left(k_1+k_2\right) \left(N_{0,1}\left(n_1,n_2,n\right)
   \left(\left(k_1+k_2\right) \varpi _{k_2,n_2}+k_2 \varpi
   _{k_1+k_2,n}\right)-N_{0,1}\left(n_2,n_1,n\right) \left(\left(k_1+k_2\right) \varpi
   _{k_1,n_1}+k_1 \varpi _{k_1+k_2,n}\right)-N_{1,1}\left(n_1,n_2,n\right) \left(k_2 \varpi
   _{k_1,n_1}-k_1 \varpi _{k_2,n_2}\right)\right)}{8 k_1 k_2 \varpi _{k_1+k_2,n}^2} &
   \text{$\Omega $B}_+\left(n,k_1+k_2\right) \\
 -\frac{\left(k_1-k_2\right) \left(\varpi _{k_1-k_2,n} \left(k_2
   N_{0,1}\left(n_1,n_2,n\right)+k_1 N_{0,1}\left(n_2,n_1,n\right)\right)-\varpi
   _{k_2,n_2} \left(\left(k_1-k_2\right) N_{0,1}\left(n_1,n_2,n\right)+k_1
   N_{1,1}\left(n_1,n_2,n\right)\right)+\varpi _{k_1,n_1} \left(\left(k_1-k_2\right)
   N_{0,1}\left(n_2,n_1,n\right)-k_2 N_{1,1}\left(n_1,n_2,n\right)\right)\right)}{8 k_1
   k_2 \varpi _{k_1-k_2,n}^2} & \text{$\Omega $B}_+\left(n,k_1-k_2\right) \\
 +0 & \text{$\Omega $A}_-\left(n,k_1+k_2\right) \\
 +0 & \text{$\Omega $A}_-\left(n,k_1-k_2\right) \\
 +\frac{\left(k_1+k_2\right) \left(N_{0,1}\left(n_1,n_2,n\right)
   \left(\left(k_1+k_2\right) \varpi _{k_2,n_2}-k_2 \varpi
   _{k_1+k_2,n}\right)-N_{0,1}\left(n_2,n_1,n\right) \left(\left(k_1+k_2\right) \varpi
   _{k_1,n_1}-k_1 \varpi _{k_1+k_2,n}\right)-N_{1,1}\left(n_1,n_2,n\right) \left(k_2 \varpi
   _{k_1,n_1}-k_1 \varpi _{k_2,n_2}\right)\right)}{8 k_1 k_2 \varpi _{k_1+k_2,n}^2} &
   \text{$\Omega $B}_-\left(n,k_1+k_2\right) \\
 +\frac{\left(k_1-k_2\right) \left(\varpi _{k_1-k_2,n} \left(k_2
   N_{0,1}\left(n_1,n_2,n\right)+k_1 N_{0,1}\left(n_2,n_1,n\right)\right)+\varpi
   _{k_2,n_2} \left(\left(k_1-k_2\right) N_{0,1}\left(n_1,n_2,n\right)+k_1
   N_{1,1}\left(n_1,n_2,n\right)\right)+\varpi _{k_1,n_1} \left(\left(k_2-k_1\right)
   N_{0,1}\left(n_2,n_1,n\right)+k_2 N_{1,1}\left(n_1,n_2,n\right)\right)\right)}{8 k_1
   k_2 \varpi _{k_1-k_2,n}^2} & \text{$\Omega $B}_-\left(n,k_1-k_2\right) \\
\end{array} \nonumber\\
\end{eqnarray}
\begin{eqnarray}
\label{table3}
  && \text{$\Omega $A}_+[n_1,k_1]\diamond \text{$\Omega $A}_-[n_2,k_2]\, = \, \nonumber\\
  &&\begin{array}{rl}
 -\frac{\left(k_1+k_2\right) \left(N_{0,1}\left(n_1,n_2,n\right)
   \left(\left(k_1+k_2\right) \varpi _{k_2,n_2}-k_2 \varpi
   _{k_1+k_2,n}\right)+N_{0,1}\left(n_2,n_1,n\right) \left(\left(k_1+k_2\right) \varpi
   _{k_1,n_1}+k_1 \varpi _{k_1+k_2,n}\right)+N_{1,1}\left(n_1,n_2,n\right) \left(k_2 \varpi
   _{k_1,n_1}+k_1 \varpi _{k_2,n_2}\right)\right)}{8 k_1 k_2 \varpi _{k_1+k_2,n}^2} &
   \text{$\Omega $A}_+\left(n,k_1+k_2\right) \\
 +\frac{\left(k_1-k_2\right) \left(N_{0,1}\left(n_1,n_2,n\right) \left(k_2 \varpi
   _{k_1-k_2,n}+\left(k_1-k_2\right) \varpi
   _{k_2,n_2}\right)+N_{0,1}\left(n_2,n_1,n\right) \left(\left(k_1-k_2\right) \varpi
   _{k_1,n_1}+k_1 \varpi _{k_1-k_2,n}\right)-N_{1,1}\left(n_1,n_2,n\right) \left(k_2 \varpi
   _{k_1,n_1}-k_1 \varpi _{k_2,n_2}\right)\right)}{8 k_1 k_2 \varpi _{k_1-k_2,n}^2} &
   \text{$\Omega $A}_+\left(n,k_1-k_2\right) \\
 +0 & \text{$\Omega $B}_+\left(n,k_1+k_2\right) \\
 +0 & \text{$\Omega $B}_+\left(n,k_1-k_2\right) \\
 -\frac{\left(k_1+k_2\right) \left(N_{0,1}\left(n_1,n_2,n\right)
   \left(\left(k_1+k_2\right) \varpi _{k_2,n_2}+k_2 \varpi
   _{k_1+k_2,n}\right)+N_{0,1}\left(n_2,n_1,n\right) \left(\left(k_1+k_2\right) \varpi
   _{k_1,n_1}-k_1 \varpi _{k_1+k_2,n}\right)+N_{1,1}\left(n_1,n_2,n\right) \left(k_2 \varpi
   _{k_1,n_1}+k_1 \varpi _{k_2,n_2}\right)\right)}{8 k_1 k_2 \varpi _{k_1+k_2,n}^2} &
   \text{$\Omega $A}_-\left(n,k_1+k_2\right) \\
 -\frac{\left(k_1-k_2\right) \left(\varpi _{k_1-k_2,n} \left(k_2
   N_{0,1}\left(n_1,n_2,n\right)+k_1 N_{0,1}\left(n_2,n_1,n\right)\right)-\varpi
   _{k_2,n_2} \left(\left(k_1-k_2\right) N_{0,1}\left(n_1,n_2,n\right)+k_1
   N_{1,1}\left(n_1,n_2,n\right)\right)+\varpi _{k_1,n_1} \left(\left(k_2-k_1\right)
   N_{0,1}\left(n_2,n_1,n\right)+k_2 N_{1,1}\left(n_1,n_2,n\right)\right)\right)}{8 k_1
   k_2 \varpi _{k_1-k_2,n}^2} & \text{$\Omega $A}_-\left(n,k_1-k_2\right) \\
 +0 & \text{$\Omega $B}_-\left(n,k_1+k_2\right) \\
 +0 & \text{$\Omega $B}_-\left(n,k_1-k_2\right) \\
\end{array}
\end{eqnarray}
\begin{eqnarray}
\label{table4}
  && \text{$\Omega $A}_+[n_1,k_1]\diamond \text{$\Omega $B}_{-}[n_2,k_2] \, = \, \nonumber\\
  && \begin{array}{rl}
 +0 & \text{$\Omega $A}_+\left(n,k_1+k_2\right) \\
 +0 & \text{$\Omega $A}_+\left(n,k_1-k_2\right) \\
 -\frac{\left(k_1+k_2\right) \left(N_{0,1}\left(n_1,n_2,n\right)
   \left(\left(k_1+k_2\right) \varpi _{k_2,n_2}-k_2 \varpi
   _{k_1+k_2,n}\right)+N_{0,1}\left(n_2,n_1,n\right) \left(\left(k_1+k_2\right) \varpi
   _{k_1,n_1}+k_1 \varpi _{k_1+k_2,n}\right)+N_{1,1}\left(n_1,n_2,n\right) \left(k_2 \varpi
   _{k_1,n_1}+k_1 \varpi _{k_2,n_2}\right)\right)}{8 k_1 k_2 \varpi _{k_1+k_2,n}^2} &
   \text{$\Omega $B}_+\left(n,k_1+k_2\right) \\
 -\frac{\left(k_1-k_2\right) \left(\varpi _{k_1-k_2,n} \left(k_2
   N_{0,1}\left(n_1,n_2,n\right)+k_1 N_{0,1}\left(n_2,n_1,n\right)\right)+\varpi
   _{k_2,n_2} \left(\left(k_1-k_2\right) N_{0,1}\left(n_1,n_2,n\right)+k_1
   N_{1,1}\left(n_1,n_2,n\right)\right)+\varpi _{k_1,n_1} \left(\left(k_1-k_2\right)
   N_{0,1}\left(n_2,n_1,n\right)-k_2 N_{1,1}\left(n_1,n_2,n\right)\right)\right)}{8 k_1
   k_2 \varpi _{k_1-k_2,n}^2} & \text{$\Omega $B}_+\left(n,k_1-k_2\right) \\
 +0 & \text{$\Omega $A}_-\left(n,k_1+k_2\right) \\
 +0 & \text{$\Omega $A}_-\left(n,k_1-k_2\right) \\
 -\frac{\left(k_1+k_2\right) \left(N_{0,1}\left(n_1,n_2,n\right)
   \left(\left(k_1+k_2\right) \varpi _{k_2,n_2}+k_2 \varpi
   _{k_1+k_2,n}\right)+N_{0,1}\left(n_2,n_1,n\right) \left(\left(k_1+k_2\right) \varpi
   _{k_1,n_1}-k_1 \varpi _{k_1+k_2,n}\right)+N_{1,1}\left(n_1,n_2,n\right) \left(k_2 \varpi
   _{k_1,n_1}+k_1 \varpi _{k_2,n_2}\right)\right)}{8 k_1 k_2 \varpi _{k_1+k_2,n}^2} &
   \text{$\Omega $B}_-\left(n,k_1+k_2\right) \\
 +\frac{\left(k_1-k_2\right) \left(\varpi _{k_1-k_2,n} \left(k_2
   N_{0,1}\left(n_1,n_2,n\right)+k_1 N_{0,1}\left(n_2,n_1,n\right)\right)-\varpi
   _{k_2,n_2} \left(\left(k_1-k_2\right) N_{0,1}\left(n_1,n_2,n\right)+k_1
   N_{1,1}\left(n_1,n_2,n\right)\right)+\varpi _{k_1,n_1} \left(\left(k_2-k_1\right)
   N_{0,1}\left(n_2,n_1,n\right)+k_2 N_{1,1}\left(n_1,n_2,n\right)\right)\right)}{8 k_1
   k_2 \varpi _{k_1-k_2,n}^2} & \text{$\Omega $B}_-\left(n,k_1-k_2\right) \\
\end{array} \nonumber\\
\end{eqnarray}
\begin{eqnarray}
\label{table5}
  && \text{$\Omega $B}_+[n_1,k_1]\diamond \text{$\Omega $A}_+[n_2,k_2] \, = \, \nonumber\\
  &&  \begin{array}{rl}
 +0 & \text{$\Omega $A}_+\left(n,k_1+k_2\right) \\
 +0 & \text{$\Omega $A}_+\left(n,k_1-k_2\right) \\
 +\frac{\left(k_1+k_2\right) \left(N_{0,1}\left(n_1,n_2,n\right)
   \left(\left(k_1+k_2\right) \varpi _{k_2,n_2}+k_2 \varpi
   _{k_1+k_2,n}\right)-N_{0,1}\left(n_2,n_1,n\right) \left(\left(k_1+k_2\right) \varpi
   _{k_1,n_1}+k_1 \varpi _{k_1+k_2,n}\right)-N_{1,1}\left(n_1,n_2,n\right) \left(k_2 \varpi
   _{k_1,n_1}-k_1 \varpi _{k_2,n_2}\right)\right)}{8 k_1 k_2 \varpi _{k_1+k_2,n}^2} &
   \text{$\Omega $B}_+\left(n,k_1+k_2\right) \\
 +\frac{\left(k_1-k_2\right) \left(N_{0,1}\left(n_1,n_2,n\right) \left(k_2 \varpi
   _{k_1-k_2,n}+\left(k_2-k_1\right) \varpi
   _{k_2,n_2}\right)+N_{0,1}\left(n_2,n_1,n\right) \left(\left(k_1-k_2\right) \varpi
   _{k_1,n_1}+k_1 \varpi _{k_1-k_2,n}\right)-N_{1,1}\left(n_1,n_2,n\right) \left(k_2 \varpi
   _{k_1,n_1}+k_1 \varpi _{k_2,n_2}\right)\right)}{8 k_1 k_2 \varpi _{k_1-k_2,n}^2} &
   \text{$\Omega $B}_+\left(n,k_1-k_2\right) \\
 +0 & \text{$\Omega $A}_-\left(n,k_1+k_2\right) \\
 +0 & \text{$\Omega $A}_-\left(n,k_1-k_2\right) \\
 +\frac{\left(k_1+k_2\right) \left(N_{0,1}\left(n_1,n_2,n\right)
   \left(\left(k_1+k_2\right) \varpi _{k_2,n_2}-k_2 \varpi
   _{k_1+k_2,n}\right)-N_{0,1}\left(n_2,n_1,n\right) \left(\left(k_1+k_2\right) \varpi
   _{k_1,n_1}-k_1 \varpi _{k_1+k_2,n}\right)-N_{1,1}\left(n_1,n_2,n\right) \left(k_2 \varpi
   _{k_1,n_1}-k_1 \varpi _{k_2,n_2}\right)\right)}{8 k_1 k_2 \varpi _{k_1+k_2,n}^2} &
   \text{$\Omega $B}_-\left(n,k_1+k_2\right) \\
 -\frac{\left(k_1-k_2\right) \left(\varpi _{k_1-k_2,n} \left(k_2
   N_{0,1}\left(n_1,n_2,n\right)+k_1 N_{0,1}\left(n_2,n_1,n\right)\right)+\varpi
   _{k_2,n_2} \left(\left(k_1-k_2\right) N_{0,1}\left(n_1,n_2,n\right)+k_1
   N_{1,1}\left(n_1,n_2,n\right)\right)+\varpi _{k_1,n_1} \left(\left(k_2-k_1\right)
   N_{0,1}\left(n_2,n_1,n\right)+k_2 N_{1,1}\left(n_1,n_2,n\right)\right)\right)}{8 k_1
   k_2 \varpi _{k_1-k_2,n}^2} & \text{$\Omega $B}_-\left(n,k_1-k_2\right) \\
\end{array} \nonumber\\
\end{eqnarray}
\begin{eqnarray}
\label{table6}
  && \text{$\Omega $B}_+[n_1,k_1]\diamond \text{$\Omega $B}_+[n_2,k_2] \, = \, \nonumber\\
  && 
\begin{array}{rl}
 -\frac{\left(k_1+k_2\right) \left(N_{0,1}\left(n_1,n_2,n\right)
   \left(\left(k_1+k_2\right) \varpi _{k_2,n_2}+k_2 \varpi
   _{k_1+k_2,n}\right)-N_{0,1}\left(n_2,n_1,n\right) \left(\left(k_1+k_2\right) \varpi
   _{k_1,n_1}+k_1 \varpi _{k_1+k_2,n}\right)-N_{1,1}\left(n_1,n_2,n\right) \left(k_2 \varpi
   _{k_1,n_1}-k_1 \varpi _{k_2,n_2}\right)\right)}{8 k_1 k_2 \varpi _{k_1+k_2,n}^2} &
   \text{$\Omega $A}_+\left(n,k_1+k_2\right) \\
 +\frac{\left(k_1-k_2\right) \left(N_{0,1}\left(n_1,n_2,n\right) \left(k_2 \varpi
   _{k_1-k_2,n}+\left(k_2-k_1\right) \varpi
   _{k_2,n_2}\right)+N_{0,1}\left(n_2,n_1,n\right) \left(\left(k_1-k_2\right) \varpi
   _{k_1,n_1}+k_1 \varpi _{k_1-k_2,n}\right)-N_{1,1}\left(n_1,n_2,n\right) \left(k_2 \varpi
   _{k_1,n_1}+k_1 \varpi _{k_2,n_2}\right)\right)}{8 k_1 k_2 \varpi _{k_1-k_2,n}^2} &
   \text{$\Omega $A}_+\left(n,k_1-k_2\right) \\
 +0 & \text{$\Omega $B}_+\left(n,k_1+k_2\right) \\
 +0 & \text{$\Omega $B}_+\left(n,k_1-k_2\right) \\
 -\frac{\left(k_1+k_2\right) \left(N_{0,1}\left(n_1,n_2,n\right)
   \left(\left(k_1+k_2\right) \varpi _{k_2,n_2}-k_2 \varpi
   _{k_1+k_2,n}\right)-N_{0,1}\left(n_2,n_1,n\right) \left(\left(k_1+k_2\right) \varpi
   _{k_1,n_1}-k_1 \varpi _{k_1+k_2,n}\right)-N_{1,1}\left(n_1,n_2,n\right) \left(k_2 \varpi
   _{k_1,n_1}-k_1 \varpi _{k_2,n_2}\right)\right)}{8 k_1 k_2 \varpi _{k_1+k_2,n}^2} &
   \text{$\Omega $A}_-\left(n,k_1+k_2\right) \\
 -\frac{\left(k_1-k_2\right) \left(\varpi _{k_1-k_2,n} \left(k_2
   N_{0,1}\left(n_1,n_2,n\right)+k_1 N_{0,1}\left(n_2,n_1,n\right)\right)+\varpi
   _{k_2,n_2} \left(\left(k_1-k_2\right) N_{0,1}\left(n_1,n_2,n\right)+k_1
   N_{1,1}\left(n_1,n_2,n\right)\right)+\varpi _{k_1,n_1} \left(\left(k_2-k_1\right)
   N_{0,1}\left(n_2,n_1,n\right)+k_2 N_{1,1}\left(n_1,n_2,n\right)\right)\right)}{8 k_1
   k_2 \varpi _{k_1-k_2,n}^2} & \text{$\Omega $A}_-\left(n,k_1-k_2\right) \\
 +0 & \text{$\Omega $B}_-\left(n,k_1+k_2\right) \\
 +0 & \text{$\Omega $B}_-\left(n,k_1-k_2\right) \\
\end{array}
 \nonumber\\
\end{eqnarray}
\begin{eqnarray}
\label{table7}
 && \text{$\Omega $B}_+[n_1,k_1]\diamond \text{$\Omega $A}_-[n_2,k_2] \, = \, \nonumber\\
 &&  \begin{array}{rl}
 +0 & \text{$\Omega $A}_+\left(n,k_1+k_2\right) \\
 +0 & \text{$\Omega $A}_+\left(n,k_1-k_2\right) \\
 -\frac{\left(k_1+k_2\right) \left(N_{0,1}\left(n_1,n_2,n\right)
   \left(\left(k_1+k_2\right) \varpi _{k_2,n_2}-k_2 \varpi
   _{k_1+k_2,n}\right)+N_{0,1}\left(n_2,n_1,n\right) \left(\left(k_1+k_2\right) \varpi
   _{k_1,n_1}+k_1 \varpi _{k_1+k_2,n}\right)+N_{1,1}\left(n_1,n_2,n\right) \left(k_2 \varpi
   _{k_1,n_1}+k_1 \varpi _{k_2,n_2}\right)\right)}{8 k_1 k_2 \varpi _{k_1+k_2,n}^2} &
   \text{$\Omega $B}_+\left(n,k_1+k_2\right) \\
 +\frac{\left(k_1-k_2\right) \left(N_{0,1}\left(n_1,n_2,n\right) \left(k_2 \varpi
   _{k_1-k_2,n}+\left(k_1-k_2\right) \varpi
   _{k_2,n_2}\right)+N_{0,1}\left(n_2,n_1,n\right) \left(\left(k_1-k_2\right) \varpi
   _{k_1,n_1}+k_1 \varpi _{k_1-k_2,n}\right)-N_{1,1}\left(n_1,n_2,n\right) \left(k_2 \varpi
   _{k_1,n_1}-k_1 \varpi _{k_2,n_2}\right)\right)}{8 k_1 k_2 \varpi _{k_1-k_2,n}^2} &
   \text{$\Omega $B}_+\left(n,k_1-k_2\right) \\
 +0 & \text{$\Omega $A}_-\left(n,k_1+k_2\right) \\
 +0 & \text{$\Omega $A}_-\left(n,k_1-k_2\right) \\
 -\frac{\left(k_1+k_2\right) \left(N_{0,1}\left(n_1,n_2,n\right)
   \left(\left(k_1+k_2\right) \varpi _{k_2,n_2}+k_2 \varpi
   _{k_1+k_2,n}\right)+N_{0,1}\left(n_2,n_1,n\right) \left(\left(k_1+k_2\right) \varpi
   _{k_1,n_1}-k_1 \varpi _{k_1+k_2,n}\right)+N_{1,1}\left(n_1,n_2,n\right) \left(k_2 \varpi
   _{k_1,n_1}+k_1 \varpi _{k_2,n_2}\right)\right)}{8 k_1 k_2 \varpi _{k_1+k_2,n}^2} &
   \text{$\Omega $B}_-\left(n,k_1+k_2\right) \\
 -\frac{\left(k_1-k_2\right) \left(\varpi _{k_1-k_2,n} \left(k_2
   N_{0,1}\left(n_1,n_2,n\right)+k_1 N_{0,1}\left(n_2,n_1,n\right)\right)-\varpi
   _{k_2,n_2} \left(\left(k_1-k_2\right) N_{0,1}\left(n_1,n_2,n\right)+k_1
   N_{1,1}\left(n_1,n_2,n\right)\right)+\varpi _{k_1,n_1} \left(\left(k_2-k_1\right)
   N_{0,1}\left(n_2,n_1,n\right)+k_2 N_{1,1}\left(n_1,n_2,n\right)\right)\right)}{8 k_1
   k_2 \varpi _{k_1-k_2,n}^2} & \text{$\Omega $B}_-\left(n,k_1-k_2\right) \\
\end{array} \nonumber\\
\end{eqnarray}
\begin{eqnarray}
\label{table8}
  && \text{$\Omega $B}_+[n_1,k_1]\diamond \text{$\Omega $}_-[n_2,k_2] \, = \, \nonumber\\
  &&  \begin{array}{rl}
 \frac{\left(k_1+k_2\right) \left(N_{0,1}\left(n_1,n_2,n\right)
   \left(\left(k_1+k_2\right) \varpi _{k_2,n_2}-k_2 \varpi
   _{k_1+k_2,n}\right)+N_{0,1}\left(n_2,n_1,n\right) \left(\left(k_1+k_2\right) \varpi
   _{k_1,n_1}+k_1 \varpi _{k_1+k_2,n}\right)+N_{1,1}\left(n_1,n_2,n\right) \left(k_2 \varpi
   _{k_1,n_1}+k_1 \varpi _{k_2,n_2}\right)\right)}{8 k_1 k_2 \varpi _{k_1+k_2,n}^2} &
   \text{$\Omega $A}_+\left(n,k_1+k_2\right) \\
 +\frac{\left(k_1-k_2\right) \left(N_{0,1}\left(n_1,n_2,n\right) \left(k_2 \varpi
   _{k_1-k_2,n}+\left(k_1-k_2\right) \varpi
   _{k_2,n_2}\right)+N_{0,1}\left(n_2,n_1,n\right) \left(\left(k_1-k_2\right) \varpi
   _{k_1,n_1}+k_1 \varpi _{k_1-k_2,n}\right)-N_{1,1}\left(n_1,n_2,n\right) \left(k_2 \varpi
   _{k_1,n_1}-k_1 \varpi _{k_2,n_2}\right)\right)}{8 k_1 k_2 \varpi _{k_1-k_2,n}^2} &
   \text{$\Omega $A}_+\left(n,k_1-k_2\right) \\
 +0 & \text{$\Omega $B}_+\left(n,k_1+k_2\right) \\
 +0 & \text{$\Omega $B}_+\left(n,k_1-k_2\right) \\
 +\frac{\left(k_1+k_2\right) \left(N_{0,1}\left(n_1,n_2,n\right)
   \left(\left(k_1+k_2\right) \varpi _{k_2,n_2}+k_2 \varpi
   _{k_1+k_2,n}\right)+N_{0,1}\left(n_2,n_1,n\right) \left(\left(k_1+k_2\right) \varpi
   _{k_1,n_1}-k_1 \varpi _{k_1+k_2,n}\right)+N_{1,1}\left(n_1,n_2,n\right) \left(k_2 \varpi
   _{k_1,n_1}+k_1 \varpi _{k_2,n_2}\right)\right)}{8 k_1 k_2 \varpi _{k_1+k_2,n}^2} &
   \text{$\Omega $A}_-\left(n,k_1+k_2\right) \\
 -\frac{\left(k_1-k_2\right) \left(\varpi _{k_1-k_2,n} \left(k_2
   N_{0,1}\left(n_1,n_2,n\right)+k_1 N_{0,1}\left(n_2,n_1,n\right)\right)-\varpi
   _{k_2,n_2} \left(\left(k_1-k_2\right) N_{0,1}\left(n_1,n_2,n\right)+k_1
   N_{1,1}\left(n_1,n_2,n\right)\right)+\varpi _{k_1,n_1} \left(\left(k_2-k_1\right)
   N_{0,1}\left(n_2,n_1,n\right)+k_2 N_{1,1}\left(n_1,n_2,n\right)\right)\right)}{8 k_1
   k_2 \varpi _{k_1-k_2,n}^2} & \text{$\Omega $A}_-\left(n,k_1-k_2\right) \\
 +0 & \text{$\Omega $B}_-\left(n,k_1+k_2\right) \\
 +0 & \text{$\Omega $B}_-\left(n,k_1-k_2\right) \\
\end{array} \nonumber\\
\end{eqnarray}
\begin{eqnarray}
\label{table9}
  && \text{$\Omega $A}_-[n_1,k_1]\diamond \text{$\Omega $A}_+[n_2,k_2] \, = \, \nonumber\\
  &&  \begin{array}{rl}
 \frac{\left(k_1+k_2\right) \left(N_{0,1}\left(n_1,n_2,n\right)
   \left(\left(k_1+k_2\right) \varpi _{k_2,n_2}+k_2 \varpi
   _{k_1+k_2,n}\right)+N_{0,1}\left(n_2,n_1,n\right) \left(\left(k_1+k_2\right) \varpi
   _{k_1,n_1}-k_1 \varpi _{k_1+k_2,n}\right)+N_{1,1}\left(n_1,n_2,n\right) \left(k_2 \varpi
   _{k_1,n_1}+k_1 \varpi _{k_2,n_2}\right)\right)}{8 k_1 k_2 \varpi _{k_1+k_2,n}^2} &
   \text{$\Omega $A}_+\left(n,k_1+k_2\right) \\
 +\frac{\left(k_1-k_2\right) \left(\varpi _{k_1-k_2,n} \left(k_2
   N_{0,1}\left(n_1,n_2,n\right)+k_1 N_{0,1}\left(n_2,n_1,n\right)\right)-\varpi
   _{k_2,n_2} \left(\left(k_1-k_2\right) N_{0,1}\left(n_1,n_2,n\right)+k_1
   N_{1,1}\left(n_1,n_2,n\right)\right)+\varpi _{k_1,n_1} \left(\left(k_2-k_1\right)
   N_{0,1}\left(n_2,n_1,n\right)+k_2 N_{1,1}\left(n_1,n_2,n\right)\right)\right)}{8 k_1
   k_2 \varpi _{k_1-k_2,n}^2} & \text{$\Omega $A}_+\left(n,k_1-k_2\right) \\
 +0 & \text{$\Omega $B}_+\left(n,k_1+k_2\right) \\
 +0 & \text{$\Omega $B}_+\left(n,k_1-k_2\right) \\
 +\frac{\left(k_1+k_2\right) \left(N_{0,1}\left(n_1,n_2,n\right)
   \left(\left(k_1+k_2\right) \varpi _{k_2,n_2}-k_2 \varpi
   _{k_1+k_2,n}\right)+N_{0,1}\left(n_2,n_1,n\right) \left(\left(k_1+k_2\right) \varpi
   _{k_1,n_1}+k_1 \varpi _{k_1+k_2,n}\right)+N_{1,1}\left(n_1,n_2,n\right) \left(k_2 \varpi
   _{k_1,n_1}+k_1 \varpi _{k_2,n_2}\right)\right)}{8 k_1 k_2 \varpi _{k_1+k_2,n}^2} &
   \text{$\Omega $A}_-\left(n,k_1+k_2\right) \\
 -\frac{\left(k_1-k_2\right) \left(\varpi _{k_1-k_2,n} \left(k_2
   N_{0,1}\left(n_1,n_2,n\right)+k_1 N_{0,1}\left(n_2,n_1,n\right)\right)+\varpi
   _{k_2,n_2} \left(\left(k_1-k_2\right) N_{0,1}\left(n_1,n_2,n\right)+k_1
   N_{1,1}\left(n_1,n_2,n\right)\right)+\varpi _{k_1,n_1} \left(\left(k_1-k_2\right)
   N_{0,1}\left(n_2,n_1,n\right)-k_2 N_{1,1}\left(n_1,n_2,n\right)\right)\right)}{8 k_1
   k_2 \varpi _{k_1-k_2,n}^2} & \text{$\Omega $A}_-\left(n,k_1-k_2\right) \\
 +0 & \text{$\Omega $B}_-\left(n,k_1+k_2\right) \\
 +0 & \text{$\Omega $B}_-\left(n,k_1-k_2\right) \\
\end{array} \nonumber\\
\end{eqnarray}
\begin{eqnarray}
\label{table10}
  && \text{$\Omega $A}_-[n_1,k_1]\diamond \text{$\Omega $B}_+[n_2,k_2] \, = \, \nonumber\\
  &&  \begin{array}{rl}
 0 & \text{$\Omega $A}_+\left(n,k_1+k_2\right) \\
 +0 & \text{$\Omega $A}_+\left(n,k_1-k_2\right) \\
 +\frac{\left(k_1+k_2\right) \left(N_{0,1}\left(n_1,n_2,n\right)
   \left(\left(k_1+k_2\right) \varpi _{k_2,n_2}+k_2 \varpi
   _{k_1+k_2,n}\right)+N_{0,1}\left(n_2,n_1,n\right) \left(\left(k_1+k_2\right) \varpi
   _{k_1,n_1}-k_1 \varpi _{k_1+k_2,n}\right)+N_{1,1}\left(n_1,n_2,n\right) \left(k_2 \varpi
   _{k_1,n_1}+k_1 \varpi _{k_2,n_2}\right)\right)}{8 k_1 k_2 \varpi _{k_1+k_2,n}^2} &
   \text{$\Omega $B}_+\left(n,k_1+k_2\right) \\
 -\frac{\left(k_1-k_2\right) \left(\varpi _{k_1-k_2,n} \left(k_2
   N_{0,1}\left(n_1,n_2,n\right)+k_1 N_{0,1}\left(n_2,n_1,n\right)\right)-\varpi
   _{k_2,n_2} \left(\left(k_1-k_2\right) N_{0,1}\left(n_1,n_2,n\right)+k_1
   N_{1,1}\left(n_1,n_2,n\right)\right)+\varpi _{k_1,n_1} \left(\left(k_2-k_1\right)
   N_{0,1}\left(n_2,n_1,n\right)+k_2 N_{1,1}\left(n_1,n_2,n\right)\right)\right)}{8 k_1
   k_2 \varpi _{k_1-k_2,n}^2} & \text{$\Omega $B}_+\left(n,k_1-k_2\right) \\
 +0 & \text{$\Omega $A}_-\left(n,k_1+k_2\right) \\
 +0 & \text{$\Omega $A}_-\left(n,k_1-k_2\right) \\
 +\frac{\left(k_1+k_2\right) \left(N_{0,1}\left(n_1,n_2,n\right)
   \left(\left(k_1+k_2\right) \varpi _{k_2,n_2}-k_2 \varpi
   _{k_1+k_2,n}\right)+N_{0,1}\left(n_2,n_1,n\right) \left(\left(k_1+k_2\right) \varpi
   _{k_1,n_1}+k_1 \varpi _{k_1+k_2,n}\right)+N_{1,1}\left(n_1,n_2,n\right) \left(k_2 \varpi
   _{k_1,n_1}+k_1 \varpi _{k_2,n_2}\right)\right)}{8 k_1 k_2 \varpi _{k_1+k_2,n}^2} &
   \text{$\Omega $B}_-\left(n,k_1+k_2\right) \\
 +\frac{\left(k_1-k_2\right) \left(N_{0,1}\left(n_1,n_2,n\right) \left(k_2 \varpi
   _{k_1-k_2,n}+\left(k_1-k_2\right) \varpi
   _{k_2,n_2}\right)+N_{0,1}\left(n_2,n_1,n\right) \left(\left(k_1-k_2\right) \varpi
   _{k_1,n_1}+k_1 \varpi _{k_1-k_2,n}\right)-N_{1,1}\left(n_1,n_2,n\right) \left(k_2 \varpi
   _{k_1,n_1}-k_1 \varpi _{k_2,n_2}\right)\right)}{8 k_1 k_2 \varpi _{k_1-k_2,n}^2} &
   \text{$\Omega $B}_-\left(n,k_1-k_2\right) \\
\end{array} \nonumber\\
\end{eqnarray}
\begin{eqnarray}
\label{table11}
  && \text{$\Omega $A}_-[n_1,k_1]\diamond \text{$\Omega
   $A}_-[n_2,k_2] \, = \, \nonumber\\
  &&  \begin{array}{rl}
 -\frac{\left(k_1+k_2\right)
   \left(N_{0,1}\left(n_1,n_2,n\right)
   \left(\left(k_1+k_2\right) \varpi _{k_2,n_2}-k_2 \varpi
   _{k_1+k_2,n}\right)-N_{0,1}\left(n_2,n_1,n\right)
   \left(\left(k_1+k_2\right) \varpi _{k_1,n_1}-k_1 \varpi
   _{k_1+k_2,n}\right)-N_{1,1}\left(n_1,n_2,n\right)
   \left(k_2 \varpi _{k_1,n_1}-k_1 \varpi
   _{k_2,n_2}\right)\right)}{8 k_1 k_2 \varpi
   _{k_1+k_2,n}^2} & \text{$\Omega
   $A}_+\left(n,k_1+k_2\right) \\
 +\frac{\left(k_1-k_2\right) \left(\varpi _{k_1-k_2,n}
   \left(k_2 N_{0,1}\left(n_1,n_2,n\right)+k_1
   N_{0,1}\left(n_2,n_1,n\right)\right)+\varpi
   _{k_2,n_2} \left(\left(k_1-k_2\right)
   N_{0,1}\left(n_1,n_2,n\right)+k_1
   N_{1,1}\left(n_1,n_2,n\right)\right)+\varpi
   _{k_1,n_1} \left(\left(k_2-k_1\right)
   N_{0,1}\left(n_2,n_1,n\right)+k_2
   N_{1,1}\left(n_1,n_2,n\right)\right)\right)}{8
   k_1 k_2 \varpi _{k_1-k_2,n}^2} & \text{$\Omega
   $A}_+\left(n,k_1-k_2\right) \\
 +0 & \text{$\Omega $B}_+\left(n,k_1+k_2\right) \\
 +0 & \text{$\Omega $B}_+\left(n,k_1-k_2\right) \\
 -\frac{\left(k_1+k_2\right)
   \left(N_{0,1}\left(n_1,n_2,n\right)
   \left(\left(k_1+k_2\right) \varpi _{k_2,n_2}+k_2 \varpi
   _{k_1+k_2,n}\right)-N_{0,1}\left(n_2,n_1,n\right)
   \left(\left(k_1+k_2\right) \varpi _{k_1,n_1}+k_1 \varpi
   _{k_1+k_2,n}\right)-N_{1,1}\left(n_1,n_2,n\right)
   \left(k_2 \varpi _{k_1,n_1}-k_1 \varpi
   _{k_2,n_2}\right)\right)}{8 k_1 k_2 \varpi
   _{k_1+k_2,n}^2} & \text{$\Omega
   $A}_-\left(n,k_1+k_2\right) \\
 -\frac{\left(k_1-k_2\right) \left(\varpi _{k_1-k_2,n}
   \left(k_2 N_{0,1}\left(n_1,n_2,n\right)+k_1
   N_{0,1}\left(n_2,n_1,n\right)\right)-\varpi
   _{k_2,n_2} \left(\left(k_1-k_2\right)
   N_{0,1}\left(n_1,n_2,n\right)+k_1
   N_{1,1}\left(n_1,n_2,n\right)\right)+\varpi
   _{k_1,n_1} \left(\left(k_1-k_2\right)
   N_{0,1}\left(n_2,n_1,n\right)-k_2
   N_{1,1}\left(n_1,n_2,n\right)\right)\right)}{8
   k_1 k_2 \varpi _{k_1-k_2,n}^2} & \text{$\Omega
   $A}_-\left(n,k_1-k_2\right) \\
 +0 & \text{$\Omega $B}_-\left(n,k_1+k_2\right) \\
 +0 & \text{$\Omega $B}_-\left(n,k_1-k_2\right) \\
\end{array} \nonumber\\
\end{eqnarray}
\begin{eqnarray}
\label{table12}
  && \text{$\Omega $A}_-[n_1,k_1]\diamond \text{$\Omega $B}_-[n_2,k_2] \, = \, \nonumber\\
  &&  \begin{array}{rl}
 0 & \text{$\Omega $A}_+\left(n,k_1+k_2\right) \\
 +0 & \text{$\Omega $A}_+\left(n,k_1-k_2\right) \\
 -\frac{\left(k_1+k_2\right) \left(N_{0,1}\left(n_1,n_2,n\right) \left(\left(k_1+k_2\right) \varpi
   _{k_2,n_2}-k_2 \varpi _{k_1+k_2,n}\right)-N_{0,1}\left(n_2,n_1,n\right) \left(\left(k_1+k_2\right) \varpi
   _{k_1,n_1}-k_1 \varpi _{k_1+k_2,n}\right)-N_{1,1}\left(n_1,n_2,n\right) \left(k_2 \varpi _{k_1,n_1}-k_1 \varpi
   _{k_2,n_2}\right)\right)}{8 k_1 k_2 \varpi _{k_1+k_2,n}^2} & \text{$\Omega $B}_+\left(n,k_1+k_2\right)
   \\
 -\frac{\left(k_1-k_2\right) \left(\varpi _{k_1-k_2,n} \left(k_2 N_{0,1}\left(n_1,n_2,n\right)+k_1
   N_{0,1}\left(n_2,n_1,n\right)\right)+\varpi _{k_2,n_2} \left(\left(k_1-k_2\right)
   N_{0,1}\left(n_1,n_2,n\right)+k_1 N_{1,1}\left(n_1,n_2,n\right)\right)+\varpi _{k_1,n_1}
   \left(\left(k_2-k_1\right) N_{0,1}\left(n_2,n_1,n\right)+k_2
   N_{1,1}\left(n_1,n_2,n\right)\right)\right)}{8 k_1 k_2 \varpi _{k_1-k_2,n}^2} & \text{$\Omega
   $B}_+\left(n,k_1-k_2\right) \\
 +0 & \text{$\Omega $A}_-\left(n,k_1+k_2\right) \\
 +0 & \text{$\Omega $A}_-\left(n,k_1-k_2\right) \\
 -\frac{\left(k_1+k_2\right) \left(N_{0,1}\left(n_1,n_2,n\right) \left(\left(k_1+k_2\right) \varpi
   _{k_2,n_2}+k_2 \varpi _{k_1+k_2,n}\right)-N_{0,1}\left(n_2,n_1,n\right) \left(\left(k_1+k_2\right) \varpi
   _{k_1,n_1}+k_1 \varpi _{k_1+k_2,n}\right)-N_{1,1}\left(n_1,n_2,n\right) \left(k_2 \varpi _{k_1,n_1}-k_1 \varpi
   _{k_2,n_2}\right)\right)}{8 k_1 k_2 \varpi _{k_1+k_2,n}^2} & \text{$\Omega $B}_-\left(n,k_1+k_2\right)
   \\
 +\frac{\left(k_1-k_2\right) \left(N_{0,1}\left(n_1,n_2,n\right) \left(k_2 \varpi
   _{k_1-k_2,n}+\left(k_2-k_1\right) \varpi _{k_2,n_2}\right)+N_{0,1}\left(n_2,n_1,n\right)
   \left(\left(k_1-k_2\right) \varpi _{k_1,n_1}+k_1 \varpi _{k_1-k_2,n}\right)-N_{1,1}\left(n_1,n_2,n\right)
   \left(k_2 \varpi _{k_1,n_1}+k_1 \varpi _{k_2,n_2}\right)\right)}{8 k_1 k_2 \varpi _{k_1-k_2,n}^2} &
   \text{$\Omega $B}_-\left(n,k_1-k_2\right) \\
\end{array} \nonumber\\
\end{eqnarray}
\begin{eqnarray}
\label{table13}
  && \text{$\Omega $B}_-[n_1,k_1]\diamond \text{$\Omega $A}_+[n_2,k_2] \, = \, \nonumber\\
  &&  \begin{array}{rl}
 0 & \text{$\Omega $A}_+\left(n,k_1+k_2\right) \\
 +0 & \text{$\Omega $A}_+\left(n,k_1-k_2\right) \\
 +\frac{\left(k_1+k_2\right) \left(N_{0,1}\left(n_1,n_2,n\right) \left(\left(k_1+k_2\right) \varpi
   _{k_2,n_2}+k_2 \varpi _{k_1+k_2,n}\right)+N_{0,1}\left(n_2,n_1,n\right) \left(\left(k_1+k_2\right) \varpi
   _{k_1,n_1}-k_1 \varpi _{k_1+k_2,n}\right)+N_{1,1}\left(n_1,n_2,n\right) \left(k_2 \varpi _{k_1,n_1}+k_1 \varpi
   _{k_2,n_2}\right)\right)}{8 k_1 k_2 \varpi _{k_1+k_2,n}^2} & \text{$\Omega $B}_+\left(n,k_1+k_2\right)
   \\
 +\frac{\left(k_1-k_2\right) \left(\varpi _{k_1-k_2,n} \left(k_2 N_{0,1}\left(n_1,n_2,n\right)+k_1
   N_{0,1}\left(n_2,n_1,n\right)\right)-\varpi _{k_2,n_2} \left(\left(k_1-k_2\right)
   N_{0,1}\left(n_1,n_2,n\right)+k_1 N_{1,1}\left(n_1,n_2,n\right)\right)+\varpi _{k_1,n_1}
   \left(\left(k_2-k_1\right) N_{0,1}\left(n_2,n_1,n\right)+k_2
   N_{1,1}\left(n_1,n_2,n\right)\right)\right)}{8 k_1 k_2 \varpi _{k_1-k_2,n}^2} & \text{$\Omega
   $B}_+\left(n,k_1-k_2\right) \\
 +0 & \text{$\Omega $A}_-\left(n,k_1+k_2\right) \\
 +0 & \text{$\Omega $A}_-\left(n,k_1-k_2\right) \\
 +\frac{\left(k_1+k_2\right) \left(N_{0,1}\left(n_1,n_2,n\right) \left(\left(k_1+k_2\right) \varpi
   _{k_2,n_2}-k_2 \varpi _{k_1+k_2,n}\right)+N_{0,1}\left(n_2,n_1,n\right) \left(\left(k_1+k_2\right) \varpi
   _{k_1,n_1}+k_1 \varpi _{k_1+k_2,n}\right)+N_{1,1}\left(n_1,n_2,n\right) \left(k_2 \varpi _{k_1,n_1}+k_1 \varpi
   _{k_2,n_2}\right)\right)}{8 k_1 k_2 \varpi _{k_1+k_2,n}^2} & \text{$\Omega $B}_-\left(n,k_1+k_2\right)
   \\
 -\frac{\left(k_1-k_2\right) \left(\varpi _{k_1-k_2,n} \left(k_2 N_{0,1}\left(n_1,n_2,n\right)+k_1
   N_{0,1}\left(n_2,n_1,n\right)\right)+\varpi _{k_2,n_2} \left(\left(k_1-k_2\right)
   N_{0,1}\left(n_1,n_2,n\right)+k_1 N_{1,1}\left(n_1,n_2,n\right)\right)+\varpi _{k_1,n_1}
   \left(\left(k_1-k_2\right) N_{0,1}\left(n_2,n_1,n\right)-k_2
   N_{1,1}\left(n_1,n_2,n\right)\right)\right)}{8 k_1 k_2 \varpi _{k_1-k_2,n}^2} & \text{$\Omega
   $B}_-\left(n,k_1-k_2\right) \\
\end{array} \nonumber\\
\end{eqnarray}
\begin{eqnarray}
\label{table14}
  && \text{$\Omega $B}_-[n_1,k_1]\diamond \text{$\Omega $B}_+[n_2,k_2] \, = \, \nonumber\\
  &&  \begin{array}{rl}
 -\frac{\left(k_1+k_2\right) \left(N_{0,1}\left(n_1,n_2,n\right) \left(\left(k_1+k_2\right) \varpi
   _{k_2,n_2}+k_2 \varpi _{k_1+k_2,n}\right)+N_{0,1}\left(n_2,n_1,n\right) \left(\left(k_1+k_2\right) \varpi
   _{k_1,n_1}-k_1 \varpi _{k_1+k_2,n}\right)+N_{1,1}\left(n_1,n_2,n\right) \left(k_2 \varpi _{k_1,n_1}+k_1 \varpi
   _{k_2,n_2}\right)\right)}{8 k_1 k_2 \varpi _{k_1+k_2,n}^2} & \text{$\Omega $A}_+\left(n,k_1+k_2\right)
   \\
 +\frac{\left(k_1-k_2\right) \left(\varpi _{k_1-k_2,n} \left(k_2 N_{0,1}\left(n_1,n_2,n\right)+k_1
   N_{0,1}\left(n_2,n_1,n\right)\right)-\varpi _{k_2,n_2} \left(\left(k_1-k_2\right)
   N_{0,1}\left(n_1,n_2,n\right)+k_1 N_{1,1}\left(n_1,n_2,n\right)\right)+\varpi _{k_1,n_1}
   \left(\left(k_2-k_1\right) N_{0,1}\left(n_2,n_1,n\right)+k_2
   N_{1,1}\left(n_1,n_2,n\right)\right)\right)}{8 k_1 k_2 \varpi _{k_1-k_2,n}^2} & \text{$\Omega
   $A}_+\left(n,k_1-k_2\right) \\
 +0 & \text{$\Omega $B}_+\left(n,k_1+k_2\right) \\
 +0 & \text{$\Omega $B}_+\left(n,k_1-k_2\right) \\
 -\frac{\left(k_1+k_2\right) \left(N_{0,1}\left(n_1,n_2,n\right) \left(\left(k_1+k_2\right) \varpi
   _{k_2,n_2}-k_2 \varpi _{k_1+k_2,n}\right)+N_{0,1}\left(n_2,n_1,n\right) \left(\left(k_1+k_2\right) \varpi
   _{k_1,n_1}+k_1 \varpi _{k_1+k_2,n}\right)+N_{1,1}\left(n_1,n_2,n\right) \left(k_2 \varpi _{k_1,n_1}+k_1 \varpi
   _{k_2,n_2}\right)\right)}{8 k_1 k_2 \varpi _{k_1+k_2,n}^2} & \text{$\Omega $A}_-\left(n,k_1+k_2\right)
   \\
 -\frac{\left(k_1-k_2\right) \left(\varpi _{k_1-k_2,n} \left(k_2 N_{0,1}\left(n_1,n_2,n\right)+k_1
   N_{0,1}\left(n_2,n_1,n\right)\right)+\varpi _{k_2,n_2} \left(\left(k_1-k_2\right)
   N_{0,1}\left(n_1,n_2,n\right)+k_1 N_{1,1}\left(n_1,n_2,n\right)\right)+\varpi _{k_1,n_1}
   \left(\left(k_1-k_2\right) N_{0,1}\left(n_2,n_1,n\right)-k_2
   N_{1,1}\left(n_1,n_2,n\right)\right)\right)}{8 k_1 k_2 \varpi _{k_1-k_2,n}^2} & \text{$\Omega
   $A}_-\left(n,k_1-k_2\right) \\
 +0 & \text{$\Omega $B}_-\left(n,k_1+k_2\right) \\
 +0 & \text{$\Omega $B}_-\left(n,k_1-k_2\right) \\
\end{array} \nonumber\\
\end{eqnarray}
\begin{eqnarray}
\label{table15}
  && \text{$\Omega $B}_-[n_1,k_1]\diamond \text{$\Omega $A}_-[n_2,k_2] \, = \, \nonumber\\
  &&  \begin{array}{rl}
 0 & \text{$\Omega $A}_+\left(n,k_1+k_2\right) \\
 +0 & \text{$\Omega $A}_+\left(n,k_1-k_2\right) \\
 -\frac{\left(k_1+k_2\right) \left(N_{0,1}\left(n_1,n_2,n\right) \left(\left(k_1+k_2\right) \varpi
   _{k_2,n_2}-k_2 \varpi _{k_1+k_2,n}\right)-N_{0,1}\left(n_2,n_1,n\right) \left(\left(k_1+k_2\right) \varpi
   _{k_1,n_1}-k_1 \varpi _{k_1+k_2,n}\right)-N_{1,1}\left(n_1,n_2,n\right) \left(k_2 \varpi _{k_1,n_1}-k_1 \varpi
   _{k_2,n_2}\right)\right)}{8 k_1 k_2 \varpi _{k_1+k_2,n}^2} & \text{$\Omega $B}_+\left(n,k_1+k_2\right)
   \\
 +\frac{\left(k_1-k_2\right) \left(\varpi _{k_1-k_2,n} \left(k_2 N_{0,1}\left(n_1,n_2,n\right)+k_1
   N_{0,1}\left(n_2,n_1,n\right)\right)+\varpi _{k_2,n_2} \left(\left(k_1-k_2\right)
   N_{0,1}\left(n_1,n_2,n\right)+k_1 N_{1,1}\left(n_1,n_2,n\right)\right)+\varpi _{k_1,n_1}
   \left(\left(k_2-k_1\right) N_{0,1}\left(n_2,n_1,n\right)+k_2
   N_{1,1}\left(n_1,n_2,n\right)\right)\right)}{8 k_1 k_2 \varpi _{k_1-k_2,n}^2} & \text{$\Omega
   $B}_+\left(n,k_1-k_2\right) \\
 +0 & \text{$\Omega $A}_-\left(n,k_1+k_2\right) \\
 +0 & \text{$\Omega $A}_-\left(n,k_1-k_2\right) \\
 -\frac{\left(k_1+k_2\right) \left(N_{0,1}\left(n_1,n_2,n\right) \left(\left(k_1+k_2\right) \varpi
   _{k_2,n_2}+k_2 \varpi _{k_1+k_2,n}\right)-N_{0,1}\left(n_2,n_1,n\right) \left(\left(k_1+k_2\right) \varpi
   _{k_1,n_1}+k_1 \varpi _{k_1+k_2,n}\right)-N_{1,1}\left(n_1,n_2,n\right) \left(k_2 \varpi _{k_1,n_1}-k_1 \varpi
   _{k_2,n_2}\right)\right)}{8 k_1 k_2 \varpi _{k_1+k_2,n}^2} & \text{$\Omega $B}_-\left(n,k_1+k_2\right)
   \\
 -\frac{\left(k_1-k_2\right) \left(\varpi _{k_1-k_2,n} \left(k_2 N_{0,1}\left(n_1,n_2,n\right)+k_1
   N_{0,1}\left(n_2,n_1,n\right)\right)-\varpi _{k_2,n_2} \left(\left(k_1-k_2\right)
   N_{0,1}\left(n_1,n_2,n\right)+k_1 N_{1,1}\left(n_1,n_2,n\right)\right)+\varpi _{k_1,n_1}
   \left(\left(k_1-k_2\right) N_{0,1}\left(n_2,n_1,n\right)-k_2
   N_{1,1}\left(n_1,n_2,n\right)\right)\right)}{8 k_1 k_2 \varpi _{k_1-k_2,n}^2} & \text{$\Omega
   $B}_-\left(n,k_1-k_2\right) \\
\end{array} \nonumber\\
\end{eqnarray}
\begin{eqnarray}
\label{table16}
  && \text{$\Omega $B}_-[n_1,k_1]\diamond \text{$\Omega $B}_-[n_2,k_2]\, = \, \nonumber\\
  &&  \begin{array}{rl}
 \frac{\left(k_1+k_2\right) \left(N_{0,1}\left(n_1,n_2,n\right) \left(\left(k_1+k_2\right) \varpi
   _{k_2,n_2}-k_2 \varpi _{k_1+k_2,n}\right)-N_{0,1}\left(n_2,n_1,n\right) \left(\left(k_1+k_2\right) \varpi
   _{k_1,n_1}-k_1 \varpi _{k_1+k_2,n}\right)-N_{1,1}\left(n_1,n_2,n\right) \left(k_2 \varpi _{k_1,n_1}-k_1 \varpi
   _{k_2,n_2}\right)\right)}{8 k_1 k_2 \varpi _{k_1+k_2,n}^2} & \text{$\Omega $A}_+\left(n,k_1+k_2\right)
   \\
 \frac{\left(k_1-k_2\right) \left(\varpi _{k_1-k_2,n} \left(k_2 N_{0,1}\left(n_1,n_2,n\right)+k_1
   N_{0,1}\left(n_2,n_1,n\right)\right)+\varpi _{k_2,n_2} \left(\left(k_1-k_2\right)
   N_{0,1}\left(n_1,n_2,n\right)+k_1 N_{1,1}\left(n_1,n_2,n\right)\right)+\varpi _{k_1,n_1}
   \left(\left(k_2-k_1\right) N_{0,1}\left(n_2,n_1,n\right)+k_2
   N_{1,1}\left(n_1,n_2,n\right)\right)\right)}{8 k_1 k_2 \varpi _{k_1-k_2,n}^2} & \text{$\Omega
   $A}_+\left(n,k_1-k_2\right) \\
 0 & \text{$\Omega $B}_+\left(n,k_1+k_2\right) \\
 0 & \text{$\Omega $B}_+\left(n,k_1-k_2\right) \\
 \frac{\left(k_1+k_2\right) \left(N_{0,1}\left(n_1,n_2,n\right) \left(\left(k_1+k_2\right) \varpi
   _{k_2,n_2}+k_2 \varpi _{k_1+k_2,n}\right)-N_{0,1}\left(n_2,n_1,n\right) \left(\left(k_1+k_2\right) \varpi
   _{k_1,n_1}+k_1 \varpi _{k_1+k_2,n}\right)-N_{1,1}\left(n_1,n_2,n\right) \left(k_2 \varpi _{k_1,n_1}-k_1 \varpi
   _{k_2,n_2}\right)\right)}{8 k_1 k_2 \varpi _{k_1+k_2,n}^2} & \text{$\Omega $A}_-\left(n,k_1+k_2\right)
   \\
 -\frac{\left(k_1-k_2\right) \left(\varpi _{k_1-k_2,n} \left(k_2 N_{0,1}\left(n_1,n_2,n\right)+k_1
   N_{0,1}\left(n_2,n_1,n\right)\right)-\varpi _{k_2,n_2} \left(\left(k_1-k_2\right)
   N_{0,1}\left(n_1,n_2,n\right)+k_1 N_{1,1}\left(n_1,n_2,n\right)\right)+\varpi _{k_1,n_1}
   \left(\left(k_1-k_2\right) N_{0,1}\left(n_2,n_1,n\right)-k_2
   N_{1,1}\left(n_1,n_2,n\right)\right)\right)}{8 k_1 k_2 \varpi _{k_1-k_2,n}^2} & \text{$\Omega
   $A}_-\left(n,k_1-k_2\right) \\
 0 & \text{$\Omega $B}_-\left(n,k_1+k_2\right) \\
 0 & \text{$\Omega $B}_-\left(n,k_1-k_2\right) \\
\end{array} \nonumber\\
\end{eqnarray}
\end{landscape}
\section{A couple of examples of the ri-expansion of $\mathcal{G}^{[n]}_{0,1}(r)$ products in the same basis}
\label{impressiveprod}
As it was discussed in the main body of the paper the relevant
radial integrals are those displayed in eq.s(\ref{calendula}).
Here we want to illustrate the use of the integrals in order to re-expand the two relevant product cases in the necessary functions, namely:
\begin{eqnarray}
 \mathcal{G}_0^{[n_1]}(r)\, \mathcal{G}_1^{[n_2]}(r) &=& \sum_{n=1}^{\infty} \, c_{01}[n] \, \mathcal{G}_1^{[n]}(r)  \label{01case}\\
  \mathcal{G}_1^{[n_1]}(r)\, \mathcal{G}_1^{[n_2]}(r) &=& \sum_{n=1}^{\infty} \, c_{11}[n] \, \mathcal{G}_0^{[n]}(r) 
  \label{11case} 
\end{eqnarray}
Using the result in eq.(\ref{concistoro}) for the norms of the two type of functions we have:
\begin{eqnarray}
\label{bisacca}
  c_{01}[n] &=& \frac{1}{2} \, \int_{0}^{1} \, \mathrm{dr} \,r \,\mathcal{G}_0^{[n_1]}(r)\, \mathcal{G}_1^{[n_2]}(r) \, \mathcal{G}_1^{[n]}(r) \nonumber\\
  c_{11}[n] &=& \frac{2}{j^2_{1,n}} \, \int_{0}^{1} \, \mathrm{dr} \, r \,\mathcal{G}_1^{[n_1]}(r)\, \mathcal{G}_1^{[n_2]}(r) \, \mathcal{G}_0^{[n]}(r) 
\end{eqnarray}
We illustrate the rapid convergence of the series in eq.s (\ref{01case},\ref{11case}) and the experimental fact that we always reach an excellent approximation when truncating the series at the order $n=n_1+n_2$ with two examples, one for each of the two cases
\subsection{Case $01$}
\begin{figure}[!hbt]
\begin{center}
\includegraphics[width=70mm]{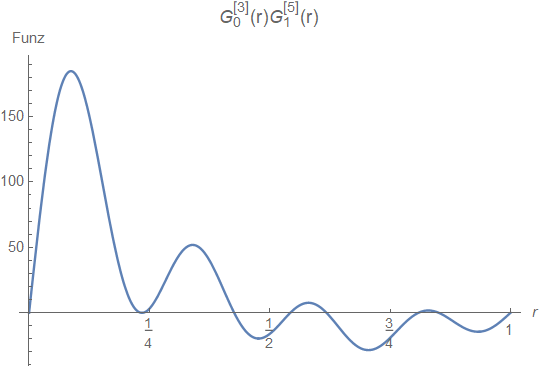}
\includegraphics[width=70mm]{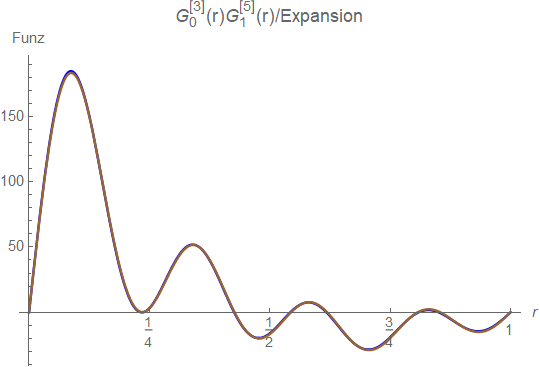}
\caption{\it In this figure I compare the  plot of the product
 $\mathcal{G}_0^{[3]}(r) \,\mathcal{G}_1^{[5]}(r)$ with that of its truncated expansion (\ref{robinia}). On the left the plot only of the product. On the right the product and the expansion compared. They overlap almost completely.
} \label{pansecco}
\end{center}
\end{figure}

As an illustration of this case, we choose $n_1=3,n_2=5$. The truncation order is therefore $n=8$. Calculating numerically the integrals we obtain:
\begin{equation}\label{agilulfo}
  c_{01}[n] \, = \, \underbrace{\{0.212722,8.25571,4.5796,3.39697,3.50714,3.62814,4.99722,8.
   56999\}}_{n=1\,\dots\,8}
\end{equation}
We define the truncated series
\begin{equation}\label{robinia}
  X_{01}^{3|5}(r) \, =\,\sum_{n=1}^{8} \,c_{01}[n]\,\mathcal{G}_1^{[n]}(r)  
\end{equation}
and in fig.\ref{pansecco} we compare the plot of the product function with its truncated expansion revealing an impressive precision.
\subsection{Case $11$}
\begin{figure}[!hbt]
\begin{center}
\includegraphics[width=70mm]{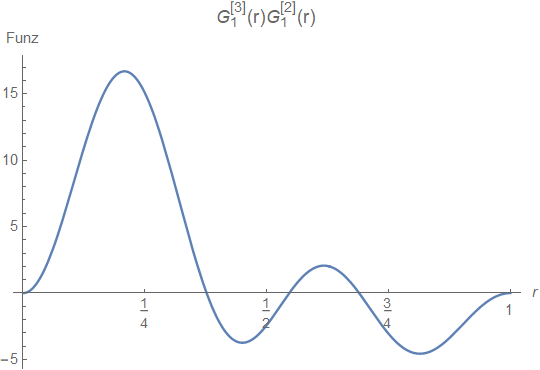}
\includegraphics[width=70mm]{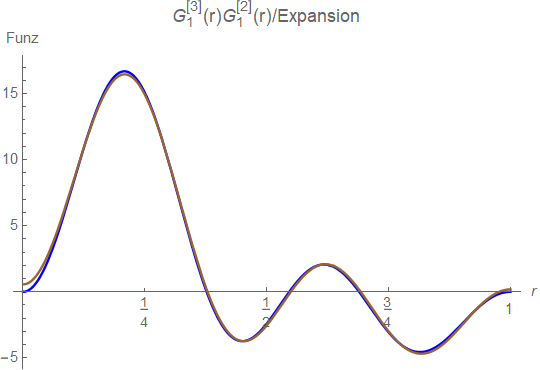}
\caption{\it In this figure I compare the  plot of the product
 $\mathcal{G}_1^{[3]}(r) \,\mathcal{G}_1^{[2]}(r)$ with that of its truncated expansion (\ref{gardenia}). On the left the plot only of the product. On the right the product and the expansion compared. They overlap almost completely.
} \label{polenta}
\end{center}
\end{figure}

As an illustration of this case, we choose $n_1=3,n_2=2$. The truncation order is therefore $n=5$. Calculating numerically the integrals we obtain:
\begin{equation}\label{desiderio}
  c_{11}[n] \, = \, \underbrace{\{0.886938,0.254267,0.0654351,-0.0306236,-0.174328\}}_{n=1\,\dots\,5}
\end{equation}
We define the truncated series
\begin{equation}\label{gardenia}
  X_{11}^{3|2}(r) \, =\,\sum_{n=1}^{5} \,c_{11}[n]\,\mathcal{G}_0^{[n]}(r)  
\end{equation}
and in fig.\ref{polenta} we compare the plot of the product function with its truncated expansion revealing once again an impressive precision.
The result is so good that I asked myself whether it could be proven analytically but so far I could not find an argument. In any case the result is very good, but not exact because the further coefficients although drastically smaller are not zero.
%%%%%%%%%%%%%%%%%%%%%%%%%%%

\end{document}